\documentclass[10pt,journal]{IEEEtran}
\usepackage{amscd}
\usepackage{amssymb}
\usepackage{cite}
\usepackage{subfigure}
\usepackage[cmex10]{amsmath}
\usepackage{algpseudocode}
\usepackage{amsmath}
\usepackage{pifont}
\usepackage{color}
\usepackage{enumerate}
\usepackage{xcolor}
\usepackage{multirow}
\usepackage{makecell}
\usepackage{colortbl,booktabs}
\usepackage{url}
\usepackage[ruled,vlined]{algorithm2e}

\ifCLASSINFOpdf
  \usepackage[pdftex]{graphicx}
  \graphicspath{{./pic/}{./jpeg/}}
  \DeclareGraphicsExtensions{.pdf,.jpeg,.png}
\else
  \usepackage[dvips]{graphicx}
 \graphicspath{{./eps/}}
  \DeclareGraphicsExtensions{.eps}
\fi

\usepackage{array}
\usepackage{subfigure}

\begin{document}

\title{Leveraging Coupled BBR and Adaptive Packet Scheduling to Boost MPTCP}

\author{Jiangping Han,
        ~Kaiping Xue,~\IEEEmembership{Senior Member,~IEEE,}
        ~Yitao Xing,
        ~Jian Li,
        ~Wenjia Wei,
        ~David S.L. Wei,~\IEEEmembership{Senior Member,~IEEE,}
        ~Guoliang Xue,~\IEEEmembership{Fellow,~IEEE}

\thanks{J. Han, K. Xue and W. Wei are with the Department of Electronic Engineering and Information Science, University of Science and
Technology of China, Hefei 230027, China.}
\thanks{Y. Xing and J. Li are with the School of Cyber Security, University of Science and Technology of China, Hefei 230027, China.}
\thanks{K. Xue is also with the School of Cyber Security, University of Science and Technology of China, Hefei 230027, China.}
\thanks{D. Wei is with the Computer and Information Science Department, Fordham University, Bronx 10458, USA.}
\thanks{G. Xue is with the School of Computing, Informatics, and Decision Systems  Engineering, Arizona State University, Tempe, AZ 85287, USA.}
\thanks{Corresponding Author: K. Xue (e-mail: kpxue@ustc.edu.cn).}
}

\maketitle

\begin{abstract}
Multipath TCP (MPTCP) utilizes multiple paths for simultaneous data transmission to enhance performance. However, existing MPTCP protocols are still far from satisfactory in wireless networks because of their loss-based congestion control and the difficulty of managing multiple subflows. To overcome these problems, we redesign the coupled congestion control algorithm and scheduler to boost MPTCP in wireless heterogeneous networks. The main purpose is to promote transmission rate under lossy networks, while also provide stability when networks suffer physical link changes and asymmetric links. In this paper, inspired by Bottleneck Bandwidth and Round-trip propagation time (BBR), we first propose Coupled BBR that utilizes detected bandwidth to adjust the sending rate within an MPTCP connection. Coupled BBR provides high loss tolerance as well as balanced congestion among MPTCP subflows. Then, to further improve the performance, we propose an Adaptively Redundant and Predictive packet (AR\&P) scheduler to improve adaptability and keep in-order packet delivery in highly dynamic network scenarios. Based on Linux kernel implementation and experiments in both testbed and real network scenarios, we show that the proposed scheme not only provides high throughput in wireless networks, but also improves robustness and reduces out-of-order packets in some harsh circumstances.
\end{abstract}

\begin{IEEEkeywords}
 MPTCP, Congestion control, Scheduler, Wireless networks.
\end{IEEEkeywords}

\IEEEpeerreviewmaketitle

\section{Introduction}

Multipath TCP (MPTCP) \cite{ietf-mptcp-rfc6824bis-12} is an emerging transport protocol, which enables the full use of the device's multiple interfaces and transmits data via multiple paths concurrently \cite{raiciu2011improving,li2016multipath}. MPTCP establishes subflows on available paths such that each subflow acts as a separate TCP flow. Based on TCP, MPTCP aims at providing higher transmission efficiency, stronger robustness, and better mobility support \cite{shi2018stms}. Till now,  MPTCP has got some deployment \cite{rfc8041} in real networks, and there have been some devices and applications, such as Apple Siri~\cite{zhang2018research}, in support of MPTCP.

Several schemes such as coupled congestion control algorithms \cite{wischik2011design, raiciu2011rfc, ferlin2016revisiting, xu2016congestion, sinky2016proactive} and scheduling algorithms \cite{choi2017optimal, guo2017accelerating, saha2019musher, lee2018raven} have been proposed to make MPTCP more practical. However, MPTCP is still not able to achieve the desired performance in wireless networks, which have a large number of random packet loss and rapidly changing link conditions. On the one hand, traditional loss-based congestion control algorithms can hardly make the best use of the full available bandwidth in lossy networks \cite{nikravesh2016depth}. On the other hand, a fixed scheduler can not meet the ever-changing network conditions, where the unpredictable degradation in a single subflow may severely degrade the performance of other subflows in an MPTCP connection \cite{pokhrel2017analytical, ferlin2018mptcp}. Based on these facts, it is hard to achieve satisfactory end-to-end transmission performance for MPTCP \cite{ li2018measurement}.

Coupled congestion control algorithms (e.g., LIA, OLIA, BALIA) in MPTCP have been designed based on traditional TCP congestion control algorithm (for example, NewReno \cite{kumar1998comparative}) and treat packet loss as an indicator of congestion and decrease their congestion window when packet loss occurs. In today's network environment where wireless links are used frequently and random packet loss caused by physical links is common, it is hard for MPTCP to achieve the desired performance. MPTCP needs to change its way for transmission control for better performance. Among some state-of-art congestion control algorithms \cite{ha2008cubic, cao2019use}, BBR shows its potential in lossy scenarios, which can make the best use of available bandwidth even when there is random packet loss. Inspired by BBR, we design a novel coupled congestion control algorithm and propose a customized scheduler for it. In this work, we mainly focus on two issues: 1) \textbf{Promote MPTCP in wireless lossy networks}, as well as provide high lossy tolerance and achieve fairness and balanced congestion, and 2) \textbf{Further improve the transmission efficiency and stability of MPTCP in ever-changing and asymmetric networks} by designing a customized scheduler that suitable for the novel congestion control algorithm that precisely controls the multipath transmission.

We first design a novel coupled congestion control algorithm for MPTCP, called Coupled BBR, which is based on TCP BBR but is modified for MPTCP to achieve better performance. Coupled BBR follows the same mechanism of periodic bandwidth detection in convention BBR to provide high bandwidth utilization. In order to achieve the goals of fairness and balanced congestion for MPTCP defined in RFCs \cite{raiciu2011rfc, ietf-mptcp-rfc6824bis-12}, Coupled BBR sets the sending rate of each subflow differently. RFC 6356 \cite{raiciu2011rfc} points out that running an uncoupled congestion control algorithm on each subflow makes an MPTCP flow unfairly take up more capacity compared with a single path TCP flow, which means aggregated bandwidth of MPTCP should be no more aggressive than that of a single path TCP flow on the best available path. To achieve this goal, which is different from previous algorithms \cite{wischik2011design, wei2020shared, thomas2020low} that modify the increase function of the Additive Increase Multiplicative Decrease (AIMD) \cite{yang2000general} scheme, Coupled BBR utilizes its measured bandwidth of all subflows to control each subflow's sending rate and achieve fairness to single-path TCP BBR flows. Besides, Coupled BBR also utilizes a data allocation rate based on the bandwidth measurement results, therefore it can better balance congestion among subflows.

 Secondly, based on the real-time measurement and steady sending rate of Coupled BBR, we propose an Adaptively Redundant and Predictive packet (AR\&P) scheduler to enhance MPTCP performance in highly dynamic and asymmetric networks. Two scheduling methods are included in AR\&P scheduler, 1) Adaptively Redundant Scheduling (AR-Scheduling), and 2) Predictive packet Scheduling (P-Scheduling). AR-Scheduling is designed to achieve high goodput and low latency in different network scenarios, and provides better adaptability in highly dynamic scenarios. It adaptively decides whether to send redundant packets on each subflow according to the real-time path conditions. By sending redundant packets on subflows with low bandwidth and high RTT, AR-Scheduling is able to provide better flexibility when the network environment changes rapidly. Besides, P-Scheduling is designed to reduce out-of-order packets in asymmetric networks. Different from previous packet schedulers which only schedule in each congestion window, P-Scheduling calculates the arrival time of packets and schedules each packet one-by-one. Taking the advantages of Coupled BBR's steady sending rate and smooth transmission, P-Scheduling could accurately control the arrival time of each packet, thereby reducing out-of-order packets significantly.

 To summarize, in this paper we present Coupled BBR and AR\&P Scheduler for MPTCP. With our scheme, the performance of MPTCP is enhanced in lossy, dynamic, and asymmetric networks. The main contributions of this paper are as follows:

\begin{itemize}
  \item We propose Coupled BBR as a coupled congestion control algorithm for MPTCP to obtain better performance in wireless lossy networks. Coupled BBR provides high bandwidth utilization and stable sending rate, while also achieving fairness to TCP BBR flows and balancing congestion among MPTCP subflows.
  \item Based on Coupled BBR, AR\&P Scheduler is proposed to further help MPTCP for managing multipath transmission. It includes two scheduling methods: 1) AR-Scheduling automatically chooses whether to send redundant packets according to real-time path conditions, in order to to provide better adaptability in highly dynamic networks. 2) P-Scheduling schedules each packet according to its arrival time, which keeps packets arriving in order and reduces out-of-order packets in asymmetric networks.
  \item Coupled BBR and AR\&P Scheduler are implemented in MPTCP Linux kernel v0.94 \cite{linuxkernel}  and tested in both testbed and real networks. Extensive results show that our scheme gives MPTCP a higher elasticity, making it more feasible in today's networks.
\end{itemize}

The rest of this paper is organized as follows: Section~\ref{Motivation} introduces the background and motivation of our work. We present our design and the details of each algorithm in Section \ref{design}, Section \ref{cbbr}, and Section \ref{AR-P}. The implementation and evaluation are shown in Section \ref{evaluation}. Sections \ref{Related} and \ref{Discussion} show the related work and the discussion of our work, respectively. Finally, Section \ref{conclusion} draws the conclusion.

\section{Background and Motivation}\label{Motivation}

  We first take a brief overview of MPTCP and BBR. Then, we discuss the opportunities and challenges that BBR brings to MPTCP.

\subsection{Overview of MPTCP}

 MPTCP is a multipath transport protocol proposed by IETF \cite{ietf-mptcp-rfc6824bis-12}. As an extension of TCP, it provides reliable transmission service, while also enables multipath transmission to gain better performance. MPTCP inherits the drawbacks of conventional TCP, which are mainly caused by traditional loss-based congestion control algorithms. They treat packet loss as a signal of congestion and halve the congestion window when packet loss occurs, which leads to poor performance and causes fluctuation of sending rate in lossy scenarios such as wireless networks \cite{abouzeid2003comprehensive, zhang2019will, singh2007tcp}.

  Besides, there are some new issues introduced by multipath transmission in MPTCP. MPTCP needs to be friendly to TCP flows, which means an MPTCP flow should not be more aggressive than a single-path TCP flow on the best path \cite{raiciu2011rfc}. Also, MPTCP needs to balance congestion, which means to migrate data from congested subflows to less congested ones \cite{oh2016feedback}. Moreover, MPTCP should achieve stronger robustness. When some subflows fail, MPTCP is supported to keep running since it can transfer data on other available subflows. Meanwhile, MPTCP also needs to reduce out-of-order packets, which is caused by different RTTs among subflows in asymmetric networks.

\subsection{Overview of BBR} \label{bbroverview}

  Different from traditional loss-based congestion control, BBR measures the bandwidth and RTT of the bottleneck which a flow goes through \cite{cardwell2017bbr}. Then based on the measurement, it adjusts the sending rate to make the best use of the bottleneck bandwidth. BBR keeps high throughput in lossy networks and maintains a smooth rate during the transmission. Through other popular congestion control algorithms like Cubic make a faster recovery for high throughput in lossy scenarios, they create fluctuating sending rate, and provide much worse performance than BBR when suffering high loss rate \cite{cao2019use}. Additionally, the use of BBR stops creating queues in the network, thereby reducing RTT and leading to low transmission delay.

 \begin{figure}[!htb]
  \centering
  \includegraphics[width=0.9\linewidth]{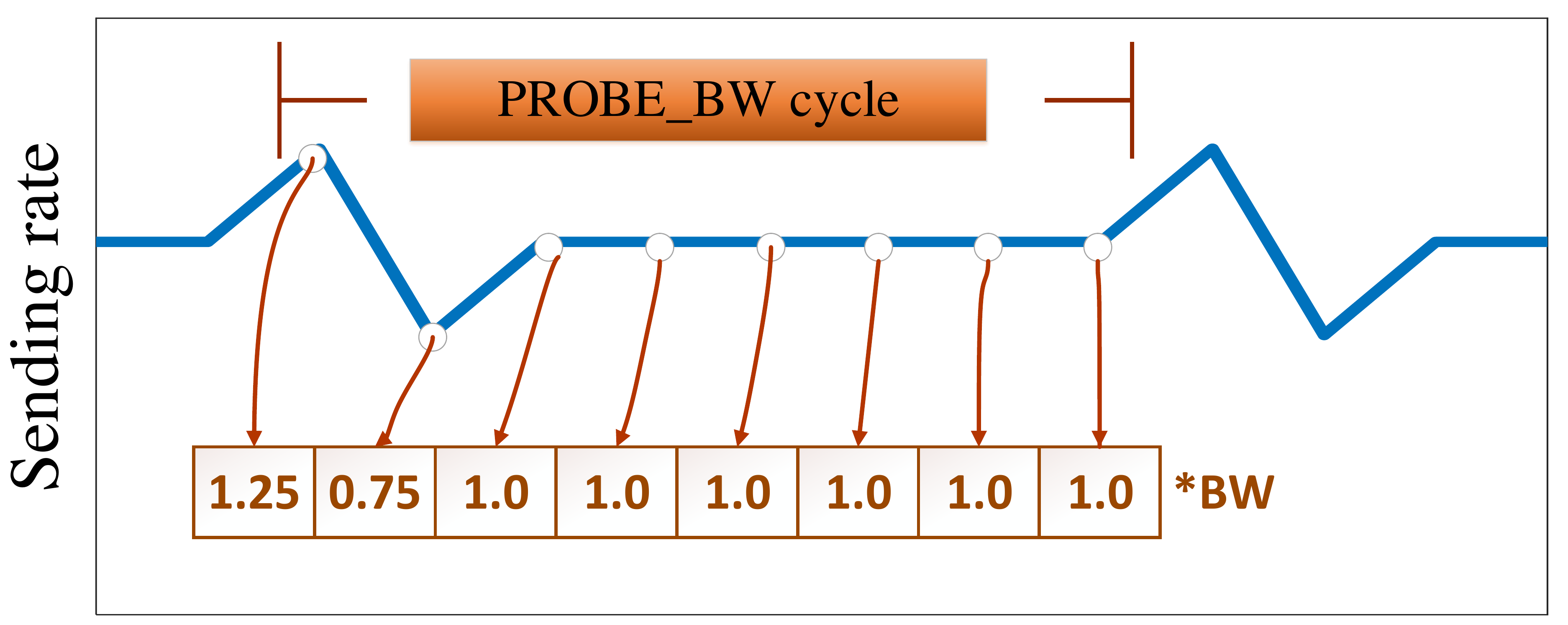}\\
  \caption{Sending rate of BBR in RROBE\_BW phase. }\label{bbr}
\end{figure}

  Specifically, BBR periodically measures bottleneck bandwidth and adjusts the transmission rate at its  PROBE$\_$BW phase, which accounts for the vast majority (i.e., almost 98\%) of its running time \cite{cardwell2017bbr}. As shown in Fig.~\ref{bbr}, BBR treats 8 RTTs as a cycle during the PROBE$\_$BW phase. In each RTT of a cycle, BBR sends data as a rate of $pacing\_rate = pacing\_gain \cdot BW$, where $pacing\_gain$ = (1.25, 0.75, 1.00, 1.00, 1.00, 1.00, 1.00, 1.00) in each RTT respectively. In this state, $BW$ is the maximum measured value of delivery rate during a period of time, which is noted as an estimated result of bottleneck bandwidth. During the first RTT, BBR increases the sending rate to $1.25 \cdot BW$ to probe the remaining available bandwidth, and during the second RTT, it reduces the rate to $0.75 \cdot BW$ to drain the queues that may be created in the previous RTT. After the first two RTTs, BBR keeps sending data smoothly using the detected bandwidth for 6 RTTs. In this process, the congestion window ($cwnd$) is no longer the deciding factor, it is $pacing\_rate$ instead. BBR sets the interval time between two packets to $packet\_size/pacing\_rate$ so as to control the sending rate and keep the transmission smooth. For each 10 s, BBR goes through a PROBE$\_$RTT phase, keeps inflight to 4 for $\max (RTT, 0.2 sec)$ to probe minimum RTT of the path.

\subsection{MPTCP over BBR: Opportunities and Challenges} \label{mptcpbbr}

\begin{figure}[!htb]
  \centering
  \includegraphics[width=0.9\linewidth]{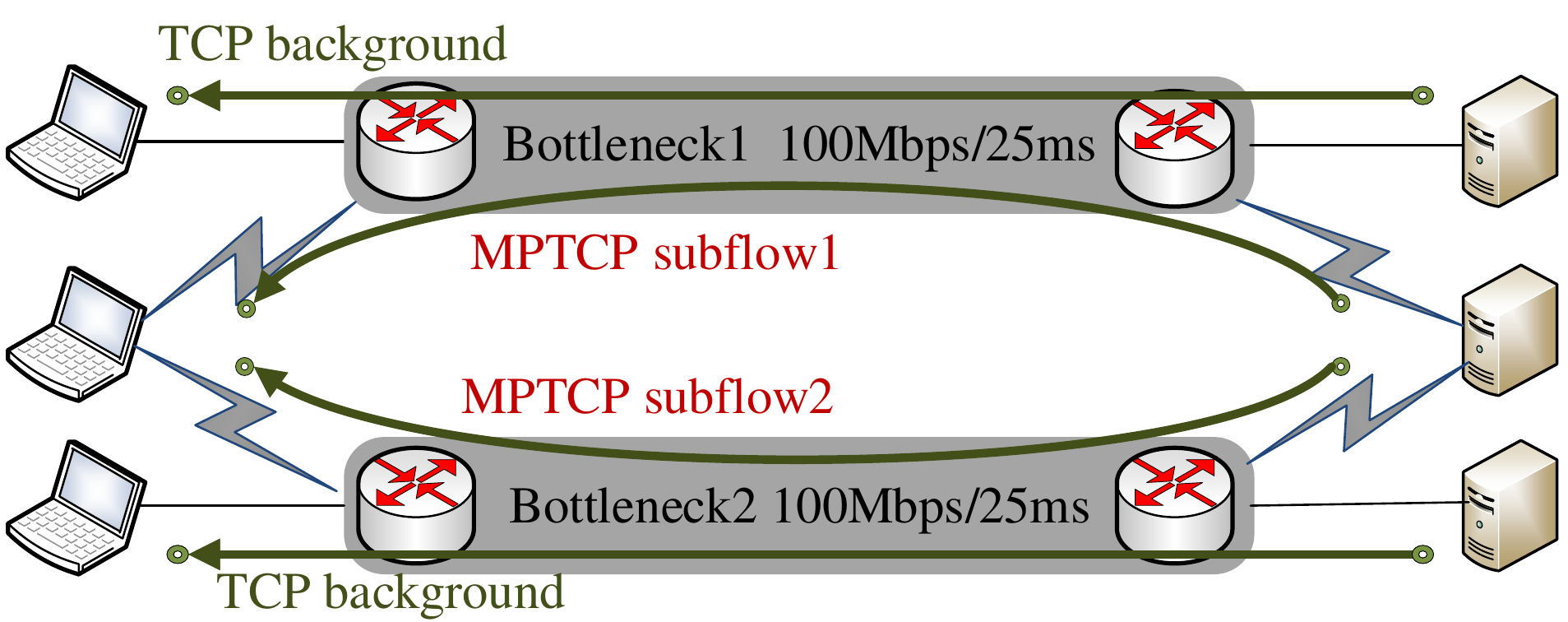}\\
  \caption{Topology of the testbed.}\label{platformtopo}
\end{figure}

  Considering the superiority of BBR, MPTCP can be promoted simply by replacing its congestion control algorithm with BBR. We measure the performance of MPTCP with conventional BBR in a lab-built platform as shown in Fig.~\ref{platformtopo}. Our testbed includes a pair of MPTCP server and client, two pairs of TCP servers and clients, and four routers within the topology shown in Fig.~\ref{platformtopo}. MPTCP connection includes two subflows, where each subflow passes through two routers. The links between two routers represent the bottleneck in the network. The links between client and router or between server and router do not affect the transmission. Both the bottleneck links have 100 Mbps bandwidth and 25 ms delay. At each bottleneck, there are two TCP background flows on each path using the same kind of congestion control algorithm as MPTCP uses.

Table \ref{bbr-table} shows the throughput performance in lossy scenarios, where the random packet loss rate of each subflow is 0.01\% (subflow$_1$) and 0.1\% (subflow$_2$), respectively. We observe that although packet loss rates of 0.01\% and 0.1\% are not too high in the actual wireless networks, the throughput of original algorithms still drops dramatically. Among them, Cubic is better than others but is unable to sustain its superiority when the packet loss rate goes up. Moreover, the bandwidth utilization of LIA, OLIA, BALIA, and Cubic is much lower than the available bandwidth that an ideal congestion control algorithm could achieve.

 \begin{table}[h]
\centering
\caption{ Average throughput \label{bbr-table}}
 \begin{tabular}{c|c|c|c|c|c}
\hline
\hline
 Throughput & BBR &Cubic & LIA & OLIA &BALIA\\
\hline
\hline
 MPTCP (Mbps)        & 55.9  & 35.1   & 20.3     & 19.9  & 20.9  \\
 Subflow 1 (Mbps)    & 30.7 & 27.6   & 19.1     & 19.6  & 20.5  \\
 Subflow 2  (Mbps)   & 25.2 & 7.5    &1.2       & 0.3    & 0.6  \\
 TCP on path 1 (Mbps)& 31.9 &  27.7  & 21.1     & 21.3   & 20.8  \\
 TCP on path 2 (Mbps)& 27.5 & 8.5    & 1.1      & 1.0   & 1.2  \\
 \hline
  Bandwidth utilization & 88\% & 54\% & 32\%  & 32\% & 33 \%  \\
\hline

\end{tabular}
\end{table}

\begin{figure}[!htb]
  \centering
  \includegraphics[width=0.9\linewidth]{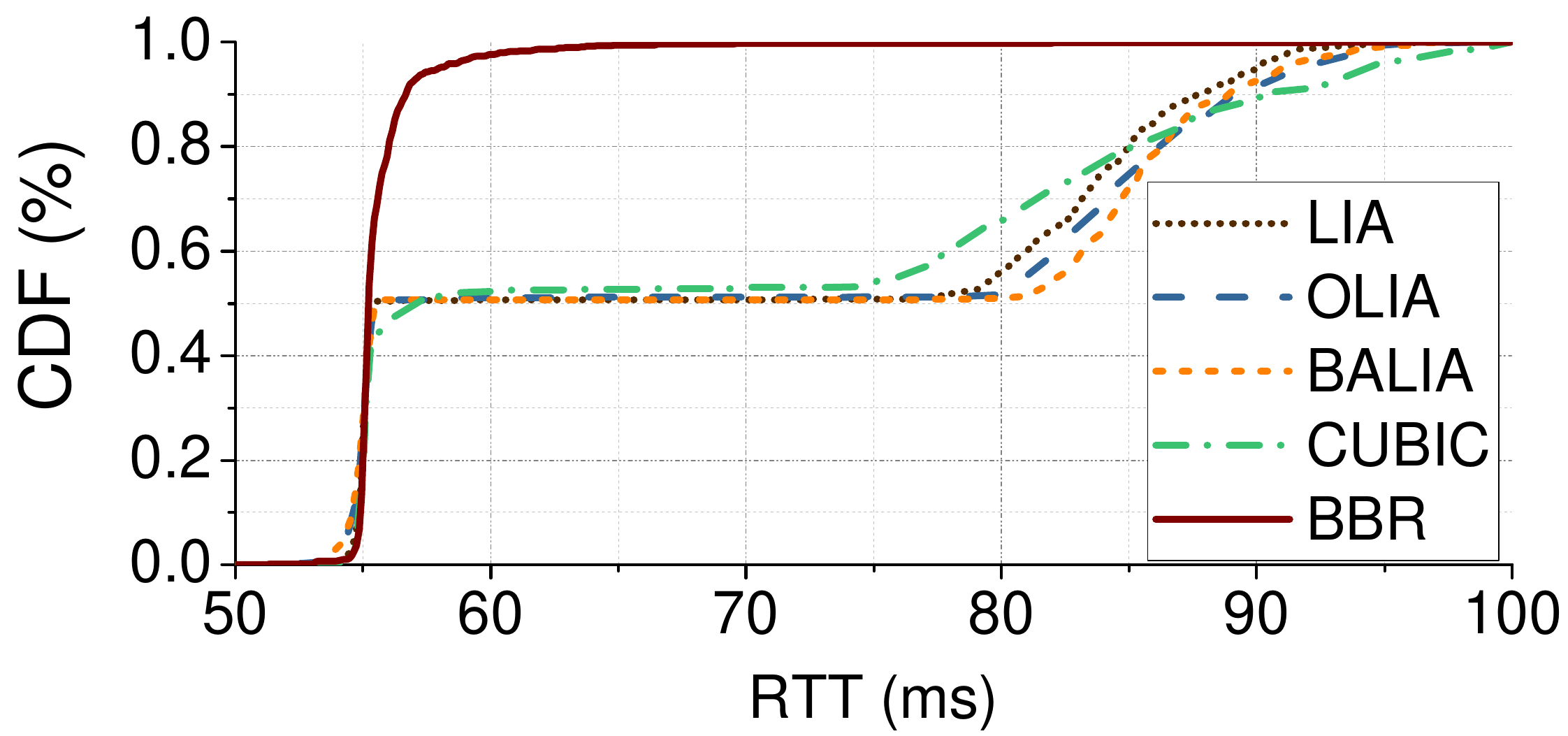}\\
  \caption{RTT distribution of MPTCP using different congestion control algorithms.
  }\label{rttloss}
\end{figure}

   Fig.~\ref{rttloss} shows the RTT distribution, where the random loss rates of subflow$_1$ and subflow$_2$ is 0 and 0.01\% respectively. BBR keeps RTT of MPTCP concentrating at around 55 ms. But half of RTTs of other algorithms are concentrated at the zone of 85 ms, which corresponds to the subflow with no random packet loss. Since traditional congestion control algorithms increase the congestion window and fill the buffer of the intermediate routers until packet loss, packets are queued at the routers for a long time, resulting in longer RTT. On the contrary, BBR does not cause network overload and keeps RTT low.

   Compare with other congestion algorithms, BBR makes each subflow obtain high throughput in lossy networks and keeps low RTT in congested networks. However, the original BBR treats multiple subflows of an MPTCP connection as separate flows that work independently rather than a unified connection. Thus the goals of fairness and balanced congestion can not be achieved. To achieve these goals, we provide a new algorithm that utilizes bandwidth detection for coupled congestion control, which is called Coupled BBR. In addition, a functional scheduler also needs to be further designed. Previous schedulers are usually based on the congestion window, while Coupled BBR changes it to smooth sending rate and makes previous schedulers no longer suitable. Thus, we propose a novel AR\&P scheduler to promote the performance in ever-changing and asymmetric networks on the basis of Coupled BBR.

   \section{System Design}\label{design}

\begin{figure}[!htb]
  \centering
  \includegraphics[width=0.9\linewidth]{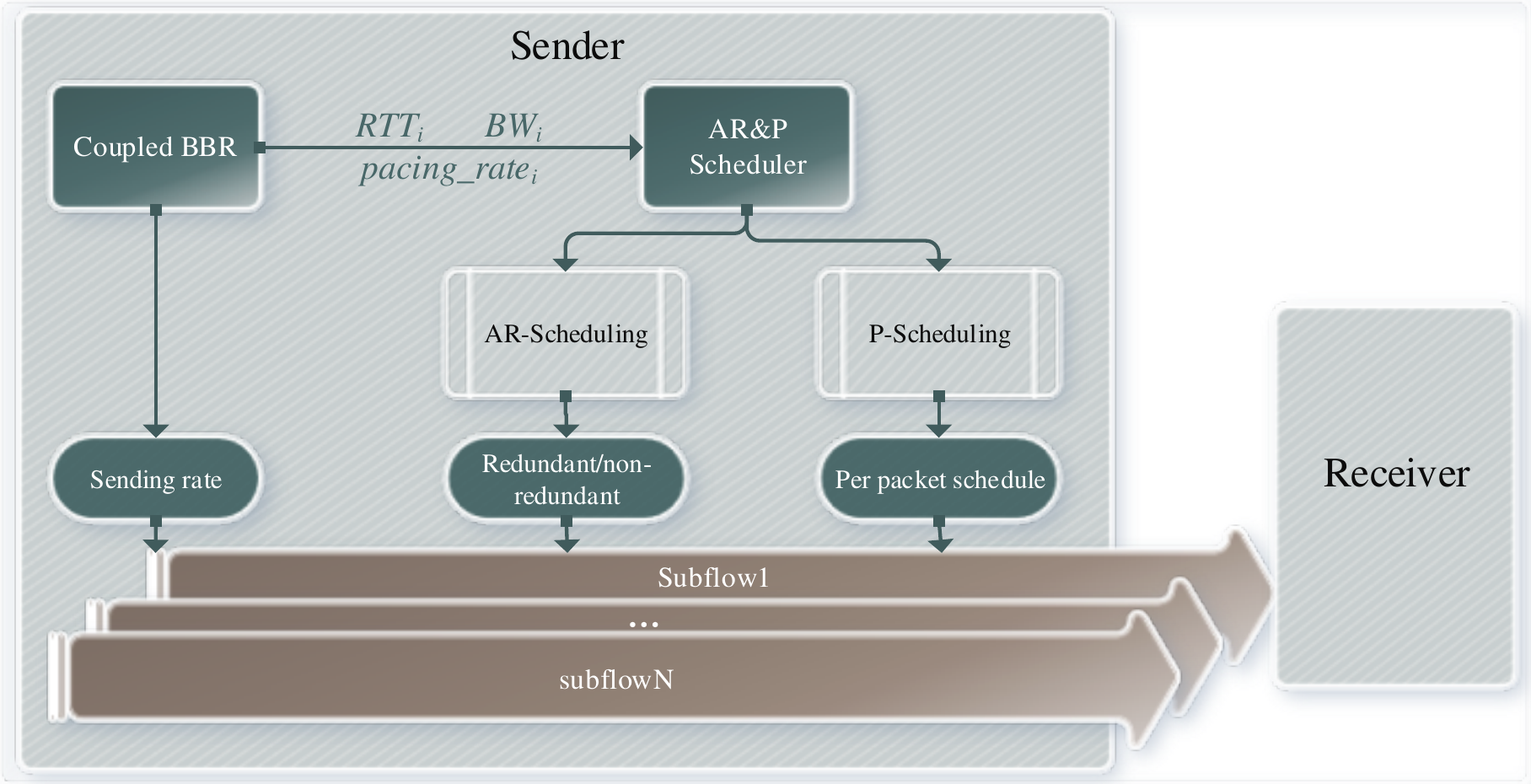}\\
  \caption{Coupled BBR and AR\&P Scheduler for MPTCP}\label{system}
\end{figure}

 In this section, we introduce our design of congestion control and scheduler for MPTCP. Fig.~\ref{system} illuminates the framework. The framework basically includes two parts: a coupled congestion control algorithm called Coupled BBR and a novel scheduler called Adaptively Redundant and Predictive packet (AR\&P) scheduler.

  Coupled BBR and AR\&P scheduler take on the functions of rate control and data scheduling, respectively. Coupled BBR performs the function of coupled congestion control for MPTCP, using measured bandwidth and RTT to control the sending rate of each subflow. It provides a steady and proper sending rate for MPTCP, ensures high throughput, and at the same time, achieves fairness to TCP BBR flows and balances congestion among MPTCP subflows. Coupled BBR shares its measured result with AR\&P scheduler for further scheduling function. AR\&P scheduler helps manage subflows by scheduling packets properly through subflows under various network conditions with the following two scheduling methods:

  1) AR-Scheduling decides the redundant/non-redundant state of each subflow. If a subflow is in poor network conditions (low bandwidth or large RTT), AR-Scheduling tends to send redundant packets via it. Otherwise, the subflow is used to transmit non-redundant packets to aggregate bandwidth resources. By adjustment based on real-time measurement, AR-Scheduling improves robustness and guarantees high throughput in highly dynamic networks.

  2) P-Scheduling schedules each packet to a target subflow according to the packet's arrival time. Each packet is scheduled to a subflow with the earliest arrival time to reduce out-of-order packets and improve performance in asymmetric scenarios.

  Coupled BBR and AR\&P scheduler are implemented at MPTCP sender for better transmission control. MPTCP receiver performs the original operation and does not need any other extra interaction with the sender. In the next sections, we present the details of each algorithm.

\section{Coupled BBR}\label{cbbr}

  Coupled BBR retains most of the operations in the conventional BBR, periodically measures the bottleneck bandwidth and uses the measured bandwidth to allocate the sending rate of each subflow. Instead of the previous way which adjusts the AIMD parameters, Coupled BBR modifies the PROBE\_BW phase and sets the sending rate directly according to the measurement results of each subflow to keep fairness and balanced congestion. And the same as the conventional BBR, Coupled BBR spends the vast majority of its time in PROBE\_BW phase (about 98 percent) \cite{cardwell2017bbr}, that makes it effective to achieve fairness and balanced congestion in the entire transmission process. In this way, Coupled BBR could keep a steady sending rate in PROBE\_BW phase and provides smooth transmission performance.

 \begin{figure}[!htb]
  \centering
  \includegraphics[width=0.95\linewidth]{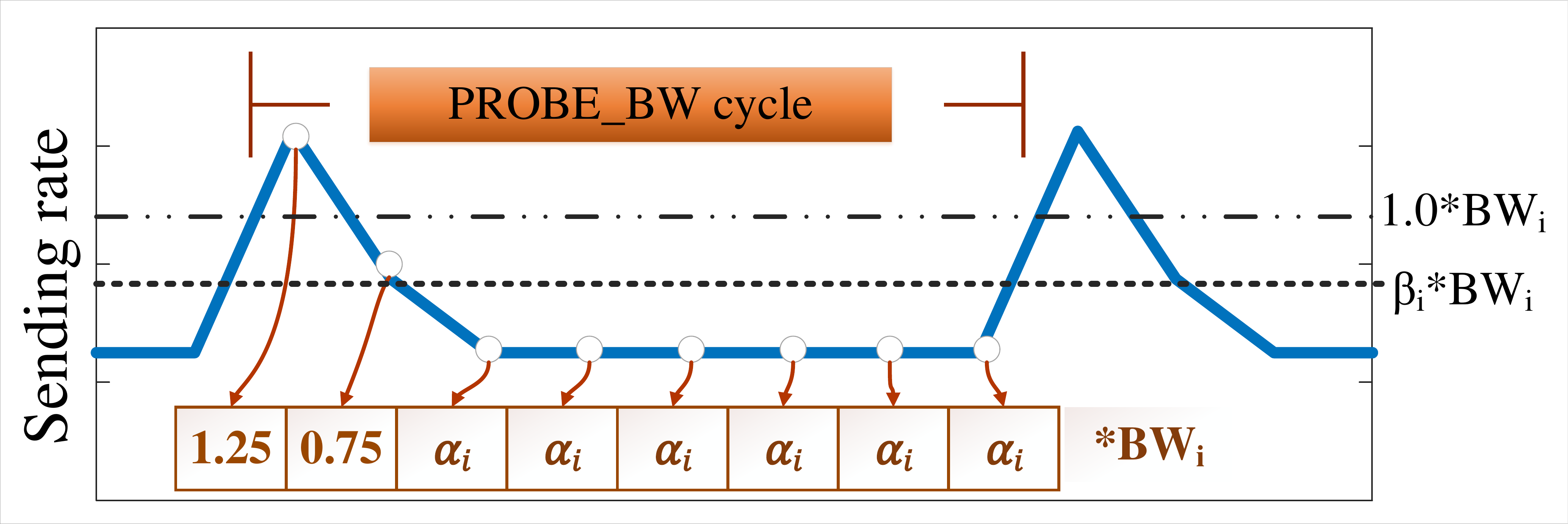}\\
   \caption{Sending rate of Coupled BBR.}\label{cbbrrate}
\end{figure}

  As shown in Fig.~\ref{cbbrrate}, Coupled BBR modifies $pacing\_gain_i$  for a subflow$_i$ to a cycle of $\{1.25, 0.75, \alpha_i, \alpha_i, \alpha_i, \alpha_i, \alpha_i, \alpha_i\} $ in the PROBE\_BW phase. The sending rate is $pacing\_gain_i \cdot BW_i = \{1.25, 0.75, \alpha_i, \alpha_i, \alpha_i, \alpha_i, \alpha_i, \alpha_i\} \cdot BW_i$, where $BW_i$ is the maximum detected available bandwidth of subflow$_i$. The $pacing\_gain_i$ of $\{1.25, 0.75\}$ for the first two RTTs is used to ensure the ability for each subflow to measure available bandwidth $BW_i$, that is the same as the conventional BBR. After that, for the next 6 RTTs, Coupled BBR replaces the $pacing\_gain$ with a smaller parameter $\alpha_i$ to achieve fairness and balanced congestion, which is related to the bandwidth of each subflow:
  \begin{align}
  & \alpha_i=\frac{4\beta_i-1}{3} ,\label{alpha}\\
  & \beta_i =\frac{BW_i \cdot \max\{ BW_i\} }{\sum_{j \in \mathcal{S}} BW_j^2},
  \end{align}
  where $\mathcal{S}$ denotes the set of all subflows.

 Coupled BBR is implemented at the sender side, and does not require interaction between the receiver and the sender. Algorithm \ref{cbbr-algorithm} shows the algorithm of Coupled BBR. Consider that $\alpha_i$ may be less than 0 because of Eq.(\ref{alpha}), Coupled BBR sets the sending rate to $4\cdot packets/RTT_i$ if $\alpha_i\leq 0$, which is similar to the PROBE\_RTT phase. If a subflow enters this state, the 4 packets in an RTT round can be utilized to probe minimum RTT and protect the activity of a subflow. Next, we will show how to use $\alpha_i$ to achieve the goals of MPTCP.

\begin{algorithm}
\caption{Coupled BBR}\label{cbbr-algorithm}
\For{each subflow$_i$}{
   $\beta_i =\frac{BW_i \cdot \max_j\{ BW_j\} }{\sum_{j \in \mathcal{S}} BW_j^2}$\;
  $\alpha_i=(4\beta_i-1)/3$\;
  pacing\_gain $=[1.25, 0.75, \alpha_i, \alpha_i, \alpha_i, \alpha_i, \alpha_i, \alpha_i] $\;
  \If{now $>=$ nextSendTime }{

    sendpacket()\;
    /* cycle\_index is the index of RTT round in a PROBE\_BW cycle */\;
    \eIf{cycle\_index $>$ 2 }{
      \eIf{$\alpha_i > 0$}{
         nextSendTime = now  + packet.size / ($\alpha_i \cdot BW_i$)\;
       }{
       nextSendTime = now  + $RTT_i/4$\;
       }

     }{
        nextSendTime = now  + packet.size / (pacing\_gain[cycle\_index] $\cdot BW_i$)\;
     }

 }
}
\end{algorithm}

   Assume that there are some TCP flows runs on subflow$_i$'s path, and throughput of a single path TCP BBR flow is $T^{TCP}_i$. Let $T^{MP}_i$ denotes the average throughput of MPTCP subflow$_i$, and $T^{MP}$ denotes the average throughput of a MPTCP connection, where $T^{MP} = \sum_{i \in \mathcal{S}} T^{MP}_i$. Coupled BBR keeps the $pacing\_rate$ of 1.25, 0.75 in the first two round and changes the $pacing\_rate$ of the last 6 RTTs to $\alpha_i$. Therefore the average sending rate of subflow$i$ is $(1.25+0.75 +6\alpha_i) BW_i/8 = \beta_i BW_i$. Therefore the average throughput of MPTCP subflow$_i$ using Coupled BBR is: $T^{MP}_i = \beta_i \cdot BW_i$. The average throughput of the overall MPTCP connection using Coupled BBR is:
   \begin{align*}
   T^{MP} &= \sum_{i \in \mathcal{S}} T^{MP}_i= \sum_{i \in \mathcal{S}} \beta_i  \cdot BW_i \\
   &= \sum_{i \in \mathcal{S}}  \frac{BW_i^2 }{\sum_{j \in \mathcal{S}} BW_j^2}\cdot \max_{j \in \mathcal{S}}\{ BW_j\}  \\
   &= \max_{j \in \mathcal{S}}\{ BW_j\}.
   \end{align*}
 This simply achieves the fairness between MPTCP and TCP BBR flows. Coupled BBR allocates a percentage of bandwidth to subflow$_i$ by the weight $\beta_i$. The overall MPTCP throughput is $T^{MP} = \max_{i \in \mathcal{S}}\{ BW_i\} =\max_{i \in \mathcal{S}} T^{TCP}_i$, which equals the throughput of TCP BBR flows on the best path.

   Moreover, Coupled BBR also has the ability to migrate data from congested paths and increase the data traffic on subflows with good path conditions. Given the information of bandwidth, Coupled BBR takes $BW_i$ as a representation of the quality of a subflow$_i$, and makes $\beta_i$ of each subflow meets the following: $\frac{ \beta_i}{BW_i} =\frac{ \beta_j}{BW_j}, \forall i,j $. The data allocation is related to each subflow's detected bandwidth, and subflows with higher bandwidth carry more data traffic.

 \section{Adaptively Redundant and Predictive Packet Scheduler}\label{AR-P}

\begin{figure}[!htb]
  \centering
  \includegraphics[width=0.95\linewidth]{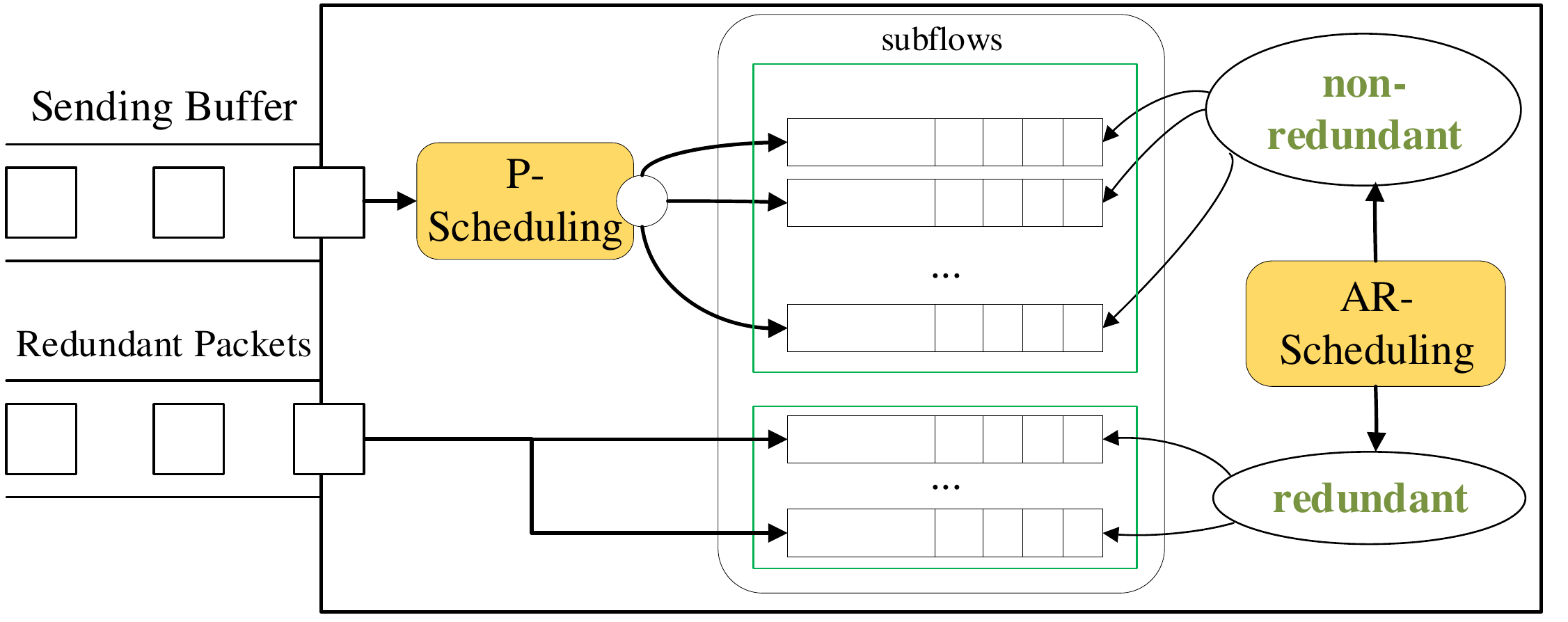}\\
   \caption{AR\&P Scheduler.}\label{scheduler}
\end{figure}

  Based on Coupled BBR, AR\&P Scheduler is further proposed to promote MPTCP in asymmetric networks, as well as in the scenarios where path conditions are changing. AR\&P scheduler includes two methods: 1) AR-Scheduling and 2) P-Scheduling. As shown in Fig.~\ref{scheduler}, AR-Scheduling first decides the redundant or non-redundant state of each subflow. Then, redundant packets are scheduled on subflows at redundant state. P-Scheduling works on the subflow in the non-redundant state, and schedules each packet according to the predictive arrival time.

\subsection{AR-Scheduling}\label{AR-Scheduler}

   By taking into account the real-time path condition measured by Coupled BBR, AR-Scheduling adaptively sends redundant packets on subflows with bad path conditions (e.g., low bandwidth and high RTT) to provide both high transmission robustness and flexibility in dynamic networks. When the user moves and the path conditions deteriorate, AR-Scheduling stops to send new packets on the subflow with bad path conditions and utilizes redundant packets to improve the reliability. The reason that utilizing redundant packets instead of stopping sending packets on the subflow with bad path conditions is to keep the activity of subflows. The redundant packets can be used to improve the robustness and detect the path conditions. Once the path conditions change for the better, AR-Scheduling returns to send non-redundant packets to improve bandwidth utilization efficiency.

The path conditions of each subflow are measured in real-time by Coupled BBR. We uses $x_i$ and $r_i$ to denote the sending rate and RTT of subflow$_i$, respectively. Let $\mathcal{N}$ denote the set of subflows in non-redundant state, and  $\mathcal{R} = \mathcal{S} -\mathcal{N}$ denote the set of subflows in redundant state. The goal of AR-Scheduling is choosing the non-redundant subflows $\mathcal{N}$ to maximize goodput ($J_1 = \log \sum_{i \in \mathcal{N}} x_i$) and minimize average RTT ($J_2 = \log \sum_{i \in \mathcal{N}} x_i r_i/\sum_{i \in \mathcal{N}} x_i$). The objective of AR-Scheduling can be expressed as a multiple objectives utility function:
\begin{align}\label{objective}
\max_{|\mathcal{N}|\geq 1}~ \log \sum_{i \in \mathcal{N}} x_i
-\log  \frac{\sum_{i \in \mathcal{N}} x_i r_i}{\sum_{i \in \mathcal{N}} x_i},
\end{align}
where $|\mathcal{N}|\geq 1$ means there should be at least one subflow to send non-redundant packets.

In order to reduce computational complexity, AR-Scheduling utilizes a greedy method:

1) Sort subflows in $\mathcal{S}$ as $i_1,\cdots, i_n$, where $\frac{x_{i_1} }{r_{i_1}}\geq \cdots \geq\frac{x_{i_n} }{r_{i_n}}$. Add subflow$_{i_1}$ to $\mathcal{N}$.

There should be at least one subflow in $\mathcal{N}$. If $\mathcal{N}$ only includes one subflow$_i$, then Eq.~(\ref{objective}) become $x_i/r_i$. AR-Scheduling chooses the subflow with the greatest $x_i/r_i$ as the initial subflow in $\mathcal{N}$, where $x_i/r_i$ denotes the subflow value.

2) For each $j$ in $2,\cdots, n$, add it to $\mathcal{N}$ if the objective after adding $i_j$ to $\mathcal{N}$ is larger than that of the original $\mathcal{N}$, which means:
\begin{align*}
 & \log (\sum_{i \in \mathcal{N}} x_i +x_{i_j})
-\log ( \frac{\sum_{i \in \mathcal{N}} x_i r_i +x_{i_j} r_{i_j}}{\sum_{i \in \mathcal{N}} x_i+x_{i_j}})\\
&~~~~~~~~~~~~~~~ > \log \sum_{i \in \mathcal{N}} x_i
-\log  \frac{\sum_{i \in \mathcal{N}} x_i r_i}{\sum_{i \in \mathcal{N}} x_i}.
\end{align*}
To simplify it, we have $\frac{x_{i_j}}{\sum_{i \in \mathcal{N}}x_i} > \frac{\sum_{i \in \mathcal{N}}x_i(r_{i_j}-2r_i)}{\sum_{i \in \mathcal{N}}x_i r_i}$. Otherwise, add it to $\mathcal{R}$.

 The greedy method gives an optimal result when there are two subflows in an MPTCP connection. Considering that mobile devices typically have two interfaces (4G/5G and Wi-Fi) to establish two subflows in practice, the greedy method can give the optimal strategy in most cases. In addition, we compare the solutions when there are three subflows in the simulation, where the result given by the greedy method is not very different from the optimal solution.

\begin{algorithm}
\caption{AR-Scheduling}\label{AR-Scheduler-algorithm}
\KwIn{Subflows, $\mathcal{S} = \{1,\cdots, n\}$.\\ }
\KwOut{Non-redundant set $\mathcal{N}$ and redundant set $\mathcal{R}$.\\ }
Sort subflows as $i_1,\cdots, i_n$, where $\frac{x_{i_1} }{r_{i_1}}\geq \cdots \geq\frac{x_{i_n} }{r_{i_n}}$\;

$\mathcal{N}=\{i_1\}$, $\mathcal{R}=\emptyset$ \;

\For{each $j \in \{2,\cdots, n\}$}{
  \eIf{$\frac{x_{i_j}}{\sum_{i \in \mathcal{N}}x_i} \leq \frac{\sum_{i \in \mathcal{N}}x_i(r_{i_j}-2r_i)}{\sum_{i \in \mathcal{N}}x_i r_i}  $ or $inflight_i<4$ }{
  $\mathcal{R} =\mathcal{R} \cup \{i_j\}$ \;
  }{
  $\mathcal{N} =\mathcal{N} \cup \{i_j\}$ \;
  }
}

\end{algorithm}

In addition, AR-Scheduling utilizes inflight packets for the auxiliary judgment. If a packet loss occurs, the sender requires 3 duplicated ACKs to start fast retransmission. If the inflight packets are less than 4, the subflow will not start fast transmission but just wait for time-out retransmission. This packet loss on the single subflow may even decrease the throughput of other subflows. So when the inflight packets of subflow$_i$ are less than 4, AR-scheduling marks that subflow$_i$ as in the redundant state. The algorithm of AR-Scheduling is shown in Algorithm~\ref{AR-Scheduler-algorithm}. To reduce computing overhead, AR-Scheduling makes decisions every $\min_{i \in \mathcal{S}}r_i$ during the transmission.

\subsection{P-Scheduling}\label{pscheduler}

\begin{figure}[!htb]
  \centering
  \includegraphics[width=0.95\linewidth]{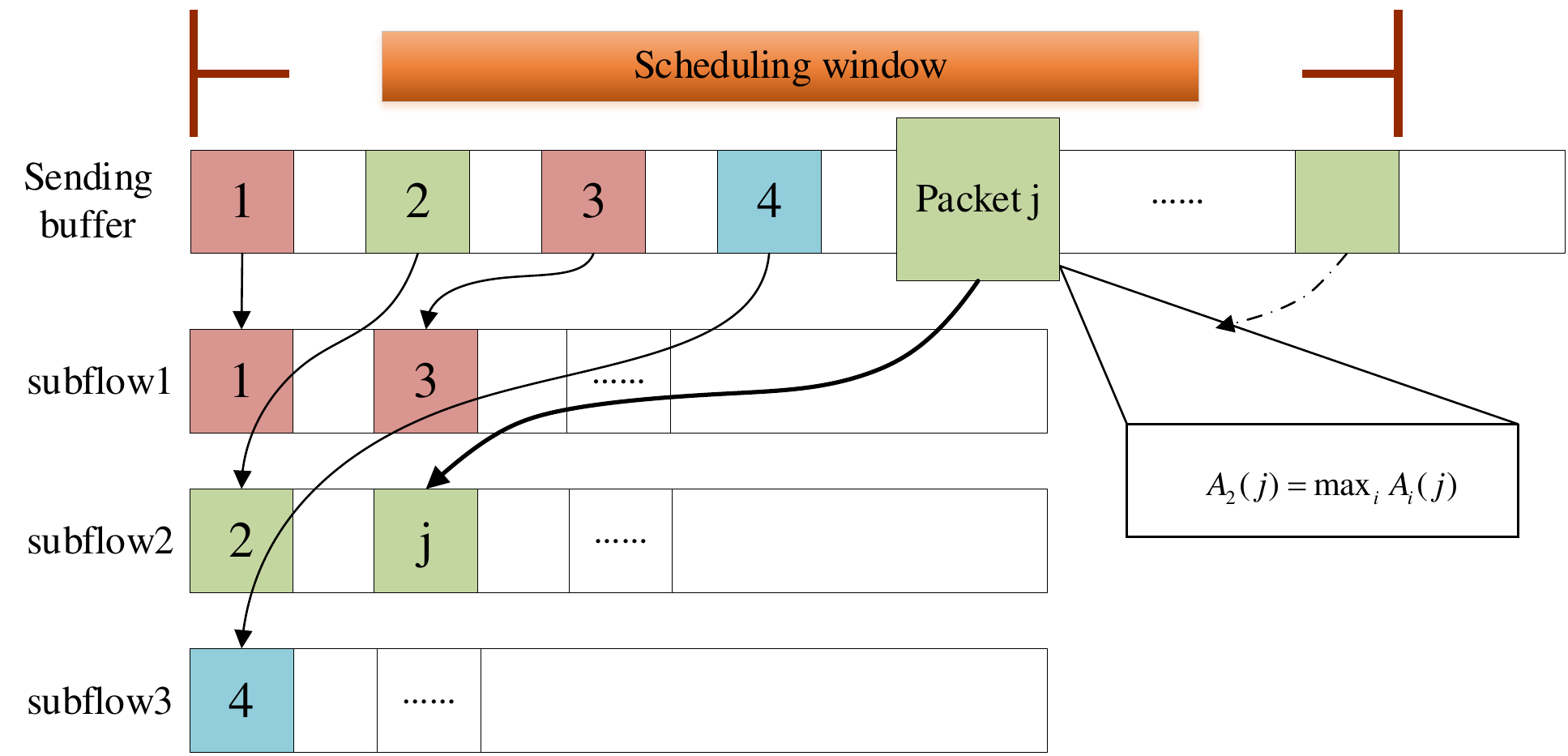}\\
  \caption{Predictive packet scheduling}\label{P-Scheduler-fig}
\end{figure}

  Based on the smooth sending rate of each subflow provided by Coupled BBR, P-Scheduling method predicts the arrival time of each packet and precisely controls each packet for better performance. P-Scheduling is a pre-scheduling method that schedules packets before they are sent. It schedules packets sequentially onto the appropriate subflow which gives the earliest arrival time of a packet to keep in-order packets arrival as well as reduce latency. As shown in Fig.~\ref{P-Scheduler-fig}, P-Scheduling pre-schedules the packets in the scheduling window to subflows for sending in the future. P-Scheduling method works as follows: When scheduling packet $j$, it calculates the arrival time of scheduling the new packet on each subflow, then chooses a subflow with the smallest arrival time and schedules packet $j$ on the subflow. In this way, P-Scheduling ensures that the packets scheduled after packet $j$ will not arrive earlier than packet $j$, therefore it keeps in-order packets arrival and the number of out-of-order packets can be significantly reduced in asymmetric networks.

P-Scheduling pre-schedules packets on subflows. Therefore how may packets should be pre-scheduled is a key parameter. Scheduling all the packets in the sending buffer is not a good idea, since network conditions are changing and old prediction may not suitable for the new environments. P-Scheduling maintains a scheduling window at the size of $\max_{i \in \mathcal{N}} RTT_i \cdot \sum_{i \in \mathcal{N}} BW_i$. The scheduling window is set to ensure in-time scheduling as well as enough packets to schedule for each available subflow. Each packet in this window is scheduled to a certain subflow according to the predicted arrival time, and the packets outside the scheduling window will not be scheduled until they are included in the scheduling window.

   Once packets in the scheduling window are sent out or the scheduling window size gets larger, new packets can be accommodated in the scheduling window and need to be scheduled by P-Scheduling.  The set of new packets are noted as \{$j_1$, $j_2$, ...,$j_m$\}. Then, P-Scheduling calculates the arrival time of each packet on available subflows, and schedules each packet to the subflow with the minimum arrival time in order.

Let $A_i(j)$ denote the predicted arrival time of sending packet $j$ on subflow$_i$, and  $t_0$ denote the current moment. When MPTCP schedules the  packet $j$, the set of packets that are already scheduled on subflow$_i$ but not sent out yet is $\mathcal{L}_i$. And the size of a packet $j$ is $s_j$. P-Scheduling predicts the arrival time as:
\begin{equation}
A_i (j)=t_0+ \frac{\sum_{j'\in \mathcal{L}_i} s_{j'}}{x_i} +\frac{r_i}{2},
\end{equation}
where the second item $\sum_{j\in \mathcal{L}_i}s_j / x_i$ on the right hand side of the equation is the waiting time for packet $j_k$ to start transmitting if it is scheduled on $subflow_i$. The third item $r_i/2$ is the transmission time of each packet. The arrival time is the current moment $t_0$ plus the waiting time and transmission time. When a packet is going to be scheduled, P-Scheduling choose a subflow with the minimum $A_i(j)$ to schedule packet on it.

\begin{algorithm}
\caption{P-Scheduling}\label{P-Scheduler-algorithm}
\KwIn{New packets in the scheduling window: $\mathcal{J}$=\{$j_{1},...,j_{m}$\}.\\ }
\For{each packet $j_k \in \mathcal{J}$}{
    \For{each $i \in \mathcal{N}$}{
    $A_i (j_k)=t_0+ \frac{\sum_{j'\in \mathcal{L}_i} s_{j'}}{x_i} +\frac{r_i}{2}$\;
    }
   $i_k = arg\min_{i \in \mathcal{N}} A_i (j_k)$\;
   schedule packet $j_k$ on the subflow $i_k$\;
}
\end{algorithm}

  Algorithm \ref{P-Scheduler-algorithm} shows the algorithm of P-Scheduling. When working with AR-Scheduling method, P-Scheduling only considers subflows in the non-redundant state and schedules new packets for them. If a subflow is in the redundant state, P-Scheduling will not schedule new packets on it. In addition, considering that the sending rate in slow-start phase changes fast and leads to high estimation error, Algorithm \ref{P-Scheduler-algorithm} is not used when all the subflows are in slow-start phase at the beginning of the connection. In this case, P-Scheduling behaves the same as Round-Robin.

\textbf{Error analysis:} Assume that the jitter of sending rate and RTT of subflow $i$ is $\Delta x_i$ ($|\Delta x_i/x_i|<\epsilon_1 $) and $\Delta r_i$ ($|\Delta r_i/r_i|<\epsilon_2 $), respectively. The prediction error of  $A_i (j)$ is:
\begin{equation*}
\Delta A_i (j) = \frac{\Delta x_i \sum_{j'\in \mathcal{L}_i} s_{j'}}{x_i(x_i+\Delta x_i )} +\frac{\Delta r_i}{2}.
\end{equation*}
Limited by the scheduling window, $\sum_{j'\in \mathcal{L}_i} s_{j'} \in [0, x_i \cdot \max_{i \in \mathcal{N}} r_i]$, and the average out-of-order packet $O$ is:
\begin{align*}
O &= \max_{i \in \mathcal{N}} \sum_{i' \in \mathcal{N}/i} x_{i'} E\{\Delta A_i (j)\} \\
& \leq  \frac{1}{2} \sum_{i \in \mathcal{N}} x_i \max_{i \in \mathcal{N}} r_i ( \left|  \frac{\max_{i \in \mathcal{N}} \Delta r_i }{\max_{i \in \mathcal{N}} r_i}\right| +\left| \frac{\Delta x_i}{ x_i}\right| )\\
& \leq  \frac{1}{2} \sum_{i \in \mathcal{N}} x_i \max_{i \in \mathcal{N}} r_i ( \epsilon_1+\epsilon_2 ) .
\end{align*}
Therefore P-Scheduling can keep the out-of-ordered packets in a low level even there are jitters in the network environment.

\section{Performance Evaluation}\label{evaluation}

 We use both the simulation and real network measurement to evaluate the performance of Coupled BBR and AR\&P scheduler. Traditional MPTCP congestion control algorithms (LIA, OLIA, BALIA) and schedulers (Round-Robin and minRTT) are using as contrasts. To verify the performance of proposed schemes in different network scenarios, we integrate Coupled BBR and AR\&P scheduler into MPTCP v0.94 implemented in Linux kernel and measure the performance in a lab-built testbed with 8 nodes and real network scenarios. In most of the tested scenarios, the performance of MPTCP can be improved by more than two-and-a-half times. In order to further evaluate the performance in larger network topologies, we also simulate the proposed algorithms under different network scenarios utilizing a packet-level simulator to show the simulation results.

\subsection{Experiments in the Testbed}\label{EvaluationTestbed}

  We integrate Coupled BBR and AR\&P scheduler into MPTCP v0.94 implemented in Linux kernel \cite{linuxkernel} and test their performance in different scenarios. The testbed topology is the same as Fig.~\ref{platformtopo}. Considering that the device usually has two interfaces (4G and Wi-Fi) in the real network, we also used two subflows in an MPTCP connection in the experiments.

\begin{figure}[!htb]
  \centering

\subfigure[Different loss rate.]{
\begin{minipage}[t]{0.70\linewidth}
\centering
\includegraphics[width=1.0\linewidth]{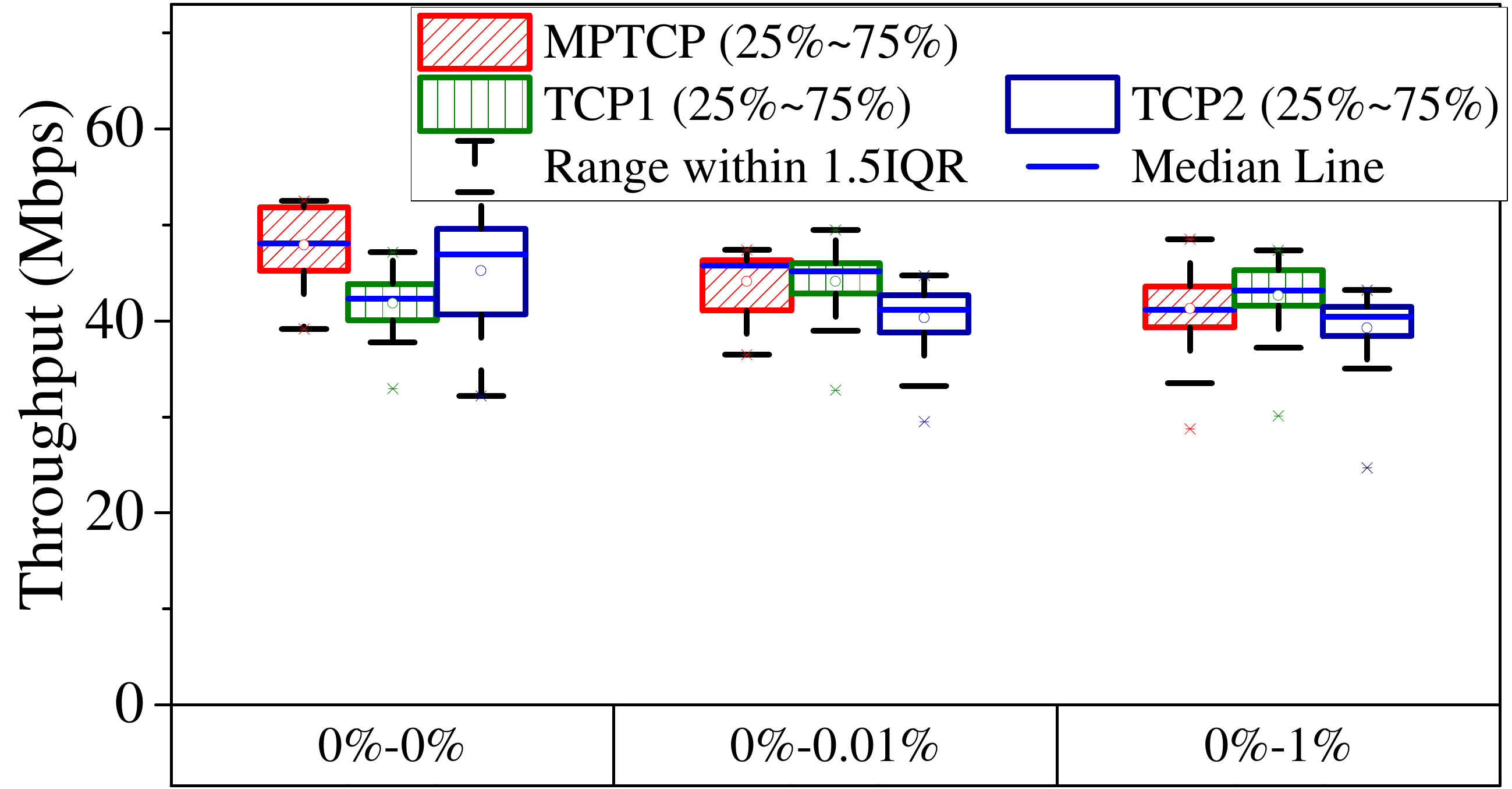} \label{cbbrthroughputloss0}
\end{minipage}
}\\
\subfigure[loss rate of two paths: 0\% - 1\%.]{
\begin{minipage}[t]{0.70\linewidth}%
\centering
\includegraphics[width=1.0\linewidth]{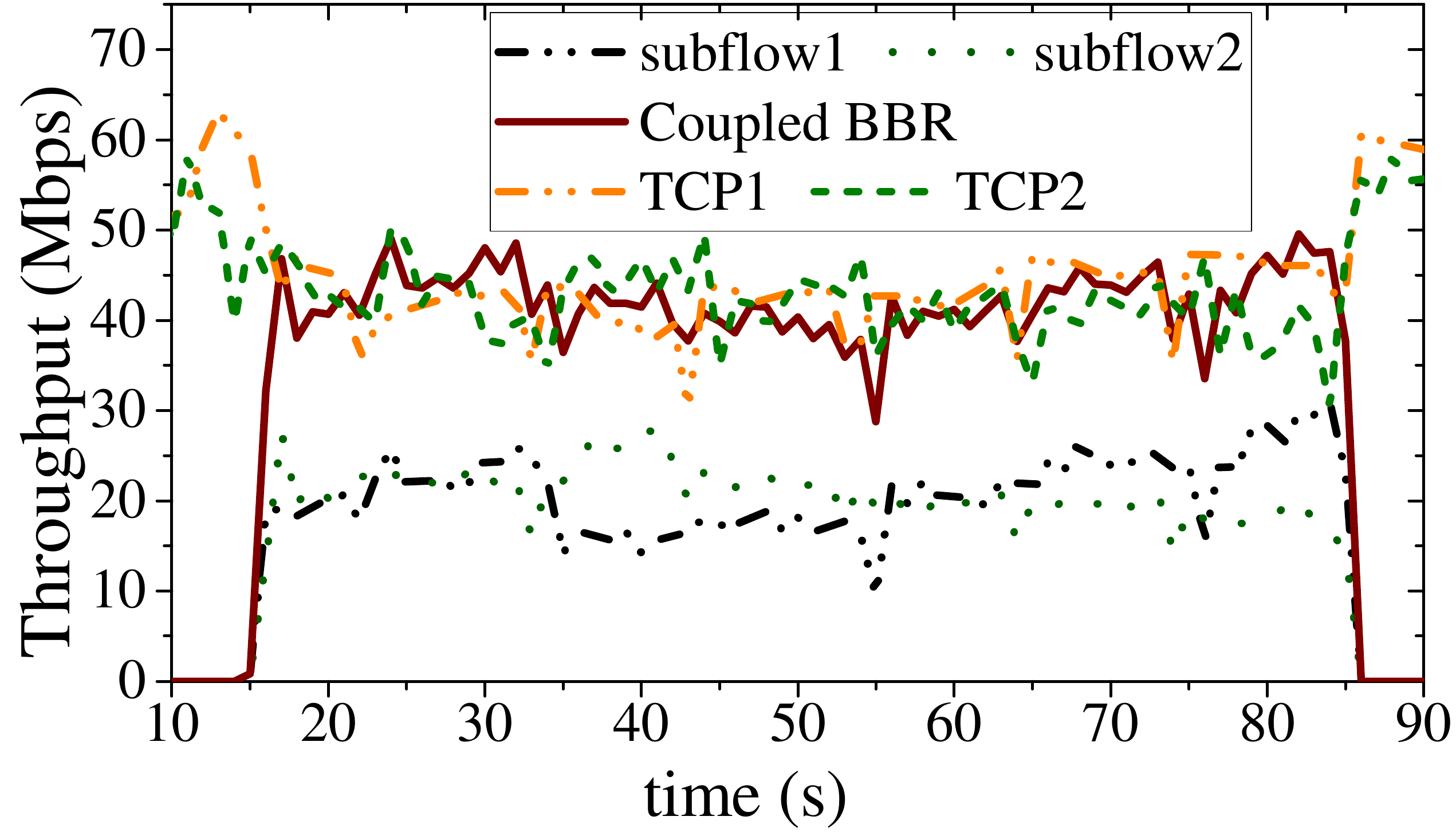}\label{cbbrthroughputloss1}
\end{minipage}
}

\caption{Coupled BBR in lossy networks.
}\label{cbbrthroughputloss}

\end{figure}

  Fig.~\ref{cbbrthroughputloss} shows the performance of Coupled BBR in lossy networks. The bandwidth and delay of both bottlenecks are 100 Mbps and 25 ms, respectively. bottleneck$_2$ suffers varying random packet loss rates of 0\%, 0.01\%, and 1\% in different scenarios, while bottleneck$_1$ does not have random packet loss. There are two TCP background flows at each bottleneck. Fig.~\ref{cbbrthroughputloss0} indicates that at different settings of path loss rate, Coupled BBR effectively achieves the goal of fairness, which gets the same throughput as that of a single-path TCP BBR flow on the best path. When the loss rate increases, the throughput of Coupled BBR decreases slightly but still achieves high throughput.  Fig.~\ref{cbbrthroughputloss1} shows the real-time throughput when the loss rate of two subflows is 0\% and 1\%, respectively. Subflow$_2$ can still get a satisfactory throughput when the loss rate reaches 1\%, and the sending rate keeps little fluctuation. In a word, MPTCP over Coupled BBR not only provides high throughput and less fluctuation in lossy networks but also achieves fairness to TCP BBR flows.

\begin{figure}[!htb]
\centering

\subfigure[Asymmetric Bandwidth.]{
\begin{minipage}[t]{0.45\linewidth}
\centering
\includegraphics[width=0.96\linewidth]{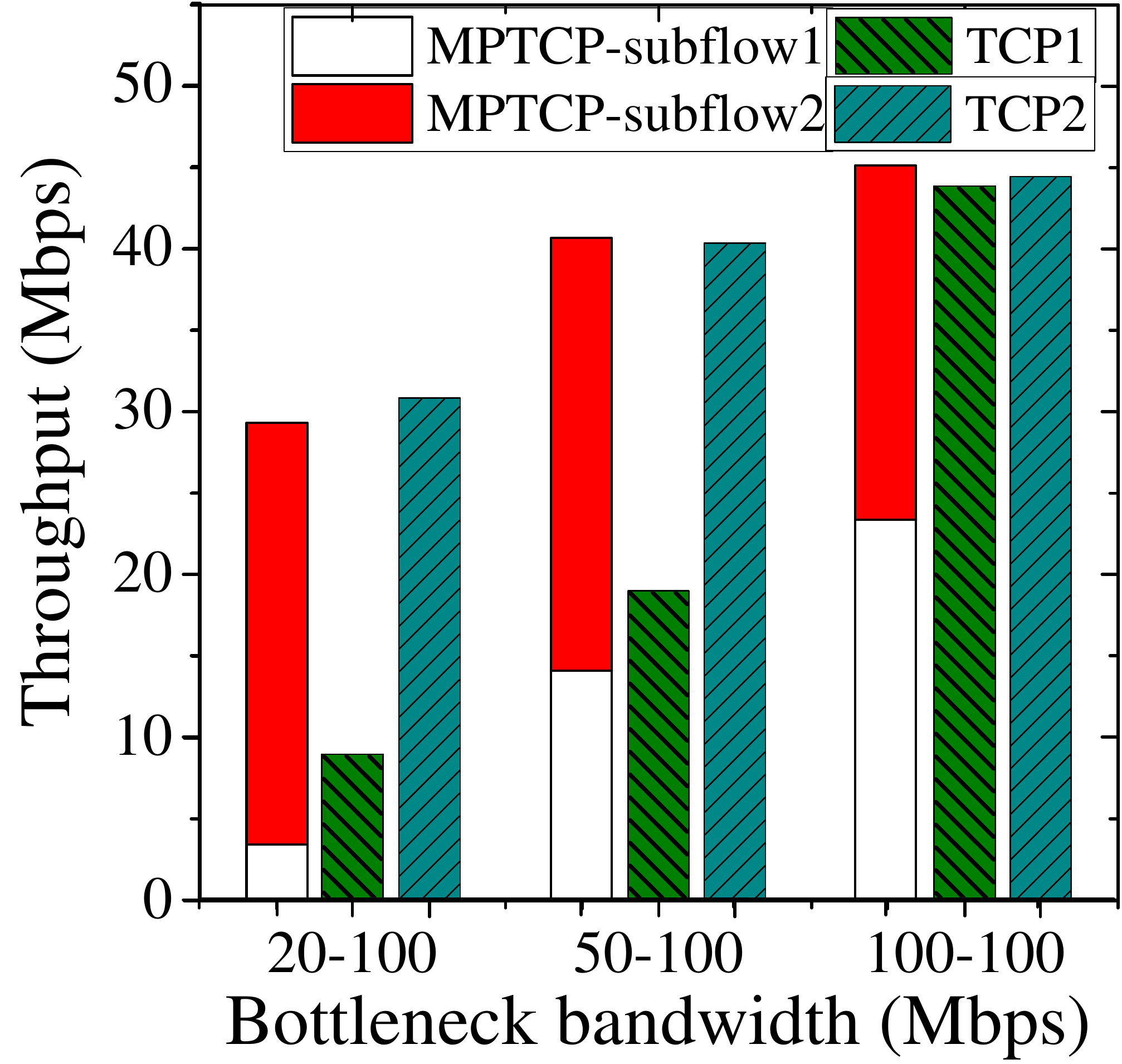}\label{cbbrthroughputbandwidth}
\end{minipage}
}
\subfigure[Asymmetric Delay.]{
\begin{minipage}[t]{0.48\linewidth}
\centering
\includegraphics[width=1.0\linewidth]{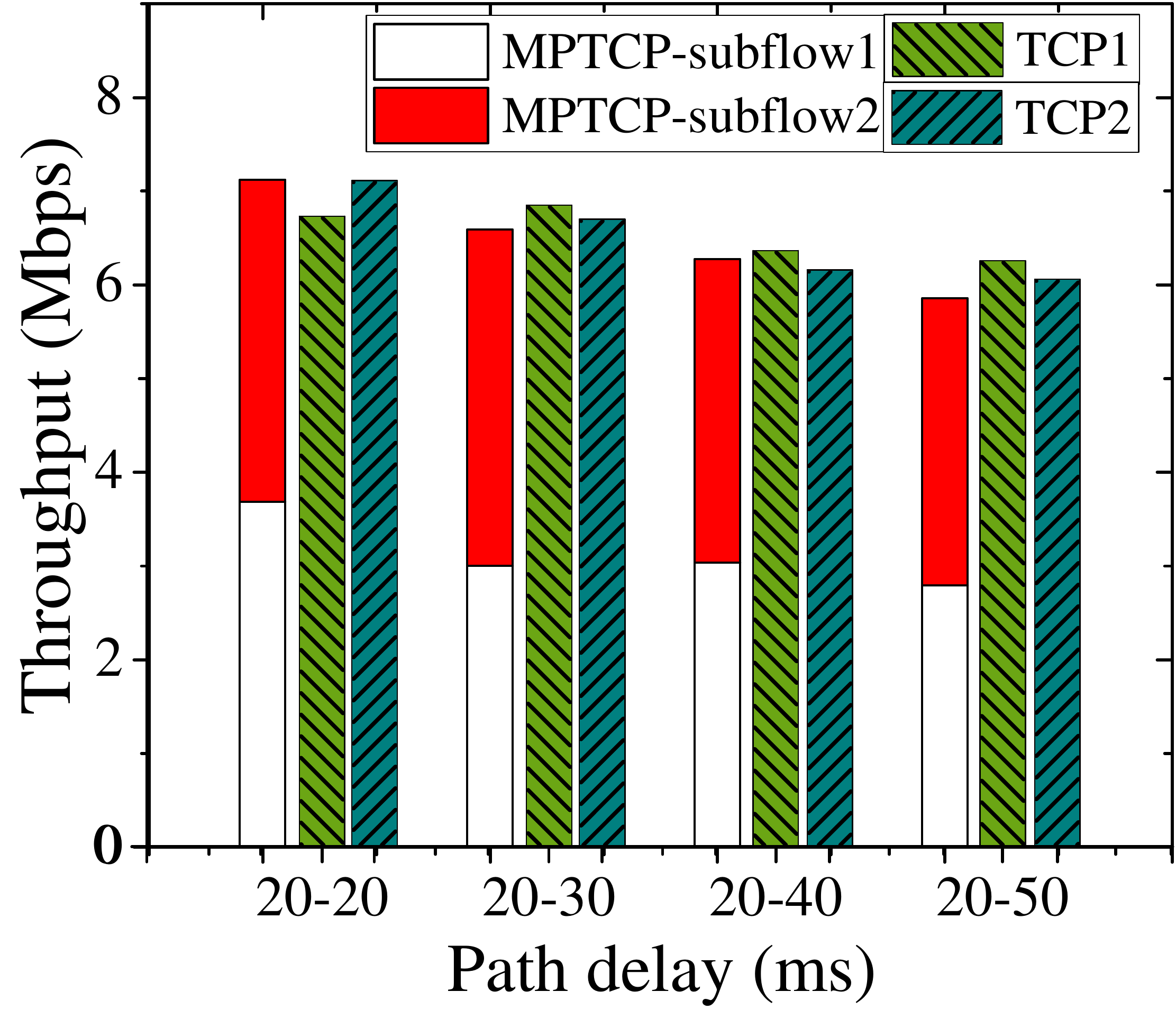}\label{cbbrthroughputRTT}
\end{minipage}
}
\caption{Coupled BBR in asymmetric networks.}\label{cbbrthroughputasymmetric}
\end{figure}

  Fig.~\ref{cbbrthroughputasymmetric} shows the performance of Coupled BBR in asymmetric networks, where Fig.~\ref{cbbrthroughputbandwidth} and Fig.~\ref{cbbrthroughputRTT} show asymmetric bandwidth scenarios and asymmetric path delay scenarios, respectively. In Fig.~\ref{cbbrthroughputbandwidth}, the loss rate and delay of both bottlenecks are 0\% and 25 ms, the bandwidth of two path changes for different scenarios. Coupled BBR can achieve the same throughput as that of a single-path TCP BBR flow on the best path, and allocate more data on the best path to balance congestion. In Fig.~\ref{cbbrthroughputRTT}, the bandwidth and loss rate of the two bottlenecks is set to 20 Mbps and 0\%, the path delay changes for different scenarios. When the delay difference becoming larger, the throughput of MPTCP decreases slightly. Meanwhile, Coupled BBR allocates the same proportion of data to each subflow and still maintains fairness to TCP BBR flows. In summary, Coupled BBR achieves better loss tolerance and steady sending rate, while also achieves fairness to TCP BBR flows and balances congestion in different scenarios.

\begin{figure}[!htb]
  \centering

\subfigure[One subflow breaks down.]{
\begin{minipage}[t]{0.70\linewidth}%
\centering
\includegraphics[width=1.0\textwidth]{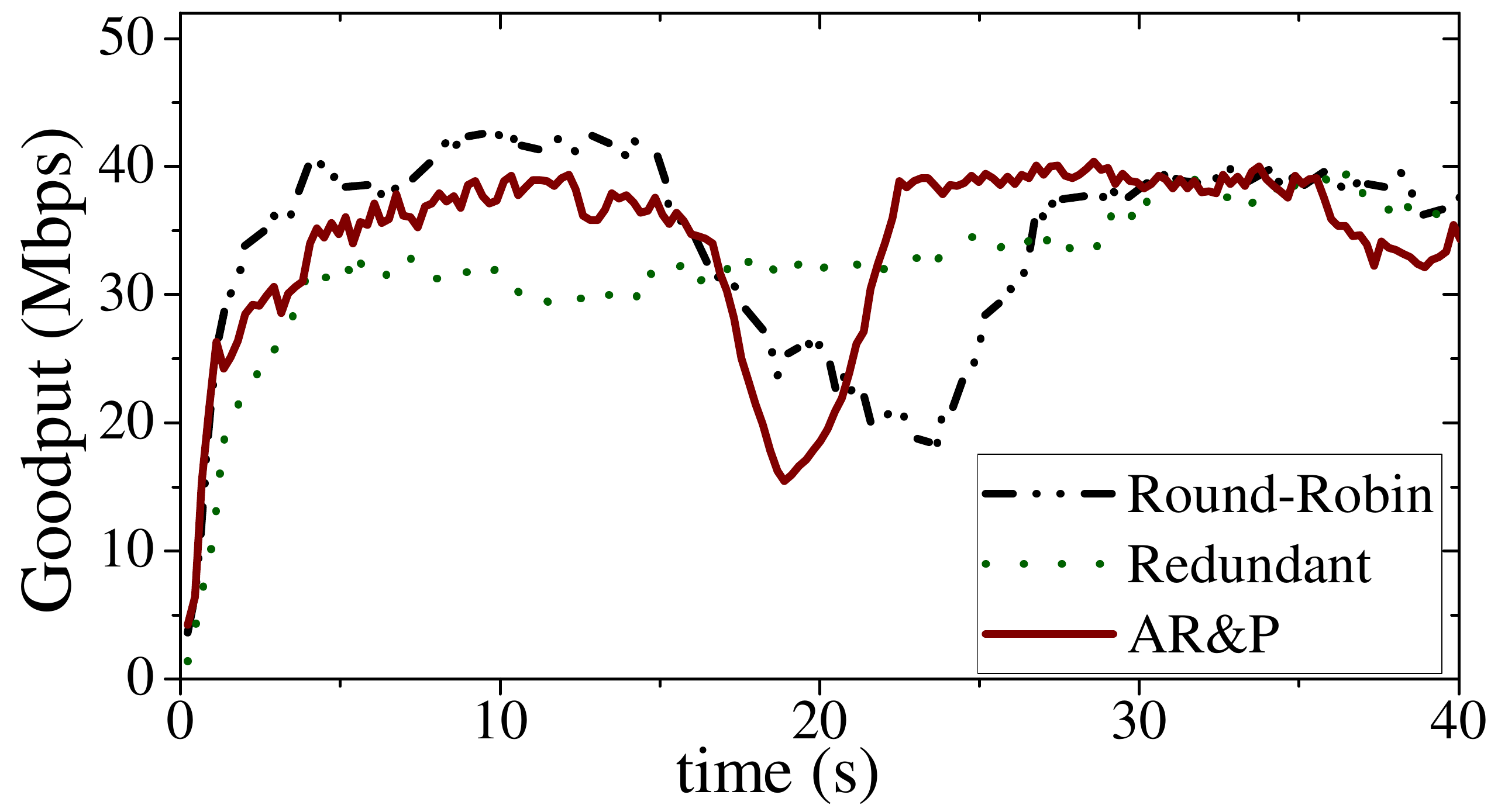}\label{schedulerloss1}
%\caption{fig1}
\end{minipage}
}\\
\subfigure[One subflow gets worse gradually.]{
\begin{minipage}[t]{0.70\linewidth}%
\centering
\includegraphics[width=1.0\textwidth]{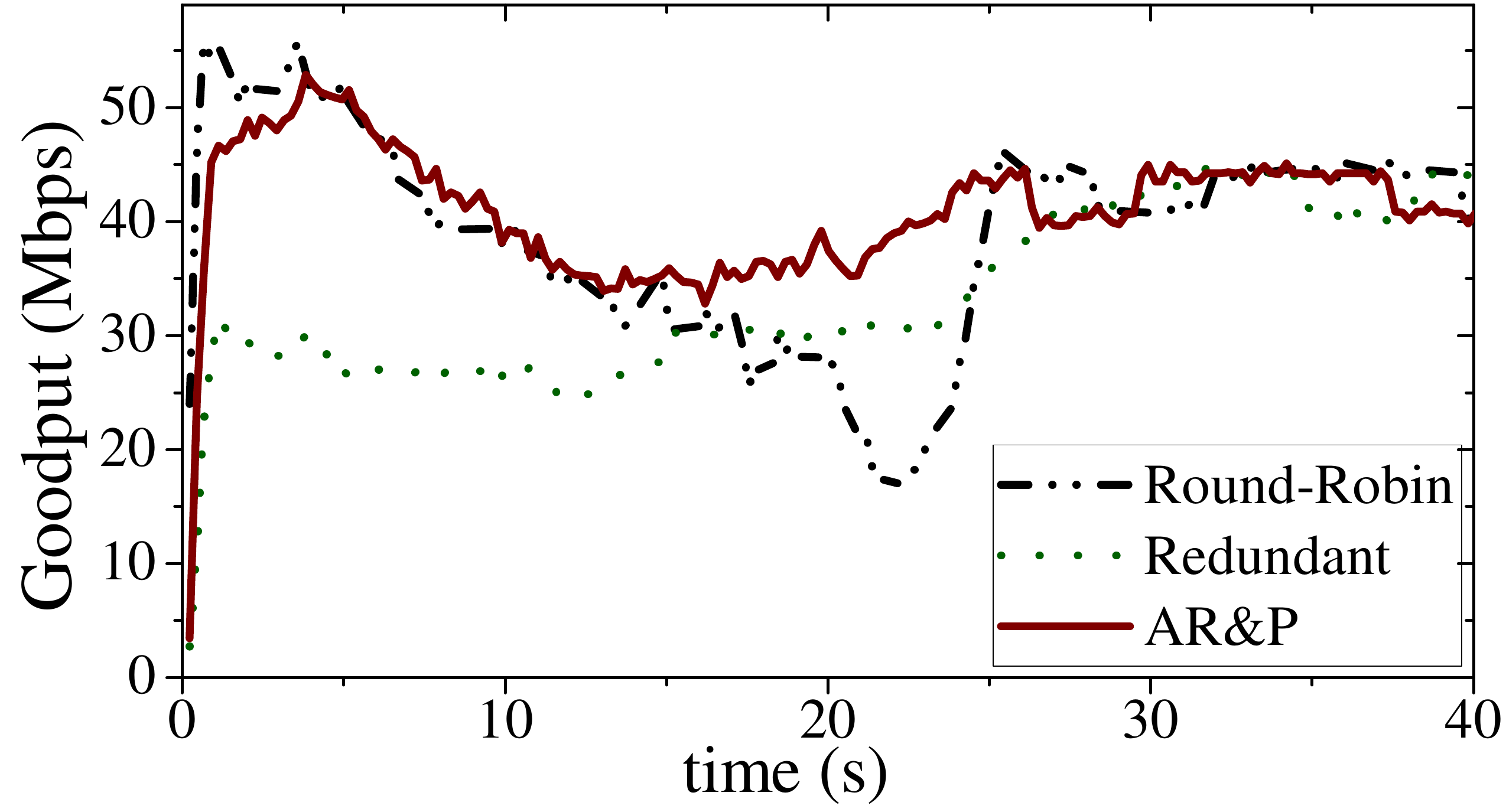}\label{schedulerloss2}
%\caption{fig2}
\end{minipage}
}
  \caption{Performance of proposed scheduler when changing the path loss rate and delay.}\label{schedulerloss}
\end{figure}

 Fig.~\ref{schedulerloss} shows the performance of AR\&P scheduler under dynamic network scenarios, where AR\&P provides faster recovery when link suddenly interrupts and provides smooth goodput under a scenario with gradually decreasing path conditions. As shown in Fig.~\ref{schedulerloss1}, in the first 15 seconds, both the paths have high bandwidth and low RTT and no packet loss occurs. AR-Scheduling finds that both paths are in good condition and their bandwidth should be aggregated for higher goodput. As a result, AR-Scheduling decides that the two subflows should both send non-redundant packets. Meanwhile, Redundant scheduler keeps sending redundant packets which results in lower goodput. At the moment of 15 seconds, one path breaks down. The throughput of Round-Robin and AR\&P drops from 40 Mbps to about 15 Mbps while redundant scheduler protects its throughput from a high packet loss rate by sending redundant packets. Although the goodput of AR\&P also drops, it recovers quickly because AR-Scheduling adaptively starts sending redundant packets on the subflow with bad path conditions and packet loss does not degrade the overall goodput. By this proactive action, AR\&P recovers much faster than Round-Robin when the path failure suddenly occurs, while also retains higher goodput than Redundant when subflows have good path conditions.

Fig.~\ref{schedulerloss2} shows another scenario in which the path conditions of one path gets worse and worse for a relatively long period until it becomes unavailable. Round-Robin and AR\&P aggregate bandwidth and outperform Redundant because both paths are in good condition at the beginning. When one of the paths gets worse and worse, the goodput of connections using Round-Robin and AR\&P starts to drop. AR-Scheduling realizes that one of the paths is no longer satisfactory and starts to send redundant packets on it for better performance at 12 s, while Round-Robin keeps sending new packets resulting in a significant throughput decrease. Besides, Redundant is not affected by the path failure. During the whole transmission, AR\&P scheduler is more adaptive to dynamic networks by adjusting its policy according to path conditions.

 \begin{figure}[!htb]
  \centering

\subfigure[Average out-of-order packets]{
\begin{minipage}[t]{0.70\linewidth}%
\centering
\includegraphics[width=1.0\textwidth]{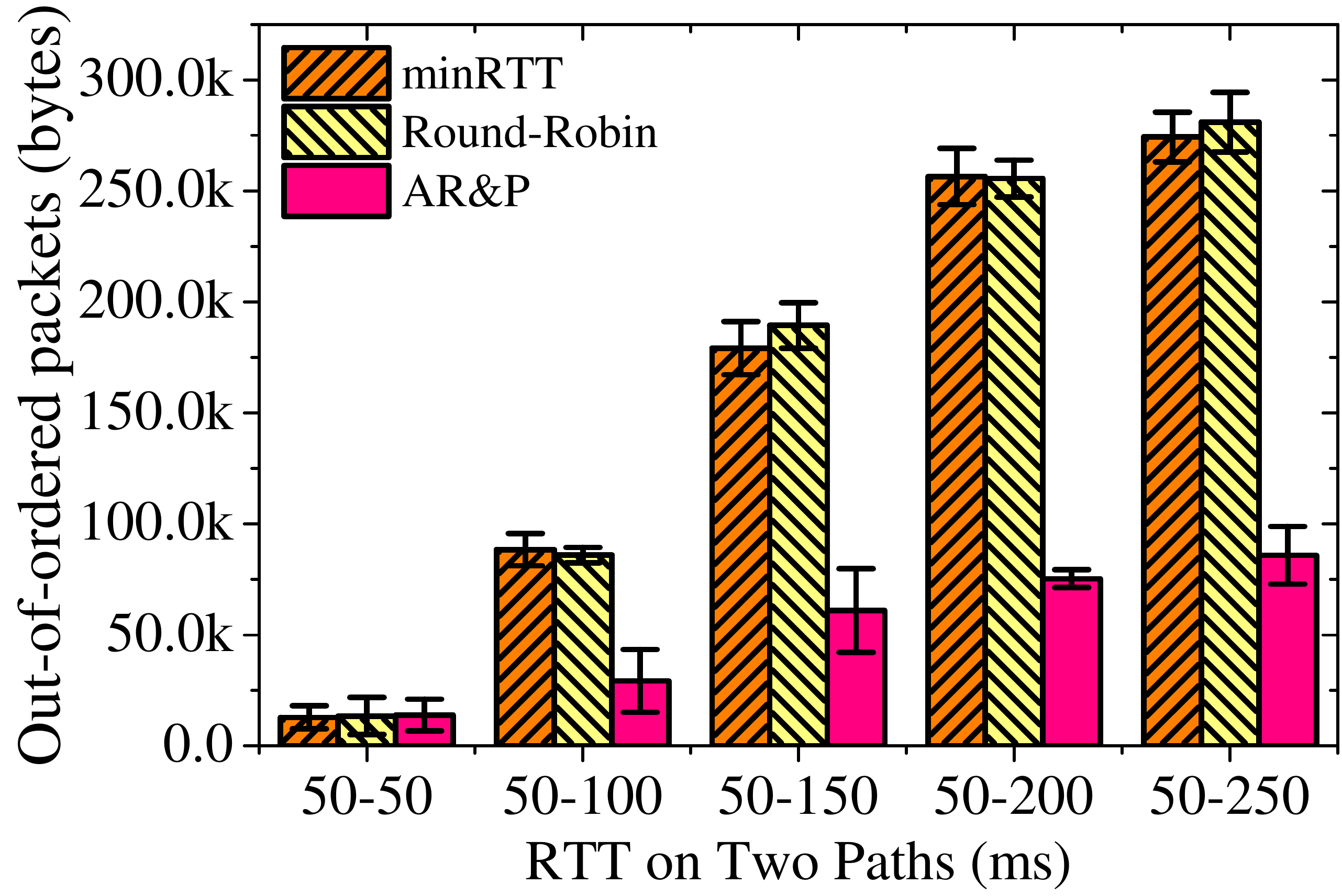}\label{ofoavg}
%\caption{fig1}
\end{minipage}
}\\
\subfigure[Real time out-of-order packets]{
\begin{minipage}[t]{0.70\linewidth}%
\centering
\includegraphics[width=1.0\textwidth]{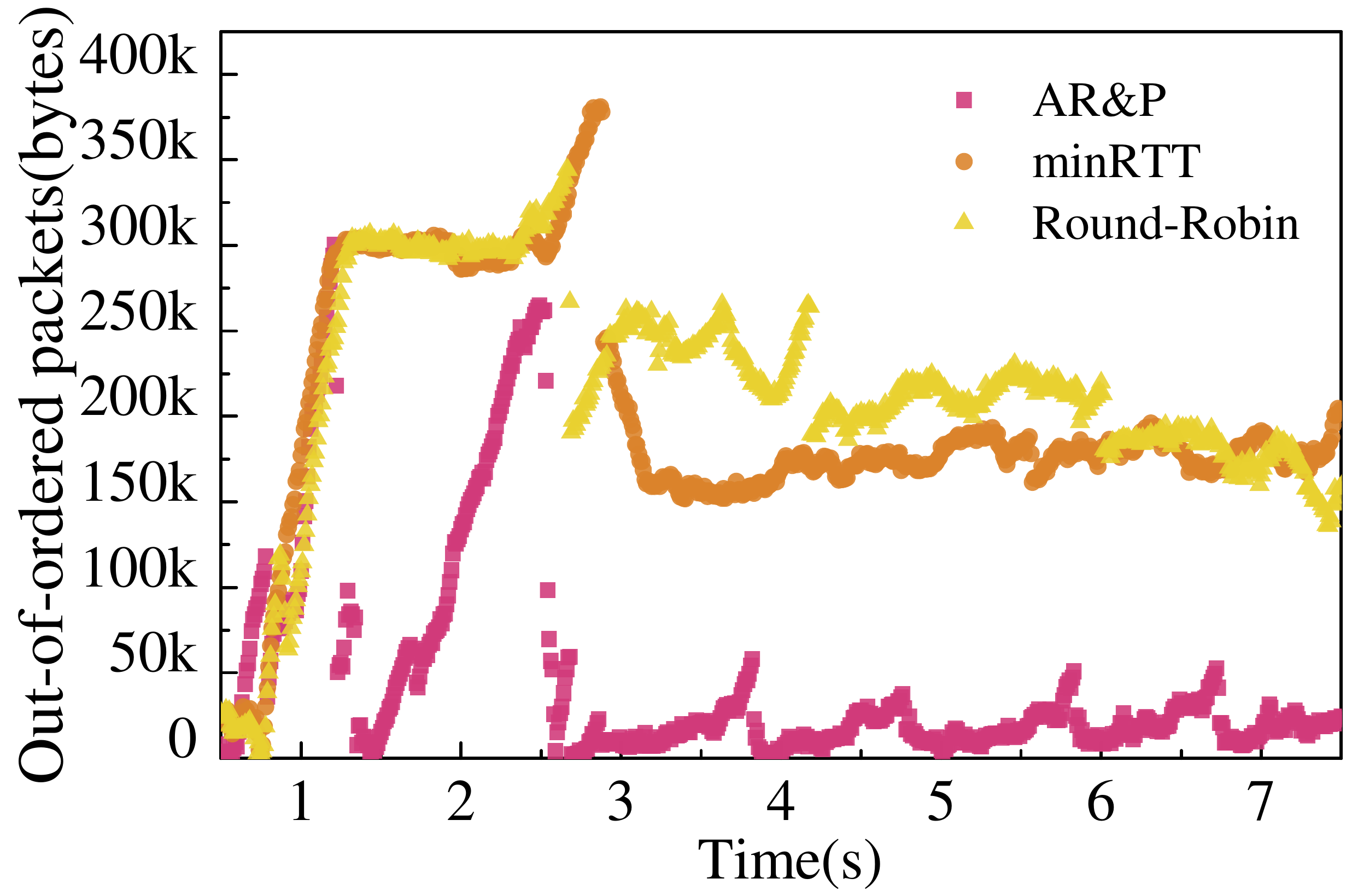}\label{oforealtime}
%\caption{fig2}
\end{minipage}
}
  \caption{Out-of-order packets.}\label{ofoevaluation}
\end{figure}

Fig.~\ref{ofoevaluation} shows the out-of-order packets in asymmetric network scenarios. We compare minRTT, and Round-Robin with AR\&P scheduler. In this experiment, both the bottlenecks have the same bandwidth. RTT of subflow$_1$ remains 50 ms, while RTT of the other one increases from 50 ms to 250 ms in different scenarios. When the two paths have the same RTT of 50 ms, the proposed scheduler creates a similar out-of-order queue to minRTT and Round-Robin. However, when the RTT of one path reaches 100 ms, we observe that both minRTT and Round-Robin increase out-of-order queues by over 300\%, which is much longer than that of AR\&P scheduler. When the RTT of one path reaches 250 ms, which means that the two paths are highly asymmetric in terms of RTT, AR\&P scheduler reduces the average out-of-order queue by 65\% compared to minRTT and Round-Robin.

To look further, Fig.~\ref{oforealtime} shows how the out-of-order queues change during data transmission when the RTT of the two paths is 150 ms and 50 ms, respectively. In the first 2 seconds, all of the schedulers create long out-of-order queues because of startup and asynchronous subflow establishment. After 2 seconds, AR\&P scheduler keeps the out-of-order queue much shorter than minRTT and Round-Robin. We observe that our scheduler empties the out-of-order queue before it gets too long, which indicates that our scheduler effectively schedules packets according to the arrival time of each packet. However, minRTT and Round-Robin are not aware of the arrival time of packets and thus create long out-of-order queues.

\subsection{Experiments in Real Networks}

We also deploy Linux kernels that support our scheme in cloud servers to conduct some tests in real networks, transmitting data from the implemented server in the cloud to the lab-built client. We use different kinds of Wi-Fi links (2.4GHz and 5GHz) and deploy our scheme in the rented cloud servers in different regions to conduct some experiments. We repeat 10-20 times of data download under different network environments (10 for using MPTCP flows and 20 for using TCP flows on different links). When we use the proposed MPTCP, the compared TCP flow uses BBR. Otherwise, the compared TCP flow uses NewReno.

\begin{figure}[htbp]
\centering
\includegraphics[width=0.7\linewidth]{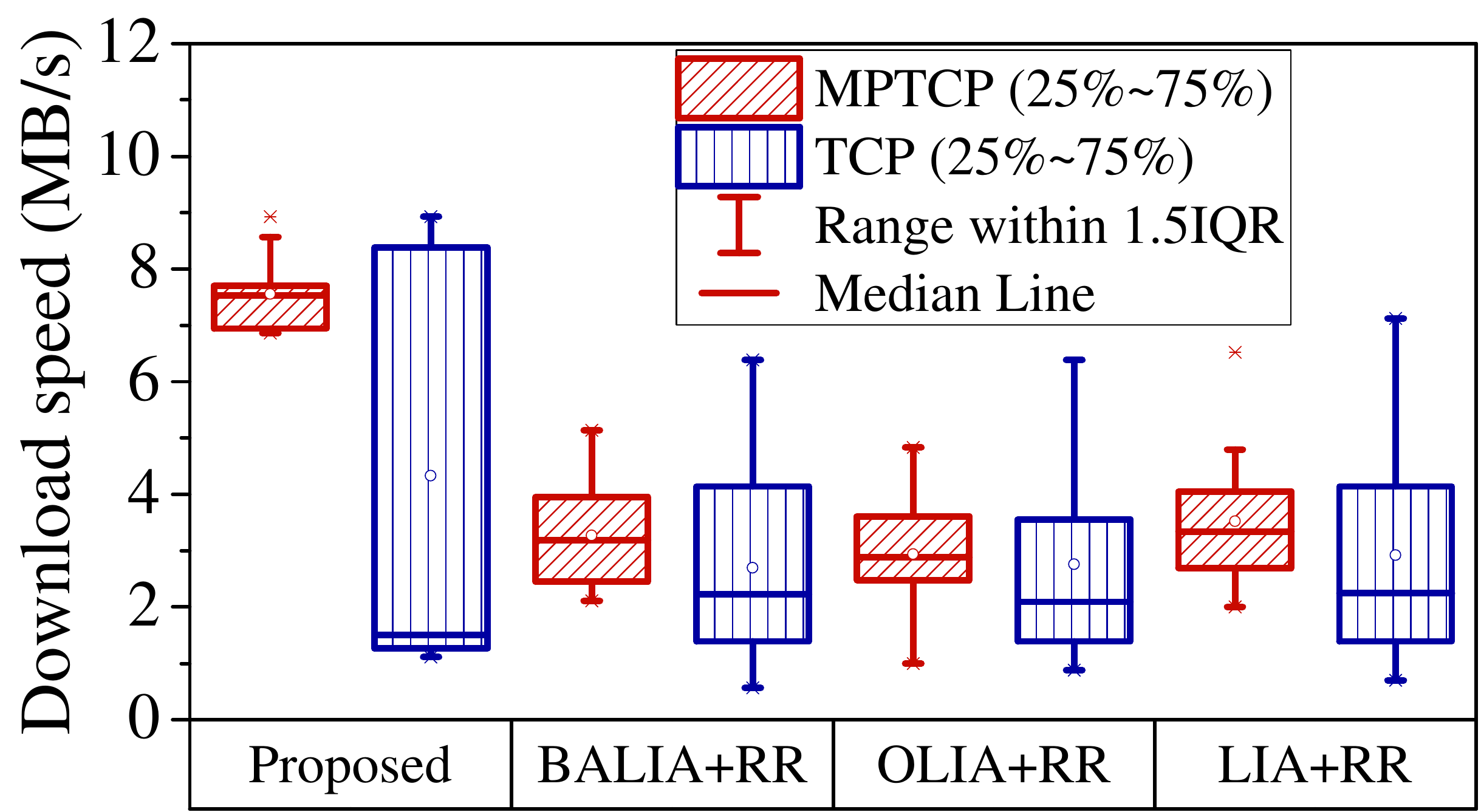}
\caption{Download data using 4G and Wi-Fi (2.4GHz).
}\label{proposedthroughputreal}
\end{figure}
\begin{figure}[htbp]
\centering
\includegraphics[width=0.7\linewidth]{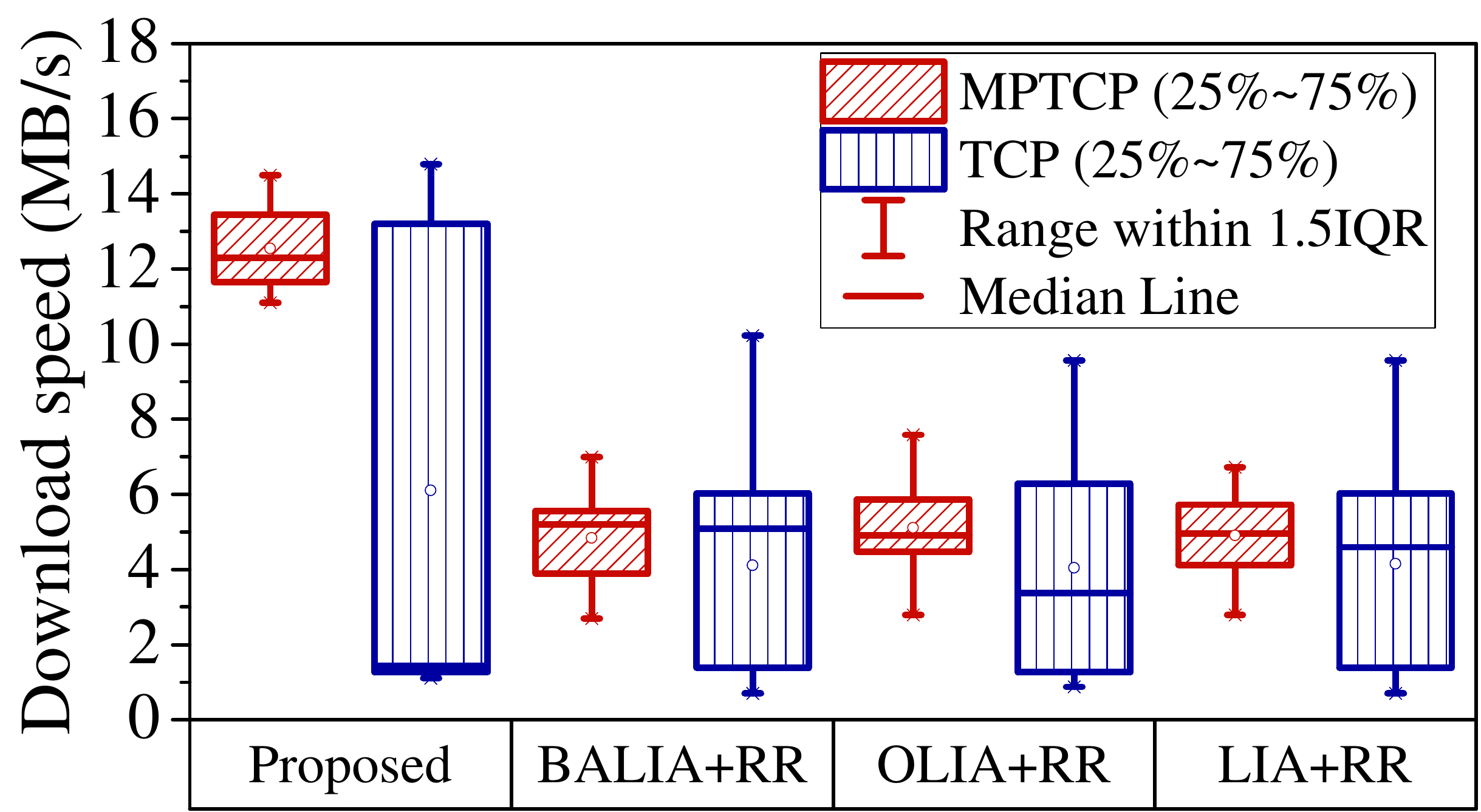}
\caption{Download data using 4G and Wi-Fi (5GHz).}\label{proposedthroughputreal5GHz}
\end{figure}

We first show the performance measurements of using different access technologies. Fig.~\ref{proposedthroughputreal} shows the throughput performance using 4G and Wi-Fi (2.4GHz). In our test environment, the bandwidth of the Wi-Fi (2.4GHz) link is twice as fast as the 4G link. Moreover, the 4G link has a higher link packet loss rate, which makes the transmission not as stable as the Wi-Fi link. The boxes show the 25\%-75\% of the download speed of each protocol and the lines show the median download speed. MPTCP flows always have a higher average speed than TCP flows, while also provides less fluctuation of performance. Among them, the proposed MPTCP scheme outperforms original MPTCP algorithms in higher throughput. The overall throughput of the proposed scheme is twice higher than that of the original MPTCP. At the same time, the proposed scheme also achieves the goal of fairness, i.e., the proposed MPTCP flow is no more aggressive than the best single TCP BBR flow.

Fig.~\ref{proposedthroughputreal5GHz} shows the download speed using 4G and Wi-Fi (5GHz). 5GHz Wi-Fi link has higher bandwidth but is not as stable as the 2.4GHz link which has a higher random loss rate. Compared with the original MPTCP, the proposed scheme brings more advantages in this scenario. The throughput of our scheme is almost 3 times higher than that of original MPTCP algorithms.

\begin{figure}[htbp]
\centering
\includegraphics[width=0.7\linewidth]{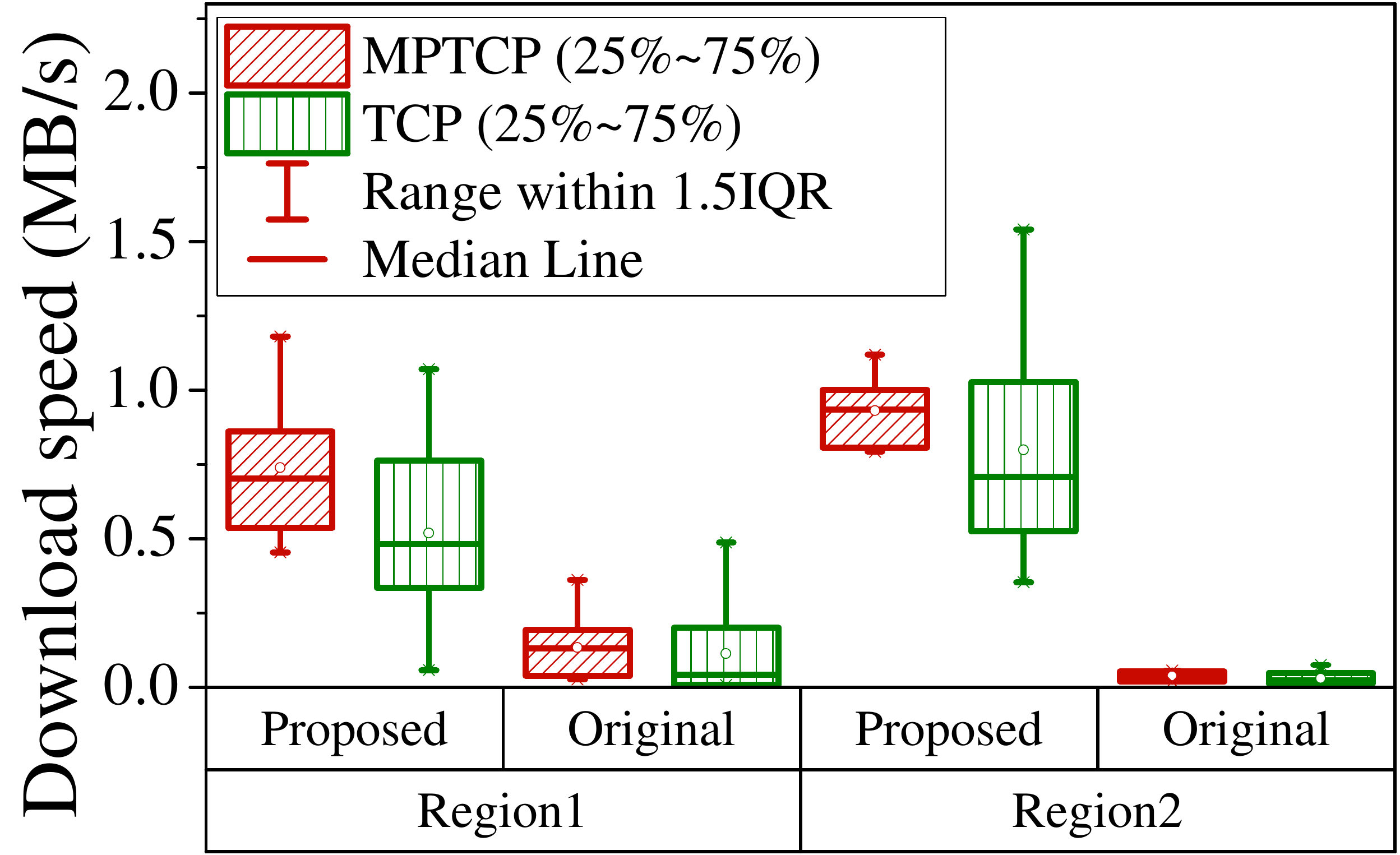}
\caption{Download data from different regions.}\label{proposedrealthroughputplace}
\end{figure}
\begin{figure}[htbp]
\centering
\includegraphics[width=0.7\linewidth]{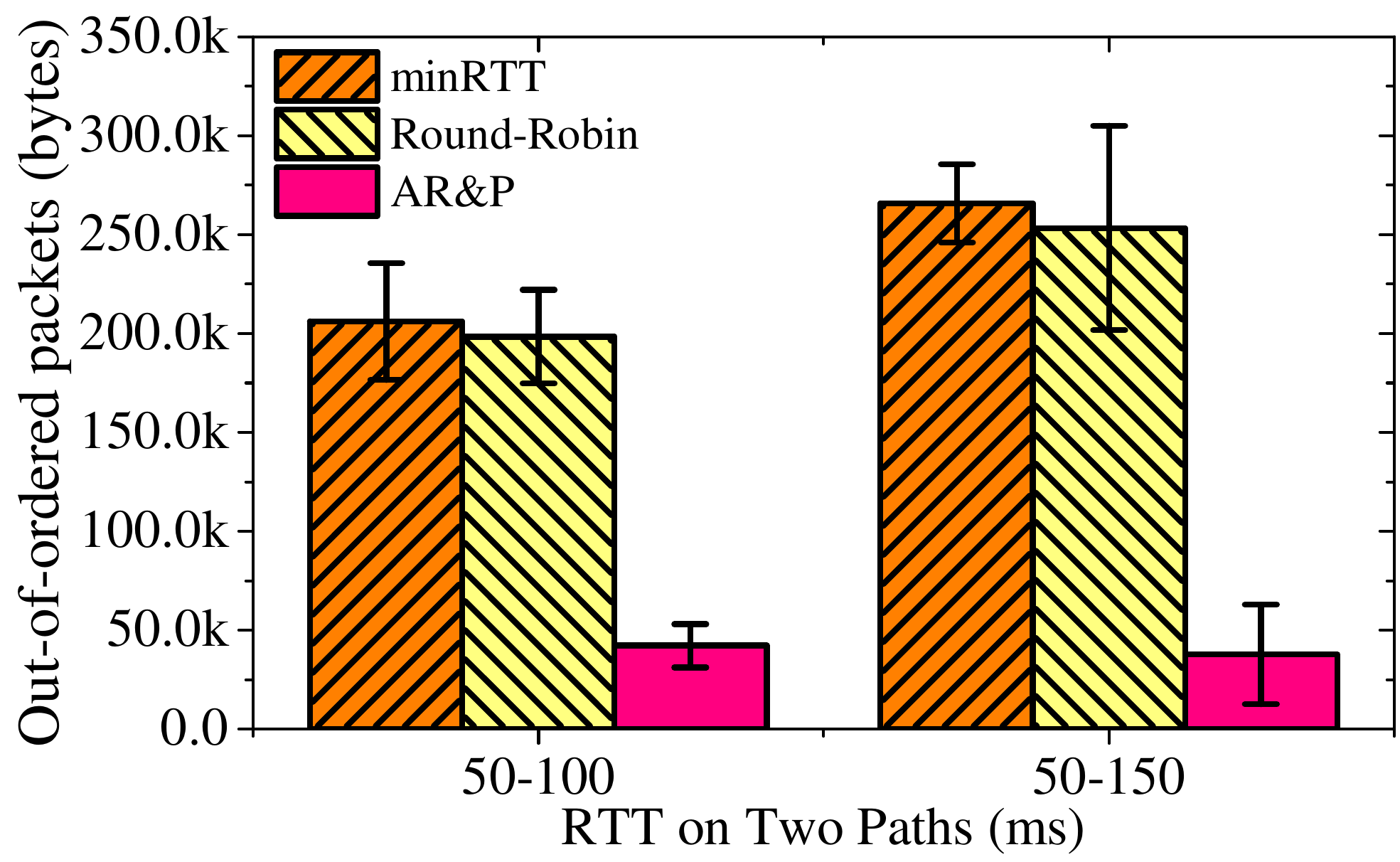}
\caption{Average out-of-order packets in the real networks.}\label{ofoavgreal}
\end{figure}

Moreover, we deploy our scheme on cloud servers in several regions, where the paths suffer large RTT and random loss rate. In this case, our scheme is more outstanding than others. Fig.~\ref{proposedrealthroughputplace} shows the performance result. In this scenario, the throughput of the original MPTCP is less than 0.2MB/s, which is far less than the available bandwidth of devices' interfaces. This is because large packet loss hinders the growth of the congestion window, and the packets in the small congestion window suffer from large RTT transmitted to the receiver. However, wherever the server is, MPTCP with our scheme achieves throughput over 10 times higher than that of original MPTCP, showing the superiority of our scheme in the networks with bad conditions.

Fig.~\ref{ofoavgreal} shows the average out-of-order packets in the real networks. MPTCP server is deployed in two cloud MPTCP servers of different regions. Our client establishes two subflows through which the two servers access 4G and Wi-Fi, respectively, and the RTT of the subflows using the two accesses are shown in Fig.~\ref{ofoavgreal}. In this experiment, our AR\&P scheduler keeps the out-of-order queue short, while minRTT and Round-Robin schedulers create up to 5 times longer out-of-order queue than AR\&P does. When the difference between the two subflows is getting larger, AR\&P does not create a longer out-of-order queue while the other two schedulers do create more out-of-order packets.

  In summary, Coupled BBR and AR\&P Scheduler make MPTCP more feasible in real networks. With our proposed schemes, MPTCP throughput can be improved by up to 2.5 times in normal wireless scenarios and more than 10 times in other scenarios with large RTT and loss. Moreover, the number of out-of-order packets can be reduced by 80\% at most in asymmetric scenarios.

\subsection{Performance Simulation}

\begin{figure}[!htb]
  \centering
  \includegraphics[width=0.88\linewidth]{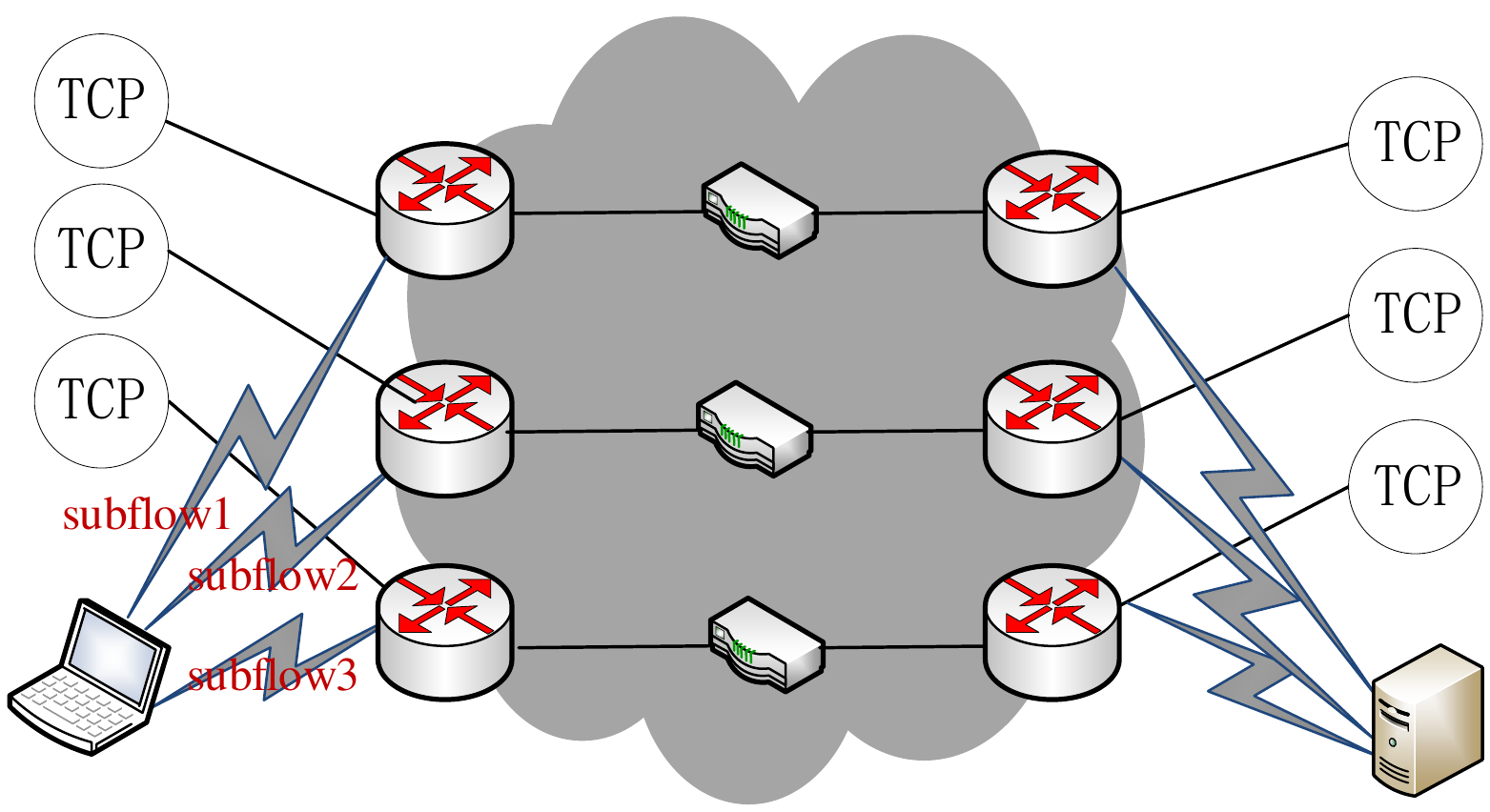}\\
  \caption{Topology of simulation.}\label{simulation-topo}

\end{figure}

We utilize a network topology shown in Fig.~\ref{simulation-topo} to test the performance of Coupled BBR and AR\&P Scheduler. The MPTCP connection includes three subflows, each of which passes through a path with bandwidth of $B_i$ on the bottleneck. There is one TCP flow that passes through the same path of each subflow. The link delay and random packet loss rate are set to $d_i$ and $p_i$ of each path $i$.

\begin{figure}[!htb]
  \centering

\subfigure[]{
\begin{minipage}[t]{0.465\linewidth}
\centering
\includegraphics[width=1.0\linewidth]{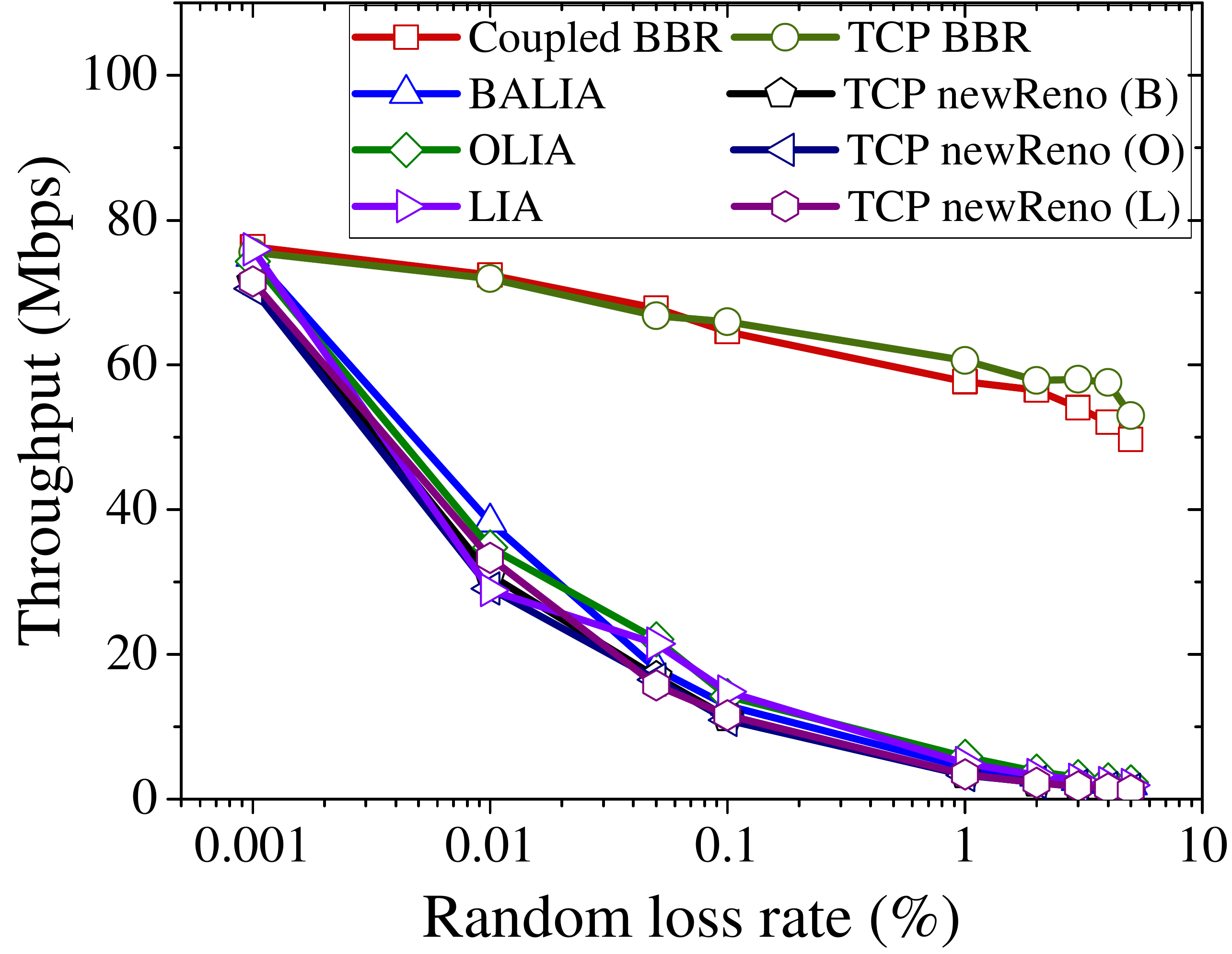}\label{sim-cbbr1}
\end{minipage}
}
\subfigure[]{
\begin{minipage}[t]{0.465\linewidth}
\centering
\includegraphics[width=1.0\linewidth]{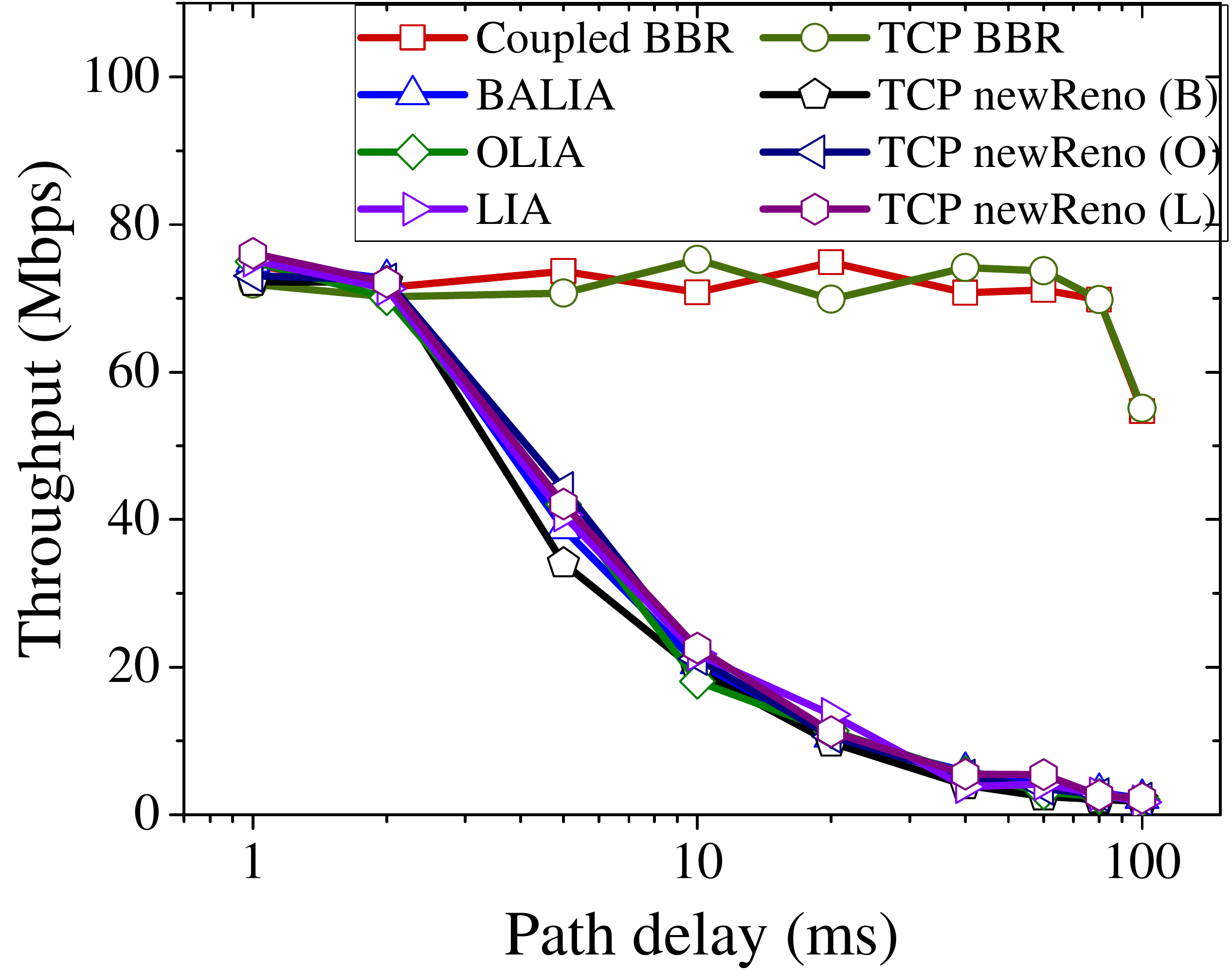}\label{sim-cbbr2}
\end{minipage}
}

\caption{Performance in lossy network scenarios, (a) Random loss rate changes, (b) Path delay changes.
}\label{sim-cbbr}
\end{figure}

By utilizing the bottleneck bandwidth detection method, Coupled BBR provides high bandwidth utilization, especially in lossy and long-delay networks. As shown in Fig.~\ref{sim-cbbr}, we compare Coupled BBR with LIA, BALIA, and OLIA in different network scenarios. TCP BBR, TCP newReno (B), (O), (L) are the background TCP flow on the MPTCP Coupled BBR, BALIA, OLIA, LIA, respectively. We set $B_1=B_2=B_3=100$ Mbps, $d_1=d_2=d_3=20$ ms and $p_1=p_2=p_3=0-0.5\%$, respectively. The overall bandwidth resource of the network is 300 Mbps. When the random loss rate is 0, both Coupled BBR and other MPTCP congestion control algorithms can achieve high bandwidth utilization. With the increase of random loss rate, the throughput of Coupled BBR almost does not decline. The throughput of other MPTCP congestion control algorithms declines significantly. In Fig.~\ref{sim-cbbr2}, we set $B_1=B_2=B_3=100$ Mbps, $d_1=d_2=d_3=2-100$ ms and $p_1=p_2=p_3=0.05\%$, respectively. When the path delay is low, the link random packet loss does not make a great effect on the original congestion algorithms. With the increase of path delay, the throughput of LIA, OLIA, BALIA declines significantly, while Coupled BBR still achieves high throughput and bandwidth utilization.

\begin{figure}[!htb]
  \centering

\subfigure{
\begin{minipage}[t]{0.465\linewidth}
\centering
\includegraphics[width=1\linewidth]{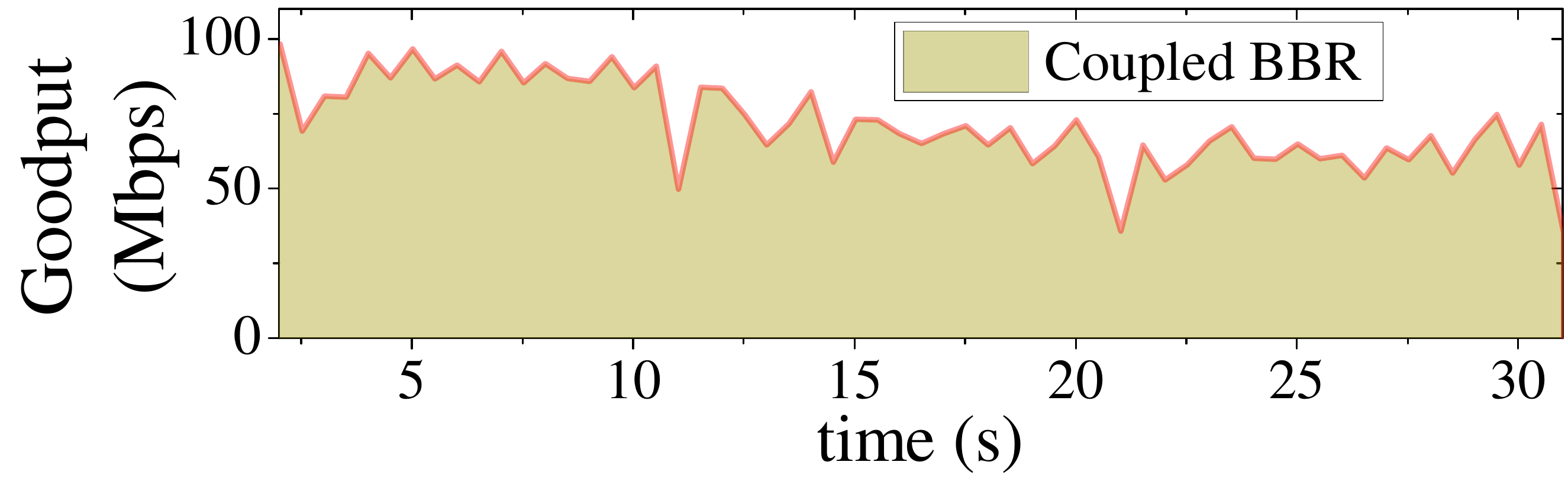}
\end{minipage}
}
\subfigure{
\begin{minipage}[t]{0.465\linewidth}
\centering
\includegraphics[width=1\linewidth]{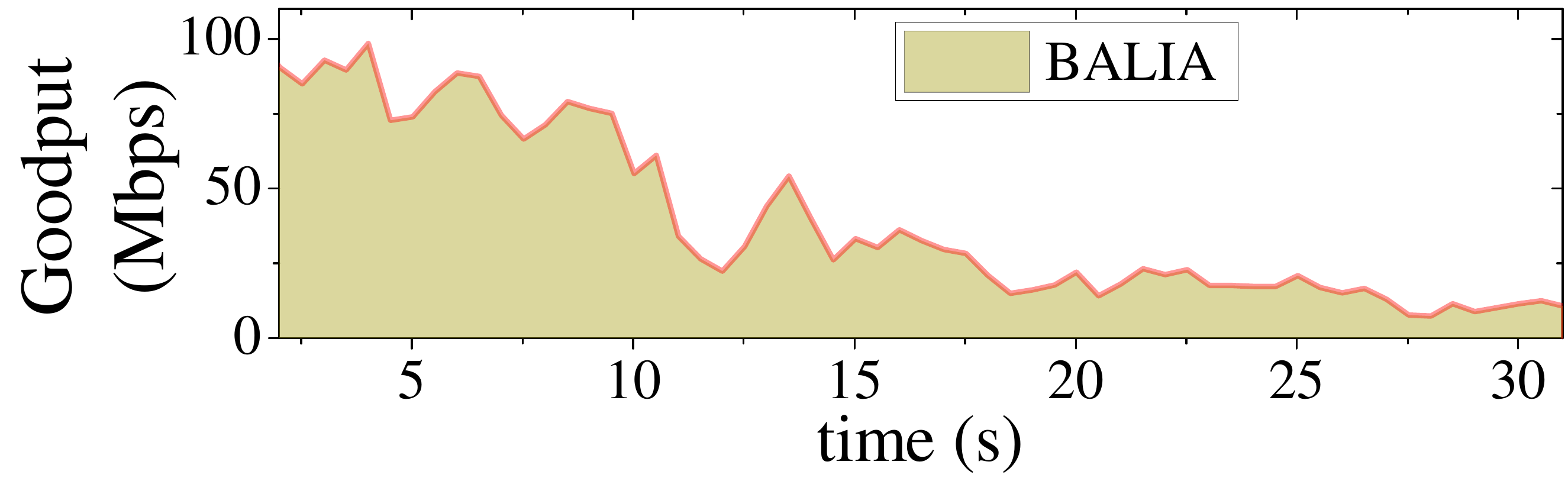}
\end{minipage}
}
\subfigure{
\begin{minipage}[t]{0.465\linewidth}
\centering
\includegraphics[width=1\linewidth]{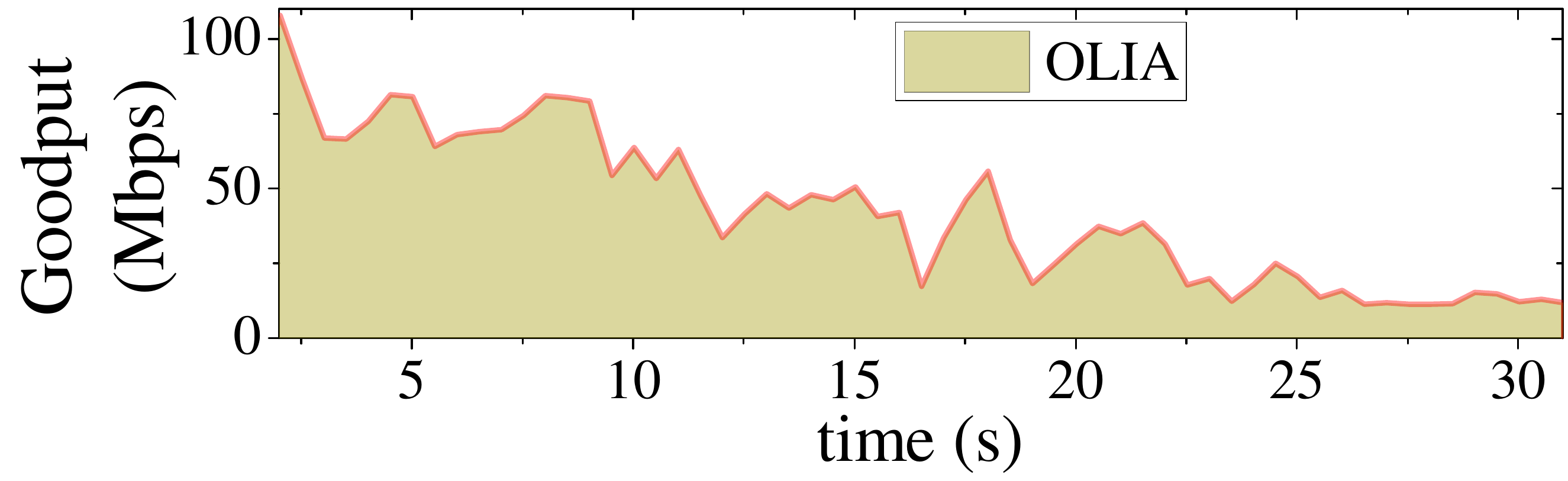}
\end{minipage}
}
\subfigure{
\begin{minipage}[t]{0.465\linewidth}
\centering
\includegraphics[width=1\linewidth]{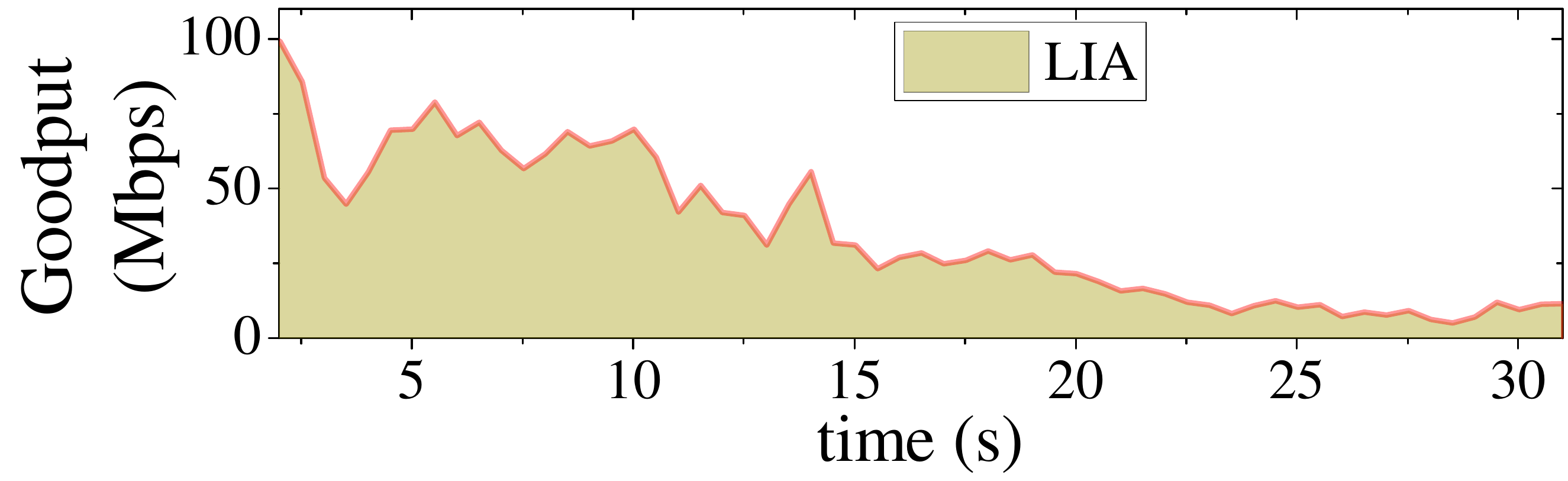}
\end{minipage}
}

\caption{Real-time throughput of different algorithms in an ever-changing lossy network scenario.
}\label{sim-cbbr-realtime}
\end{figure}

 Fig.~\ref{sim-cbbr-realtime} shows the performance of Coupled BBR in ever-changing networks. We set $B_1=B_2=B_3=100$ Mbps, $d_1=d_2=d_3=20$ ms and $p_1=p_2=p_3=0.01\%$ at beginning of the transmission. During the transmission, $p_1$, $p_2$ and $p_3$ linearly change to 1\% from 10 s to 30 s. With the increase of path loss rate, the throughput of BALIA, OLIA, LIA decreases significantly. Compare with them, Coupled BBR has always maintained high throughput.

\begin{figure}[!htb]
  \centering
\subfigure[]{
\begin{minipage}[t]{0.465\linewidth}
\centering
\includegraphics[width=1.0\linewidth]{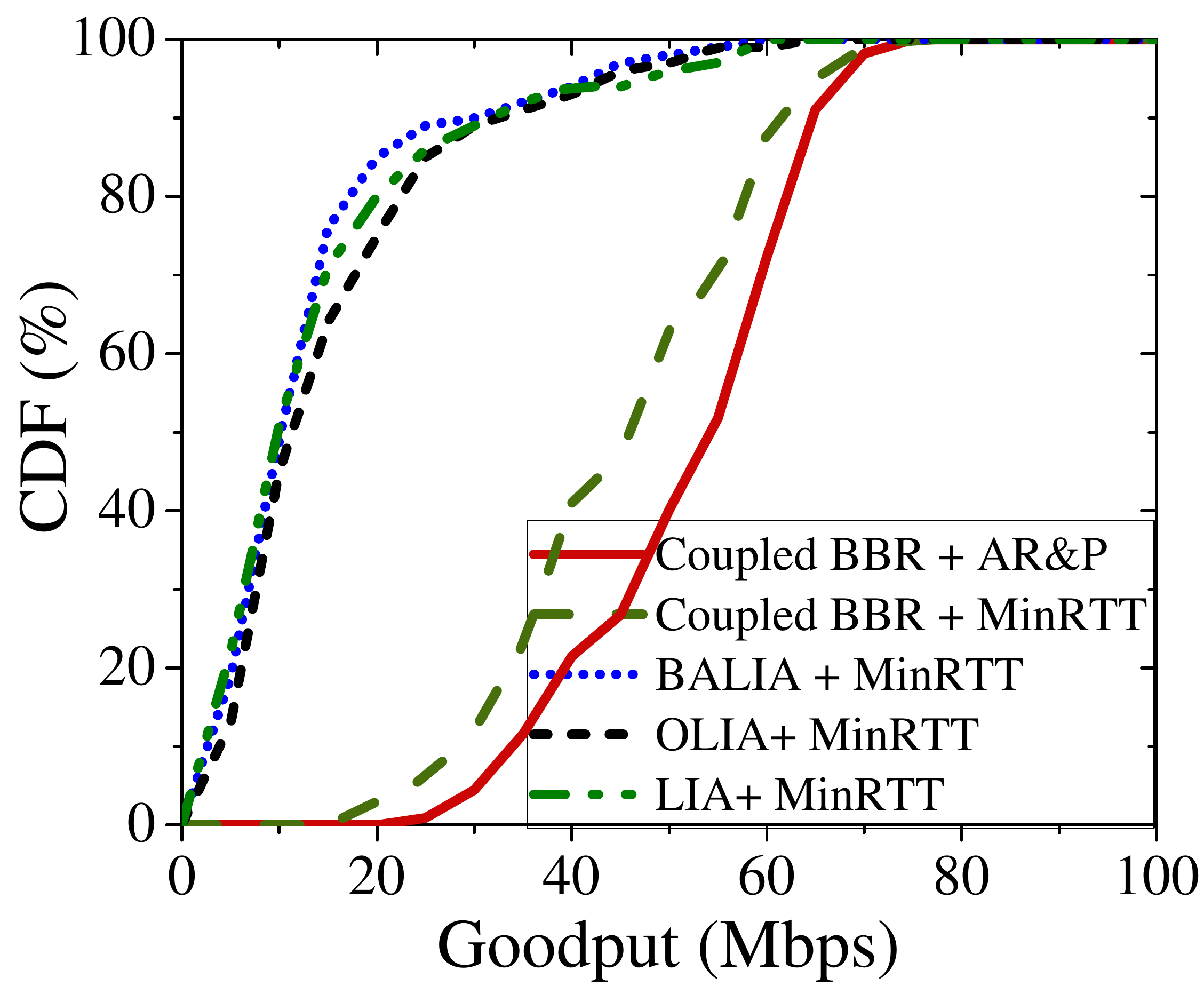}\label{sim-arp1}
\end{minipage}
}
\subfigure[]{
\begin{minipage}[t]{0.465\linewidth}
\centering
\includegraphics[width=1.0\linewidth]{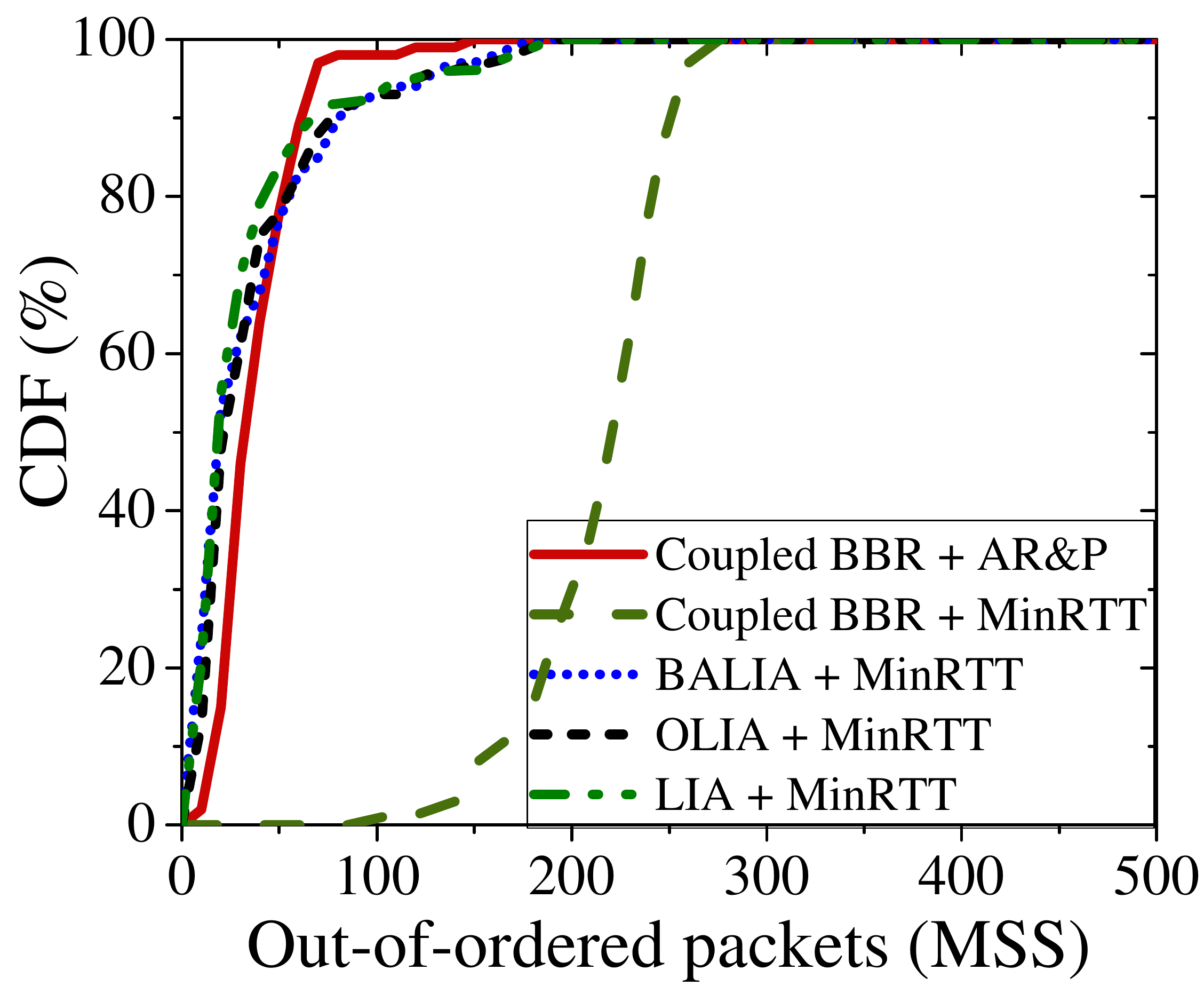}\label{sim-arp2}
\end{minipage}
}

\caption{Performance in various network scenarios, (a) Average goodput, (b) Average out-of-ordered packets.}\label{sim-arp}
\end{figure}

 Fig.~\ref{sim-arp} shows the performance of AR\&P Scheduler in various networks. We change $B_1, B_2, B_3$ from 10 to 100 Mbps, $d_1,d_2,d_3$ from 1 to 100 ms and $p_1, p_2, p_3$ from 0 to 5\%. We randomly pick up 100 points in the parameter space for simulation and analyze the simulation results. As shown in Fig.~\ref{sim-arp1}, the throughput of Coupled BBR with both minRTT and AR\&P is much higher than that of BALIA, OLIA, and LIA in random loss scenarios. Among them, AR\&P further improves the throughput based on Coupled BBR. Compare with BALIA, OLIA and LIA using the same scheduler minRTT, Coupled BBR with minRTT has much higher out-of-order packets. Because the number of out-of-order packets is related to throughput and Coupled BBR significantly improves the throughput. AR\&P scheduler further improves the goodput and reduce the number of out-of-order packets on the basis of Coupled BBR. Compare with BALIA, OLIA and LIA with minRTT, Coupled BBR with AR\&P gains much higher goodput with the same low-level out-of-order packets, which can be seen in Fig.~\ref{sim-arp2}.

\begin{figure}[!htb]
  \centering

 \subfigure{
\begin{minipage}[t]{0.465\linewidth}
\centering
\includegraphics[width=1.0\linewidth]{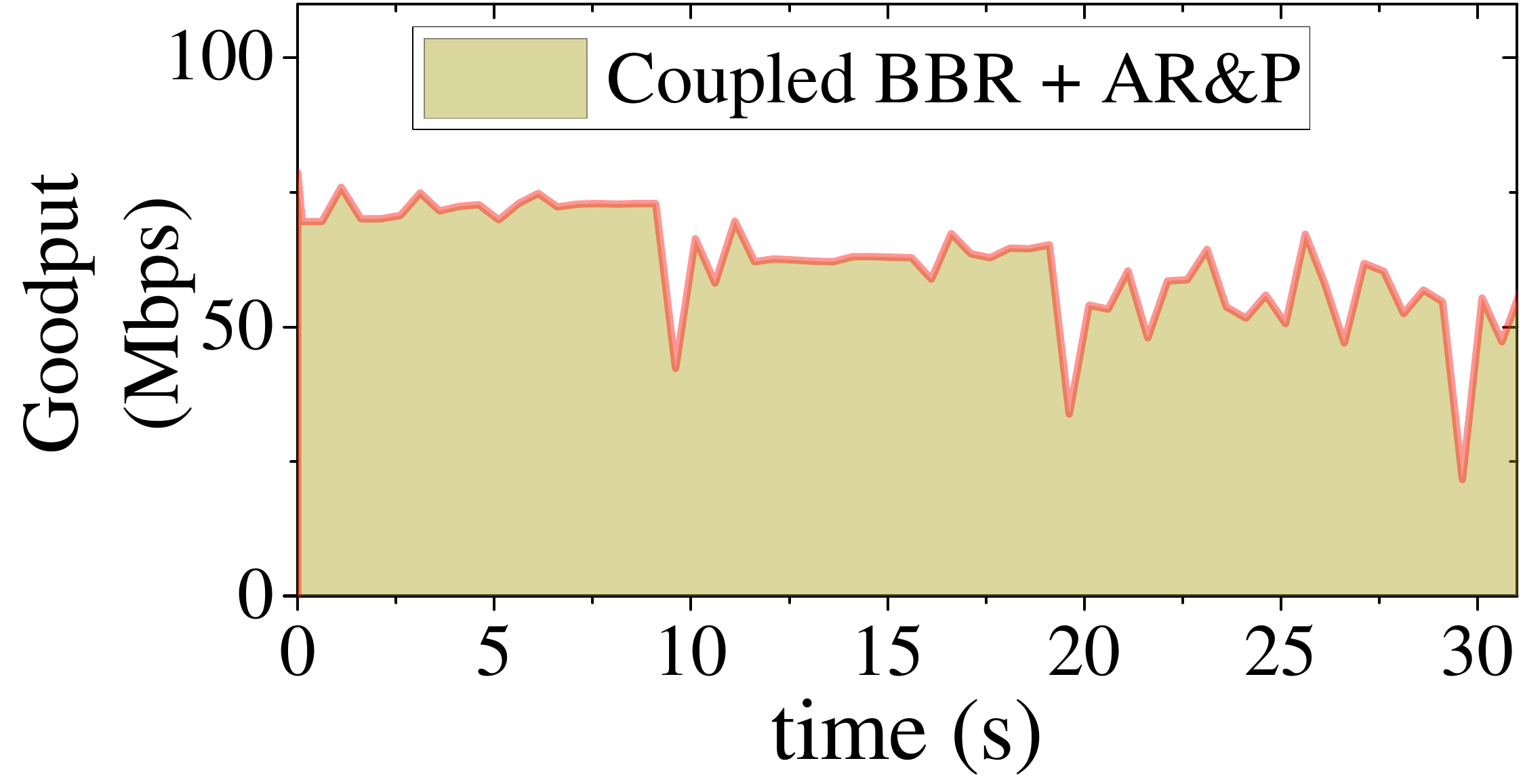}
\end{minipage}
}
\subfigure{
\begin{minipage}[t]{0.465\linewidth}
\centering
\includegraphics[width=1.0\linewidth]{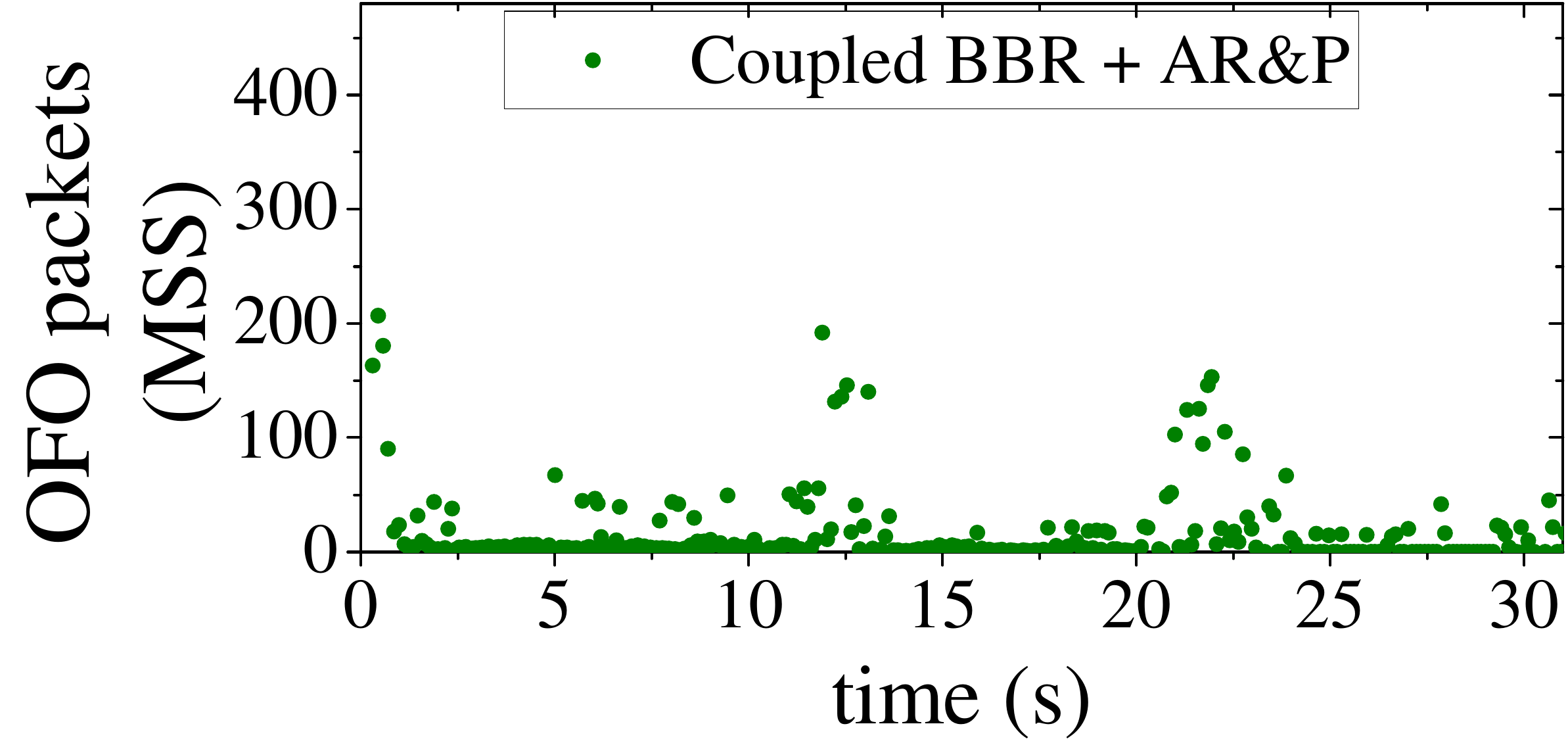}
\end{minipage}
}

\subfigure{
\begin{minipage}[t]{0.465\linewidth}
\centering
\includegraphics[width=1.0\linewidth]{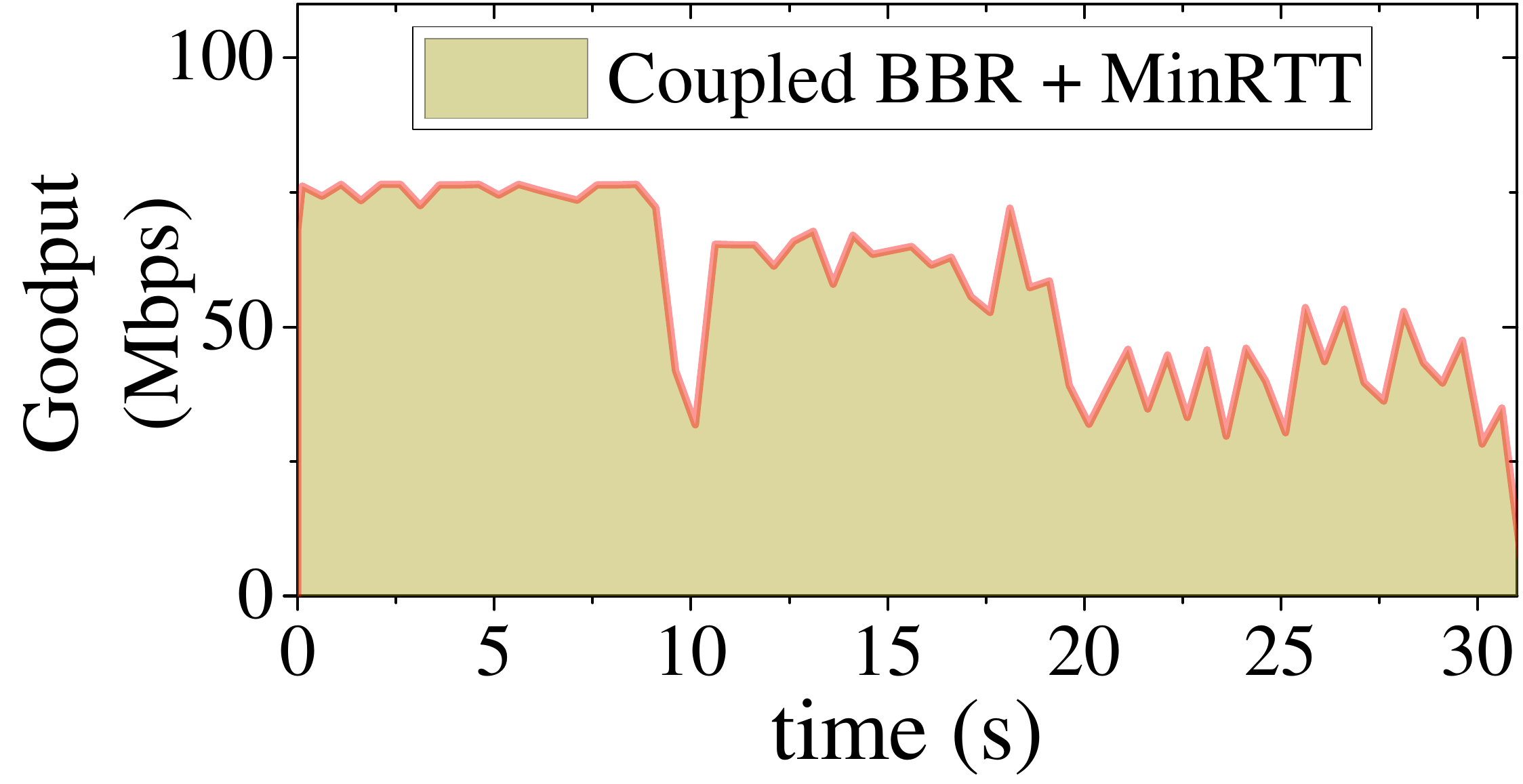}
\end{minipage}
}
\subfigure{
\begin{minipage}[t]{0.465\linewidth}
\centering
\includegraphics[width=1.0\linewidth]{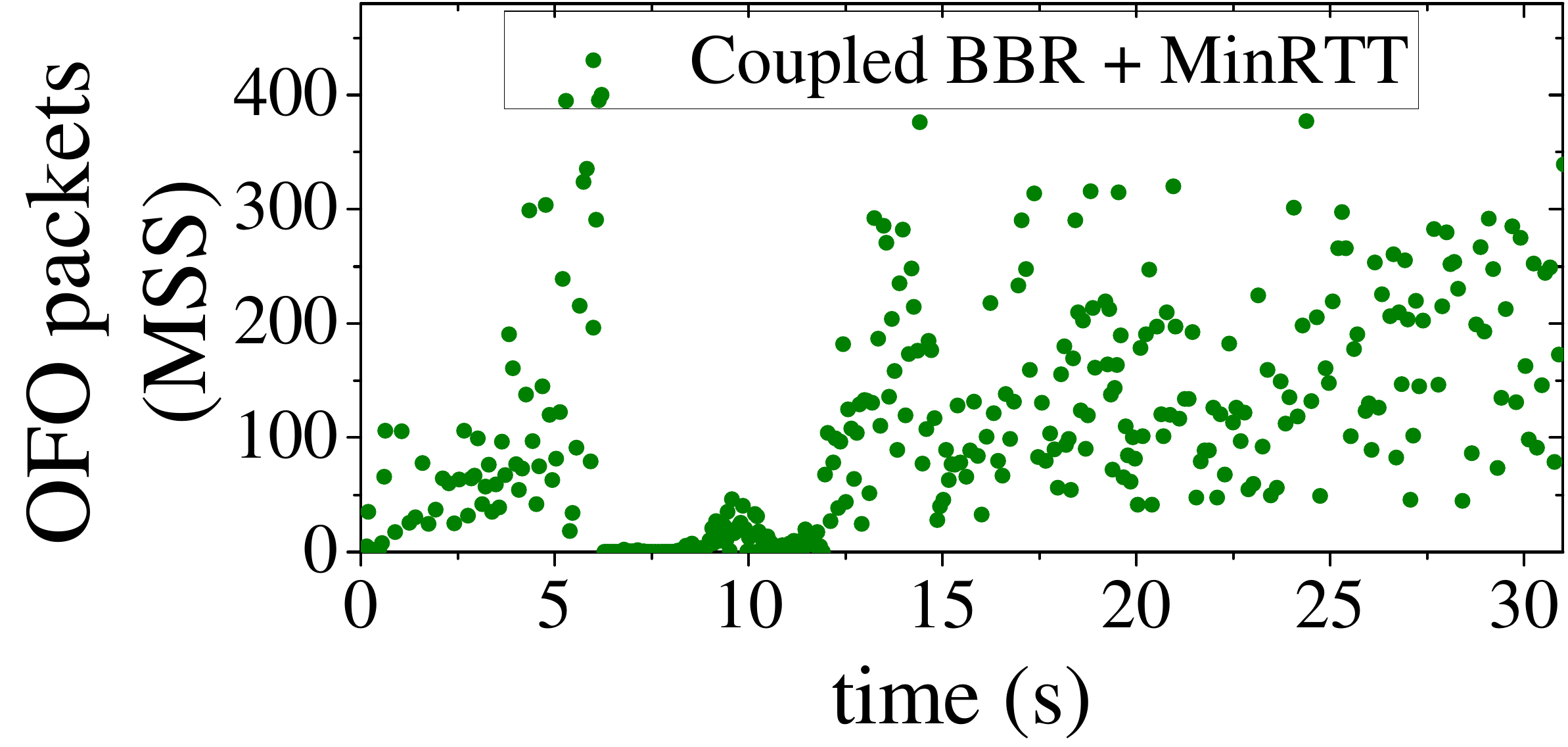}
\end{minipage}
}

\subfigure{
\begin{minipage}[t]{0.465\linewidth}
\centering
\includegraphics[width=1.0\linewidth]{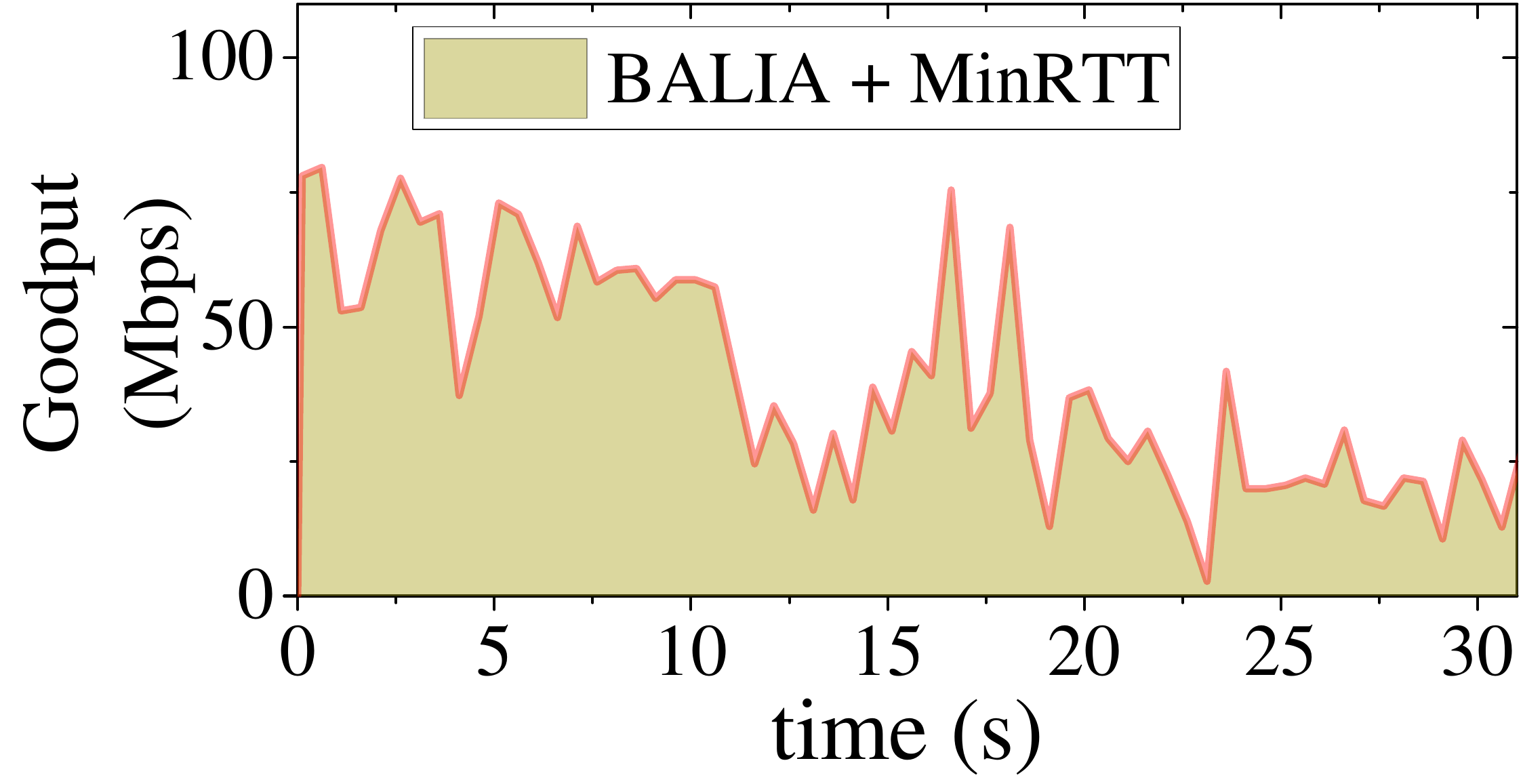}
\end{minipage}
}
\subfigure{
\begin{minipage}[t]{0.465\linewidth}
\centering
\includegraphics[width=1.0\linewidth]{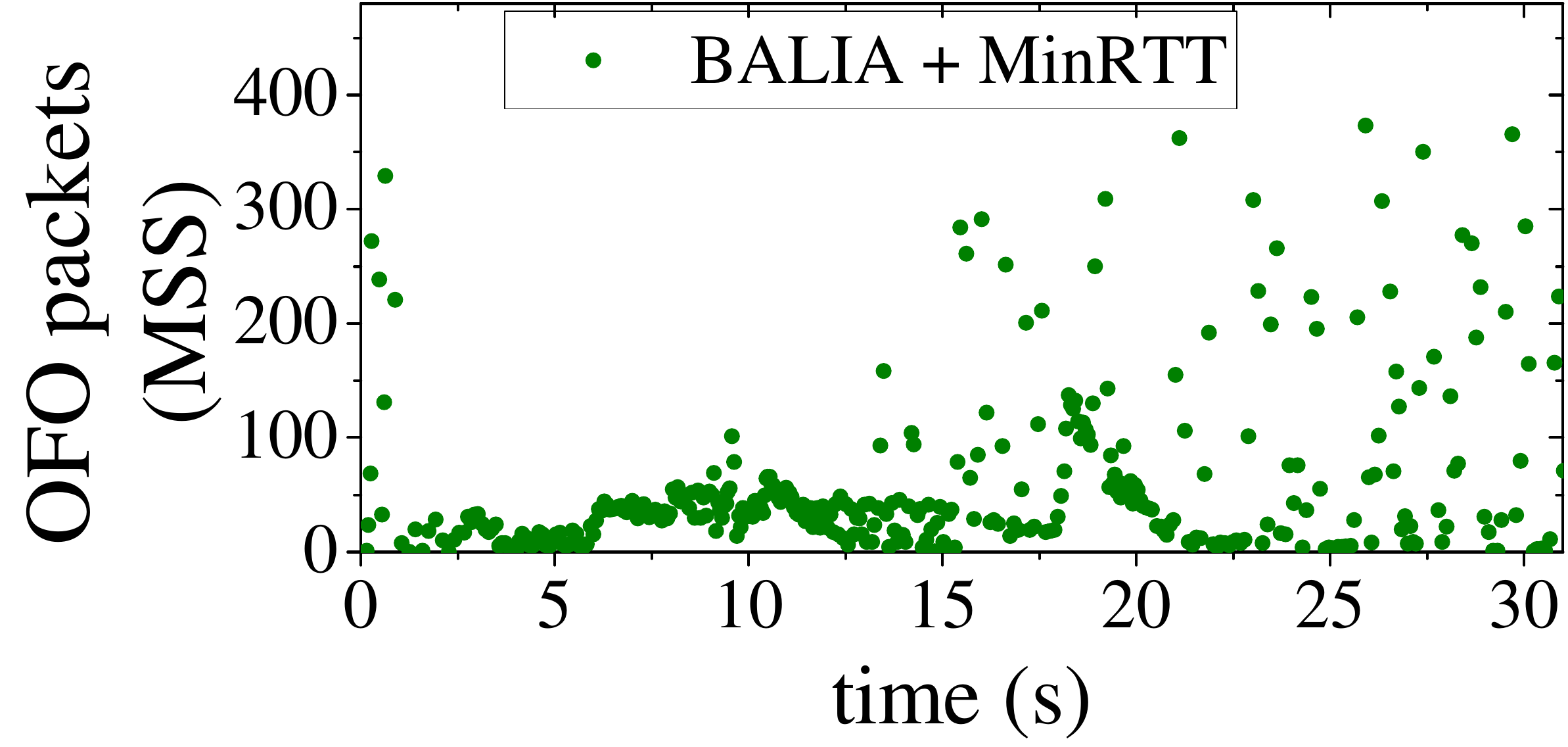}
\end{minipage}
}
\subfigure{
\begin{minipage}[t]{0.465\linewidth}
\centering
\includegraphics[width=1.0\linewidth]{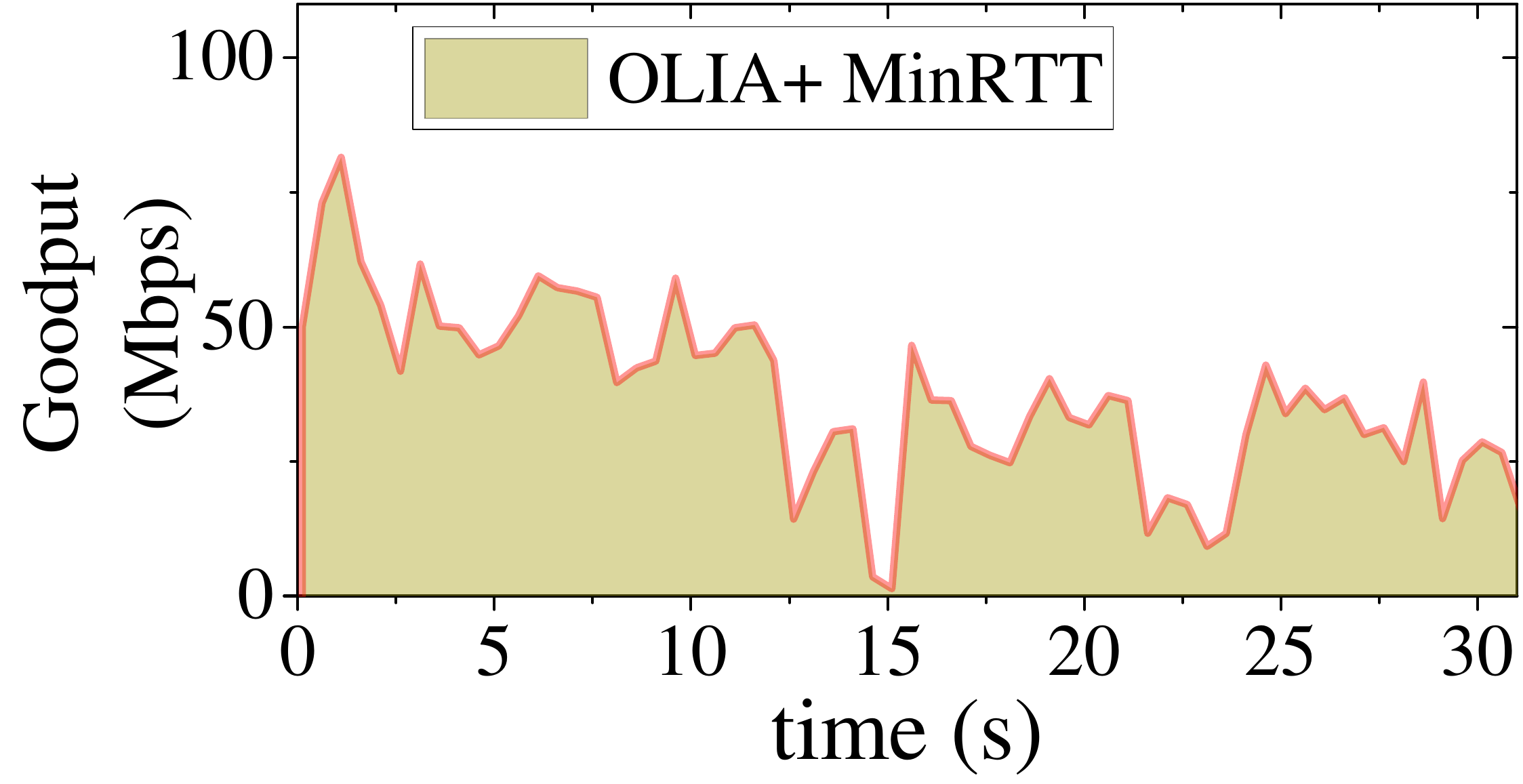}
\end{minipage}
}
\subfigure{
\begin{minipage}[t]{0.465\linewidth}
\centering
\includegraphics[width=1.0\linewidth]{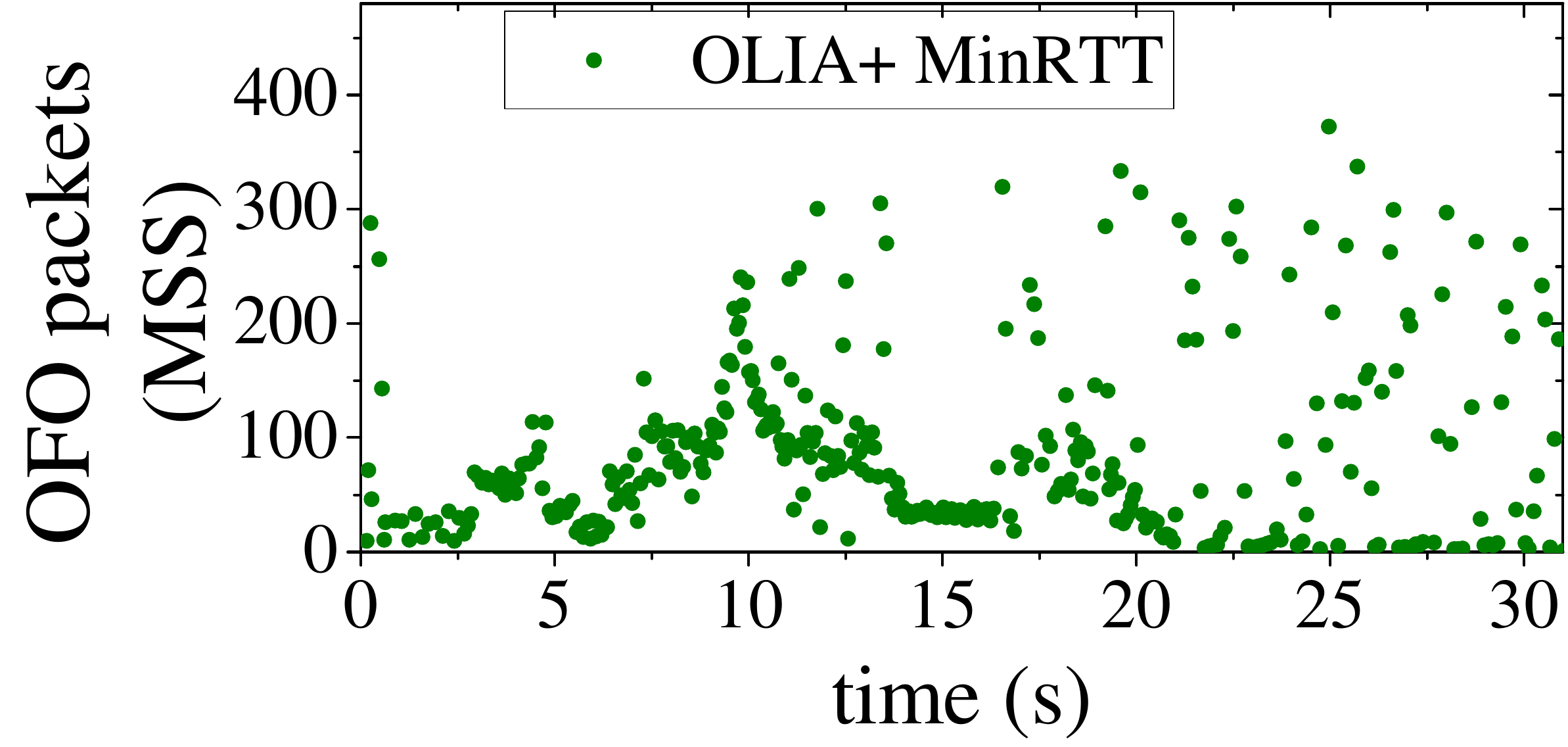}
\end{minipage}
}
\subfigure{
\begin{minipage}[t]{0.465\linewidth}
\centering
\includegraphics[width=1.0\linewidth]{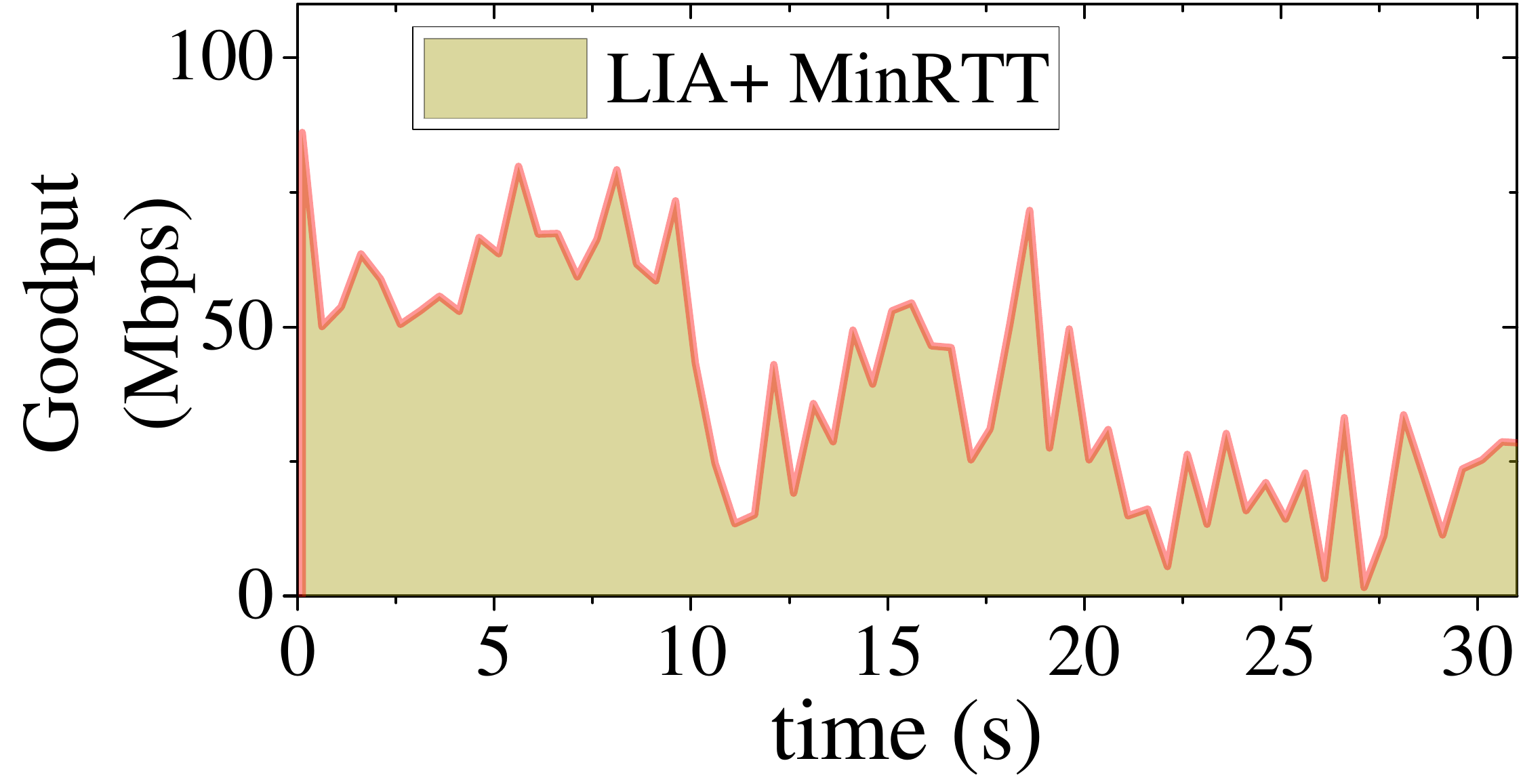}
\end{minipage}
}
\subfigure{
\begin{minipage}[t]{0.465\linewidth}
\centering
\includegraphics[width=1.0\linewidth]{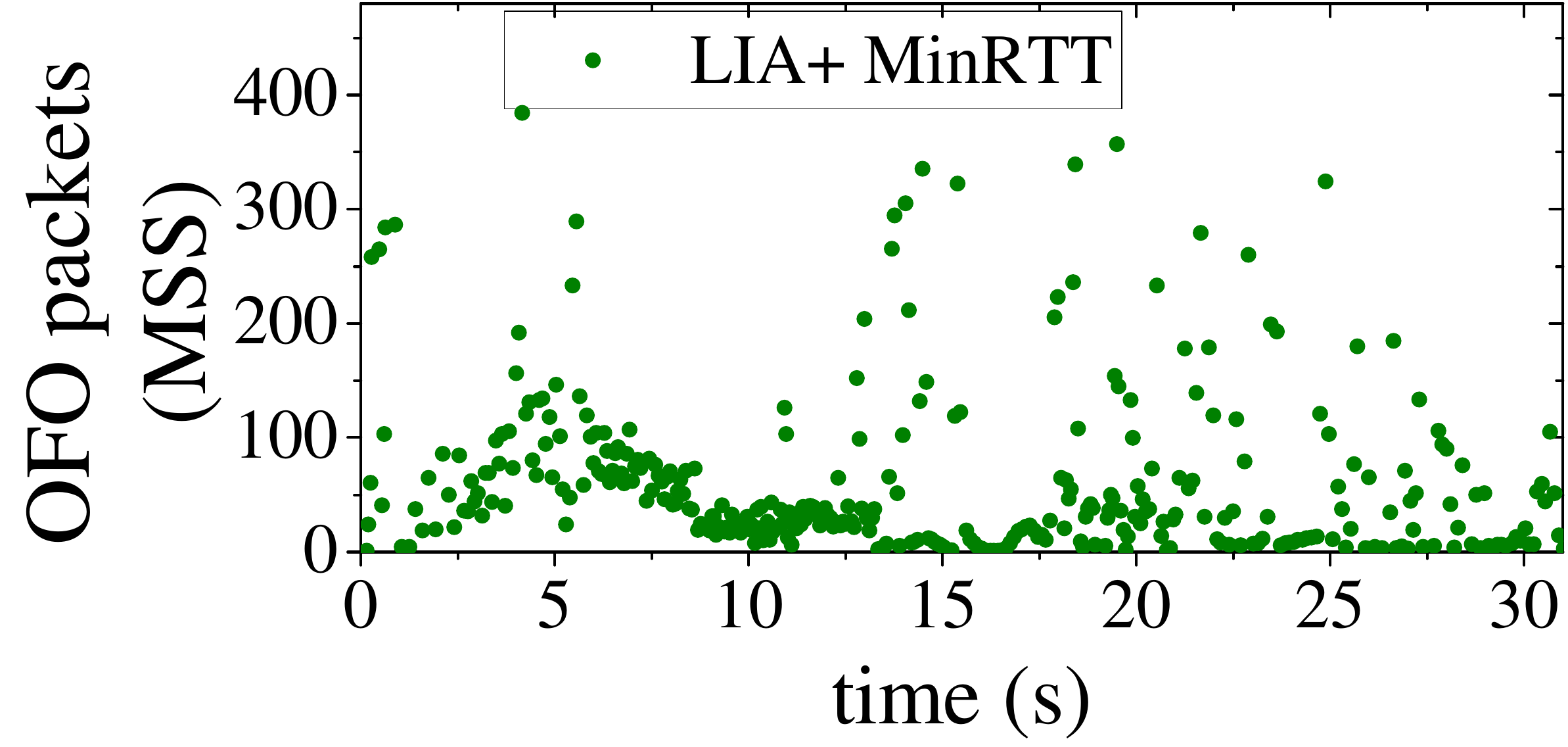}
\end{minipage}
}

\caption{Real-time goodput and out-of-ordered packets of different algorithms in an ever-changing network scenario.}\label{sim-arp-realtime}
\end{figure}

 Fig.~\ref{sim-arp-realtime} shows the performance of AR\&P Scheduler in an ever-changing network, where OFO packets denote out-of-ordered packets. We set $B_1=B_2=B_3=100$ Mbps, $d_1=d_2=d_3=20$ ms and $p_1=p_2=p_3=0.01\%$ at beginning of the transmission. During the transmission, $B_2$, $d_2$ and $p_2$ change to 10 Mbps, 100 ms and 1\% from 10 s to 15 s. $B_3$, $d_3$ and $p_3$ change to 10 Mbps, 100 ms and 1\% from 20 s to 25 s. BALIA, OLIA, and LIA with minRTT all experience low goodput when one of the paths is changing. When the delay of one of the paths increases (after 10 s), their out-of-ordered packets also increase significantly. Compare with them, Coupled BBR with minRTT improves the goodput, while the number of out-of-ordered packets also increases due to the high throughput. Coupled BBR with AR\&P further promotes the goodput and significantly reduce out-of-ordered packets when path conditions change. When the path conditions begin to change, P-Scheduling method helps to schedule packets and reduces out-of-ordered packets. When the path condition continues to deteriorate, AR-Scheduling decides to send redundant packets on the subflow with bad path conditions. Therefore the out-of-ordered packets are further reduced.

\section{Related Work}\label{Related}

\textbf{Congestion control algorithms:}  The basic goal of coupled congestion control algorithms in MPTCP is to achieve fairness with TCP flows, but it also needs to further achieve additional goals such as congestion balance. Current coupled congestion control algorithms, such as LIA \cite{wischik2011design}, OLIA \cite{khalili2013mptcp}, and BALIA \cite{peng2016multipath} couple the congestion control algorithms on different subflows by linking their increase function in AIMD based on TCP NewReno. For every RTT on subflow$_i$, coupled congestion control algorithms increase the congestion window $w_i$ by a parameter $\alpha_i$ instead of 1 in NewReno. Thus in the network with a certain loss rate, different speeds of window increasing will lead to different overall throughput. However, as BBR does not include the AIMD method, any AIMD-based scheme is not suitable for developing MPTCP over BBR congestion control.  Some RL-based algorithms, such as \cite{li2019smartcc} \cite{xu2019experience}, would provides better performance by learning from a large number of data. However, their are computationally complex and require a lot of CPU resources, which makes them are easy to be deployed.

\textbf{Scheduling algorithms:} Scheduling algorithms are mainly designed for improving robustness, reducing the out-of-order packets, or reducing latency. Based on the operating patterns, they can be divided into several categories: 1) Simple schedulers in Linux Kernel \cite{linuxkernel}, which are Round-Robin, minRTT, and Redundant. Round-Robin polls subflows and sends packets in order. minRTT always sends packets on the available subflow with the lowest RTT. Redundant sends redundant packets to ensure high robustness and low latency. 2) Schedulers acting on paths. This kind of scheduler improves MPTCP performance by controlling each path's action \cite{han2016mp, saha2019musher, lee2018raven}. For example, Musher\cite{saha2019musher} controls the allocation rate of data on each path to get better throughput. RAVEN \cite{lee2018raven} mitigates tail latency by using redundant transmission when confidence about network latency predictions is low. 3) Schedulers acting on packets \cite{xue2018dpsaf, lim2017ecf, shi2018stms, hurtig2018low}. These proposed schedulers, like ECF \cite{lim2017ecf}, STMS \cite{shi2018stms}, STTF \cite{hurtig2018low}, aim at keeping low latency and reducing out-of-order packets in asymmetric networks. They schedule packets with the larger sequence number to the subflow with larger RTT so as to keep packets delivery in order. However, existing scheduling algorithms are based on traditional congestion control algorithms, which in turn depends on the congestion window, and thus do not work for Coupled BBR.

\section{Discussion}\label{Discussion}

  In this work, we study the fairness between MPTCP Coupled BBR flows and TCP BBR flows. We consider a network that uses BBR to control all the flows so that the network is more stable and all the flows can get better performance. A full BBR network provides more advantages for developing higher performance transmission protocols in the future. Moreover, in lossy networks, traditional loss-based congestion control algorithms could not make good use of the available bandwidth of the bottleneck, which makes it meaningless to achieve fairness between BBR and other algorithms in this case. Moreover, we address network fairness. To be noted that, for bottleneck fairness, MPTCP subflows sharing one bottleneck should be coupled to achieve fairness with TCP flows in the same bottleneck, and it just needs a bottleneck detection method for Coupled BBR. Then our scheme can be easily adapted for it.

\section{Conclusion}\label{conclusion}

 In this work, we propose Coupled BBR and AR\&P scheduler to improve the performance of MPTCP in lossy or ever-changing networks. With Coupled BBR, MPTCP not only performs well in lossy circumstances but also balances congestion among subflows and achieves fairness to TCP BBR flows. AR\&P scheduler further enhances MPTCP performance in dynamic and asymmetric networks with two scheduling methods to provide better self-adaptability and reduce the out-of-order packets.

\section*{Acknowledgment}
The work of Jiangping Han, Kaiping Xue, Yitao Xing, Jian Li and Wenjia Wei was supported in part by the National Natural Science Foundation of China (NSFC) under Grant No. 61972371, and Youth Innovation Promotion Association of the Chinese Academy of Sciences (CAS) under Grant No. 2016394. The work of Guoliang Xue was supported in part by National Science Foundation (NSF) under Grant No. 1704092. The information reported here does not reflect the position or the policy of the funding agencies.

\bibliographystyle{IEEEtran}
\bibliography{reference}

\begin{IEEEbiography}
[{\includegraphics[width=1in,clip,keepaspectratio]{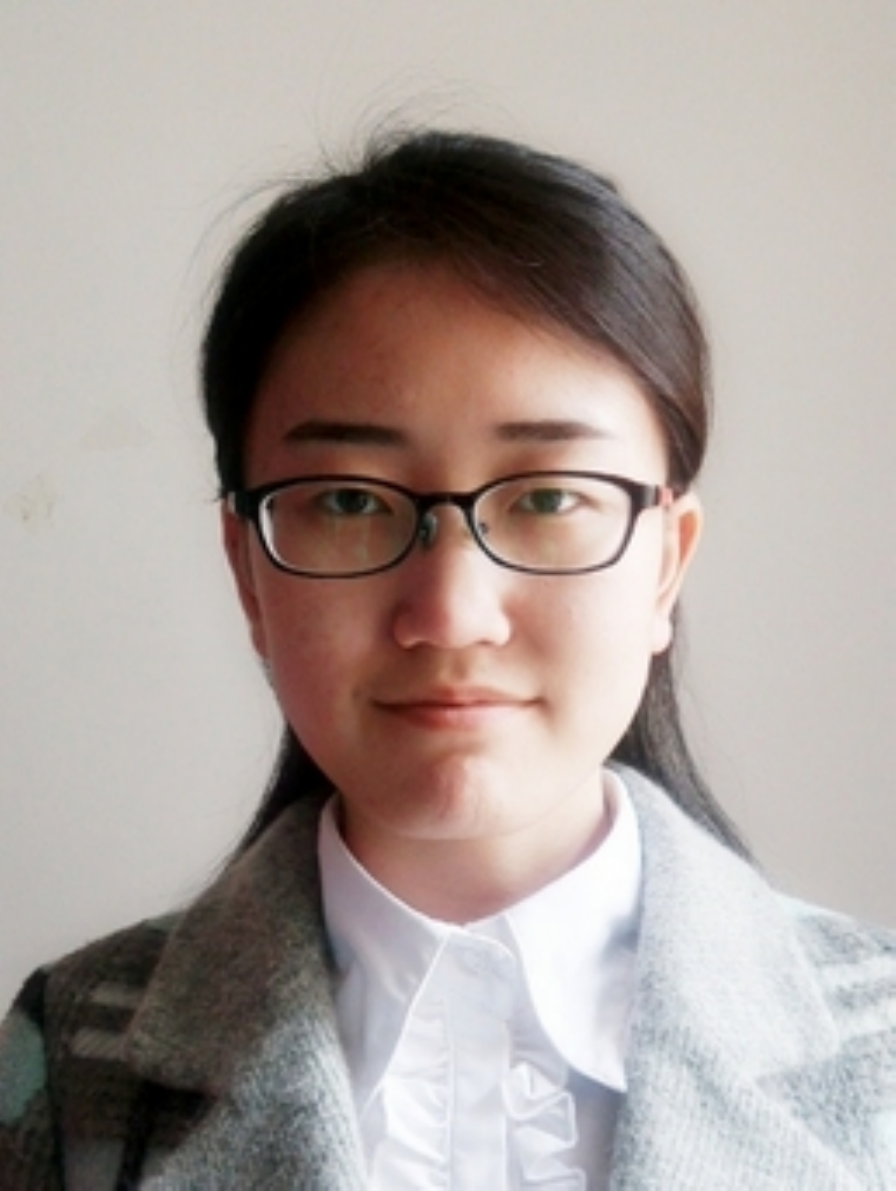}}]{Jiangping Han} receives her B.S. degree from the Department of Electronic Engineering and Information Science (EEIS), USTC, in July, 2016. She is currently working toward the Ph.D degree in communication and information systems with also the Department of EEIS, USTC. Her research interests include future Internet architecture design and transmission optimization. Part of this work was done while she was a visiting student at Arizona State University.
\end{IEEEbiography}

\begin{IEEEbiography}
[{\includegraphics[width=1in,clip,keepaspectratio]{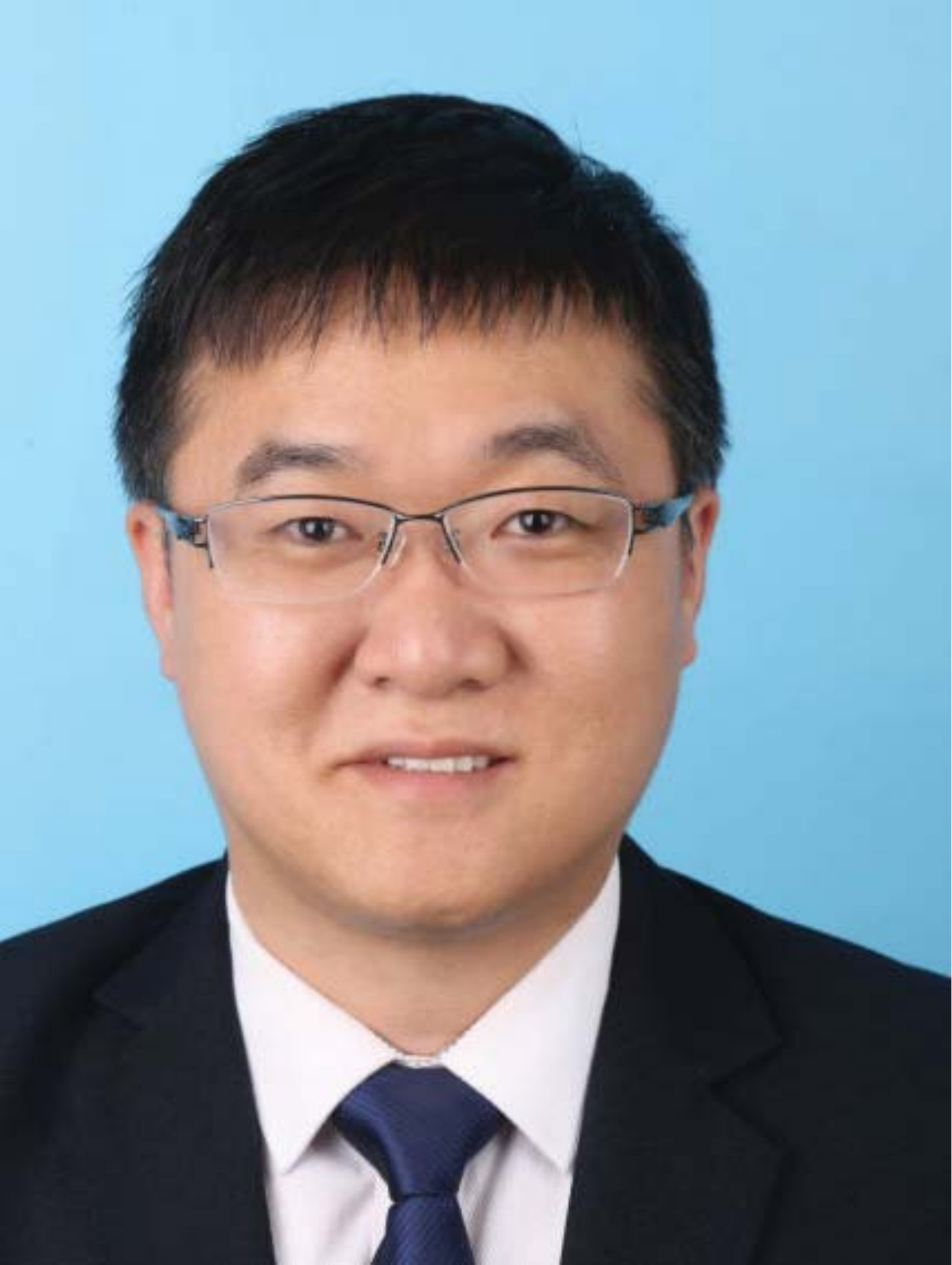}}]{Kaiping Xue}(M'09-SM'15) received his bachelor's degree from the Department of Information Security, University of Science and Technology of China (USTC), in 2003 and received his Ph.D. degree from the Department of Electronic Engineering and Information Science (EEIS), USTC, in 2007. From May 2012 to May 2013, he was a postdoctoral researcher with the Department of Electrical and Computer Engineering, University of Florida. Currently, he is a Professor in the School of Cyber Security and the Department of EEIS, USTC. His research interests include next-generation Internet, distributed networks and network security. He serves on the Editorial Board of several journals, including the IEEE Transactions on Wireless Communications (TWC), the IEEE Transactions on Network and Service Management (TNSM), and Ad Hoc Networks. He has also served as a guest editor of IEEE Journal on Selected Areas in Communications (JSAC) and a lead guest editor of IEEE Communications Magazine. He is an IET Fellow and an IEEE Senior Member. He is the corresponding author of this paper.
\end{IEEEbiography}

\begin{IEEEbiography}
[{\includegraphics[width=1in,clip,keepaspectratio]{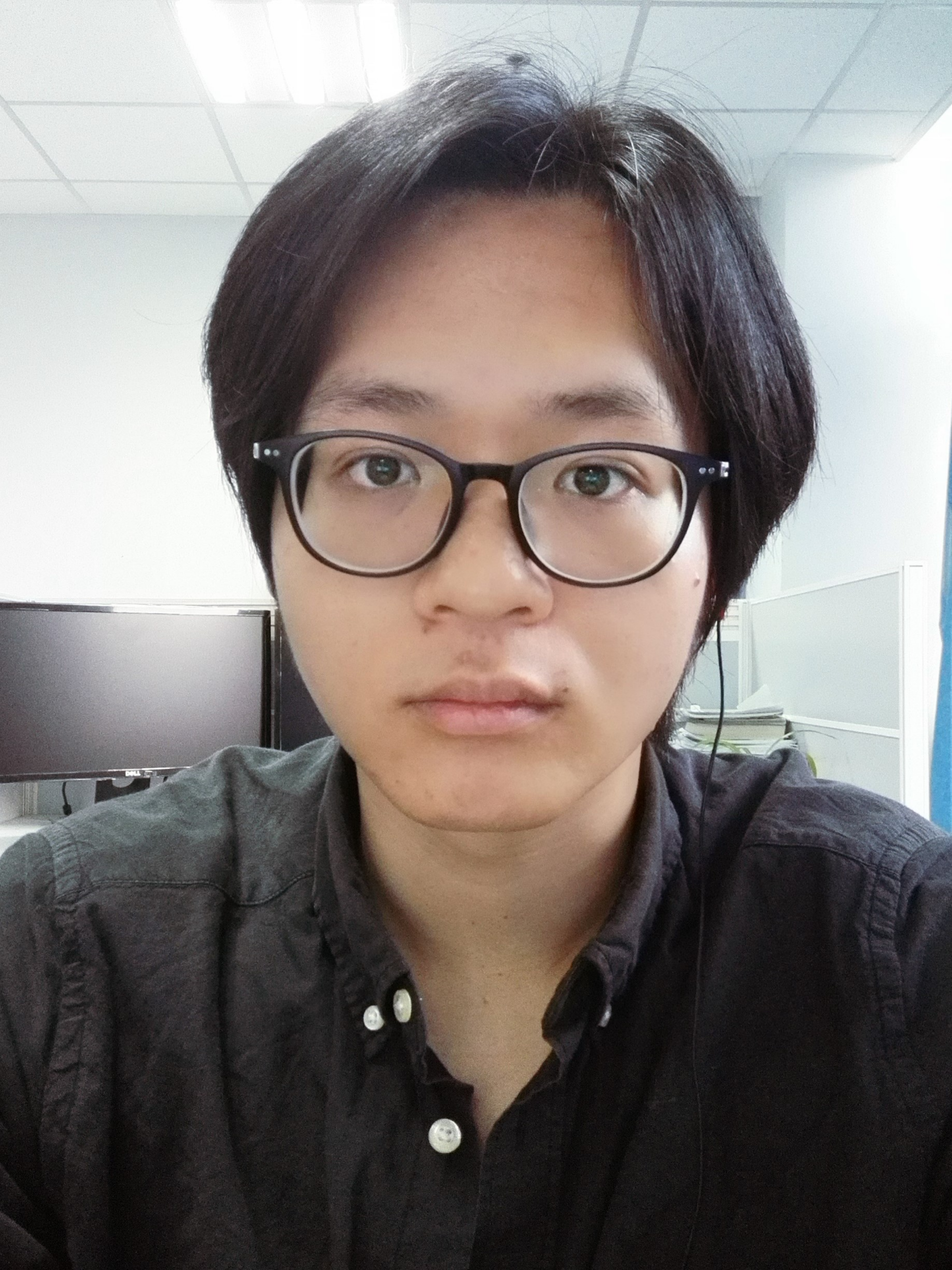}}]{Yitao Xing} received his B.S. degree in Information Security from School of the Gifted Young, University of Science and Technology of China (USTC), in 2018. He is currently a graduated student in Communication and Information System from the Department of Electronic Engineering and Information Science (EEIS), USTC. His research interests include future Internet architecture and transmission optimization.
\end{IEEEbiography}

\begin{IEEEbiography}[{\includegraphics[width=1in,height=1.25in,clip,keepaspectratio]{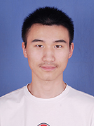}}]{Jian Li}  received his B.S. degree from the Department of Electronics and Information Engineering, Anhui University, in 2015, and received Ph.D degree from the Department of Electronic Engineering and Information Science (EEIS), University of Science and Technology of China (USTC), in 2020. From Nov. 2019 to Nov. 2020, he was a visiting scholar with the Department of Electronic and Computer Engineering, University of Florida. He is currently a Post-Doctoral researcher with the Department of EEIS, USTC. His research interests include wireless communications, satellite networks and next-generation Internet.
\end{IEEEbiography}

\begin{IEEEbiography}
[{\includegraphics[width=1in,clip,keepaspectratio]{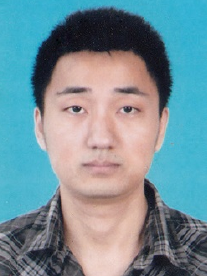}}]{Wenjia Wei} received the B.S. degree from the school of infomation  science and engineering, in 2013. He received the Ph.D. degree in Information and Communication Engineering from the Department of Electronic Engineering and Information Science (EEIS), University of Science and Technology of China (USTC), in 2020. His research interests include future Internet architecture design and transmission optimization.
\end{IEEEbiography}

\begin{IEEEbiography}[{\includegraphics[width=1in,height=1.25in,clip,keepaspectratio]{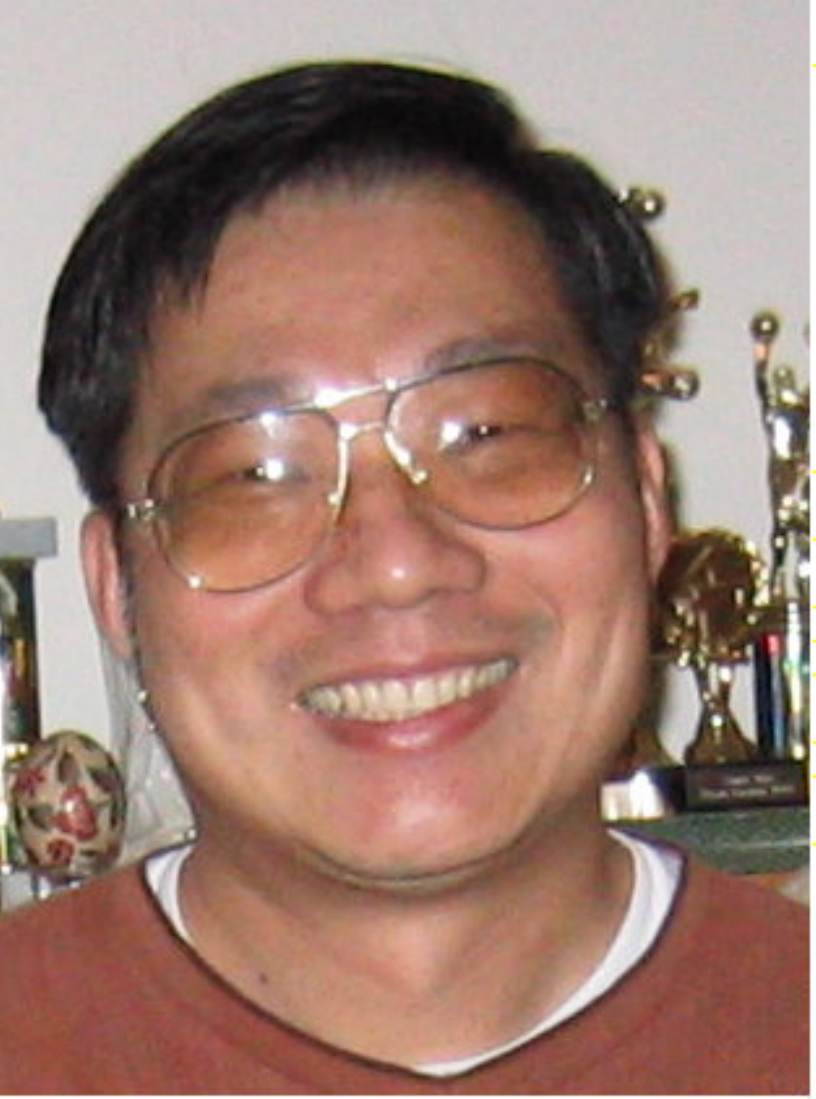}}]{David S.L. Wei} (SM'07)  received his Ph.D. degree in Computer and Information Science from the University of Pennsylvania in 1991. From May 1993 to August 1997 he was on the Faculty of Computer Science and Engineering at the University of Aizu, Japan (as an Associate Professor and then a Professor). He has authored and co-authored more than 120 technical papers in various archival journals and conference proceedings. He is currently a Professor of Computer and Information Science Department at Fordham University. His research interests include cloud computing, big data, IoT, and cognitive radio networks. He was a guest editor or a lead guest editor for several special issues in the IEEE Journal on Selected Areas in Communications, the IEEE Transactions on Cloud Computing and the IEEE Transactions on Big Data. He also served as an Associate Editor of IEEE Transactions on Cloud Computing, 2014-2018, and an Associate Editor of Journal of Circuits, Systems and Computers, 2013-2018.
\end{IEEEbiography}

\begin{IEEEbiography}[{\includegraphics[width=1in,height=1.25in,clip,keepaspectratio]{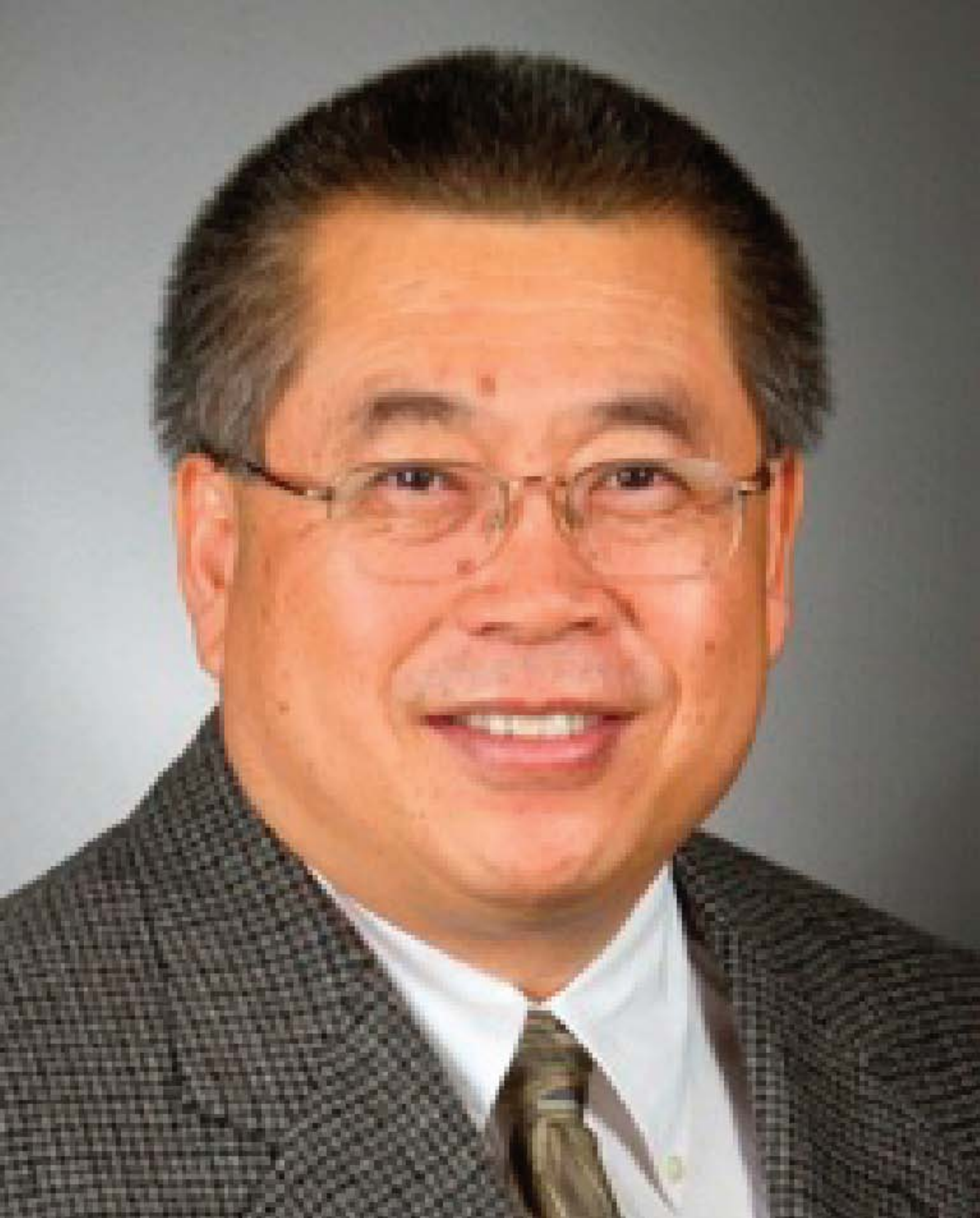}}]{Guoliang Xue} (M'96-SM'99-F'11) is a professor of Computer Science and Engineering at Arizona State University. He received the Ph.D degree in Computer Science from the University of Minnesota in 1991. His research interests span the areas of QoS provisioning, machine learning, wireless networking, and IoT. He has received the IEEE Communications Society William R. Bennett Prize in 2019. He has served as a TPC Chair of IEEE INFOCOM'2010, IEEE Globecom'2020, and IEEE IWQoS'2021, and a General Chair of IEEE CNS'2014 and IEEE/ACM IWQoS'2019. He has served on the TPC of many conferences, including ACM CCS, ACM MOBIHOC, IEEE ICNP, and IEEE INFOCOM. He served on the editorial board of IEEE/ACM Transactions on Networking and the Area Editor of IEEE Transactions on Wireless Communications, overseeing 12 editors in the Wireless Networking area. He is an IEEE Fellow, and the Steering Committee Chair of IEEE INFOCOM.
\end{IEEEbiography}

\end{document}